\documentclass[fleqn,usenatbib]{mnras}

\usepackage{newtxtext,newtxmath}

\usepackage[T1]{fontenc}

\DeclareRobustCommand{\VAN}[3]{#2}
\let\VANthebibliography\thebibliography
\def\thebibliography{\DeclareRobustCommand{\VAN}[3]{##3}\VANthebibliography}

\usepackage{graphicx}	
\usepackage{amsmath}	
\usepackage{bm}

\newcommand{\hi}{H\textsc{i}}
\newcommand{\althi}{H{\normalfont\textsc{i}}}

\title[Interferometric IM]{Towards Optimal Foreground Mitigation Strategies for Interferometric \althi\ Intensity Mapping in the Low-Redshift Universe}

\author[Z. Chen et al.]{
Zhaoting Chen,$^{1}$\thanks{E-mail: zhaoting.chen@manchester.ac.uk}
Laura Wolz,$^{1}$
and Richard Battye$^{1}$
\\
$^{1}$Jodrell Bank Centre for Astrophysics, School of Physics and Astronomy, The University of Manchester, Manchester M13 9PL, UK\\
}

\date{Accepted XXX. Received YYY; in original form ZZZ}

\pubyear{2022}

\begin{document}
\label{firstpage}
\pagerange{\pageref{firstpage}--\pageref{lastpage}}
\maketitle

\begin{abstract}
We conduct the first case study towards developing optimal foreground mitigation strategies for neutral hydrogen (\hi) intensity mapping using radio interferometers at low redshifts. A pipeline for simulation, foreground mitigation and power spectrum estimation is built, which can be used for ongoing and future surveys using \textit{MeerKAT} and Square Kilometre Array Observatory (SKAO). It simulates realistic sky signals to generate visibility data given instrument and observation specifications, which is subsequently used to perform foreground mitigation and power spectrum estimation. A quadratic estimator formalism is developed to estimate the temperature power spectrum in visibility space. Using \textit{MeerKAT} telescope specifications for observations in the redshift range, $z\sim 0.25-0.30$, corresponding to the MeerKAT International GHz Tiered Extragalactic Exploration (MIGHTEE) survey, we present a case study, where we compare different approaches of foreground mitigation. We find that component separation in visibility space provides a more accurate estimation of \hi\ clustering comparing to foreground avoidance, with the uncertainties being 30\% smaller. Power spectrum estimation from image is found to be less robust with larger bias and more information loss when compared to estimation in visibility. {We conclude that for the considered sub-band of $z\sim 0.25-0.30$, the MIGHTEE survey will be capable of measuring the \hi\ power spectrum from $k\sim 0.5\,{\rm Mpc^{-1}}$ to $k\sim 10\,{\rm Mpc^{-1}}$ with signal-to-noise ratio being $\sim 3$}. {We are the first to show that, at low redshift, component separation in visibility space suppresses foreground contamination at large line-of-sight scales, allowing measurement of \hi\ power spectrum closer to the foreground wedge, crucial for data analysis towards future detections.}
\end{abstract}




\section{Introduction}
\label{sec:intro}
Measuring the distribution of dark matter in the Universe and its evolution is one of the most important objectives of observational cosmology. The clustering of dark matter on cosmological scales, i.e. the cosmic large scale structure (LSS), can shed light on the nature of dark energy and dark matter (see e.g. \citealt{1980lssu.book.....P}). A wide range of tracers of dark matter, i.e. observables that trace the underlying density of dark matter, can be used in order to probe this distribution. These probes follow the clustering of dark matter linearly at large cosmological scales. Such probes can be galaxies, such as galaxy number count (e.g. \citealt{2021PhRvD.103h3533A}), weak lensing of galaxies (e.g. \citealt{2021arXiv210513545P}) and more. Alternatively, one can also use a relatively new technique called intensity mapping (IM, e.g. \citealt{2004MNRAS.355.1339B, 2008PhRvL.100i1303C, 2008PhRvD..78b3529M,2009MNRAS.397.1926W,2013MNRAS.434.1239B,2017arXiv170909066K}). It uses the emission lines of elements that are abundant in the Universe as tracers of dark matter, most promisingly neutral hydrogen (\hi). \hi\ is initially the most abundant element in the Universe as predicted by big bang nucleosynthesis (see e.g. \citealt{2003moco.book.....D}). The emission line caused by spin-flip transition of \hi\ has a rest wavelength of around 21\,cm, and thus can be observed in the radio band with low risks of line confusion. By mapping the distribution of the flux density across the sky, statistical inference on the underlying cosmology and astrophysics of \hi\ sources can be made (e.g. \citealt{2015ApJ...803...21B,2021MNRAS.502.5259C}).

\hi\ IM survey with large survey volume and coarse angular resolution can be used for cosmological measurements at low redshifts, such as baryon acoustic oscillations (BAO, \citealt{1998ApJ...496..605E}) and redshift-space distortions (RSD, \citealt{1987MNRAS.227....1K}). Single dish telescopes and dish/cylinder arrays operating in single dish mode can thus be powerful tools for \hi\ IM, such as \textit{Five-Hundred-Meter Aperture Spherical Radio Telescope} (FAST, \citealt{2020MNRAS.493.5854H}), \textit{MeerKAT} \citep{2017arXiv170906099S}, and future Square Kilometre Array Observatory (SKAO, \citealt{2020PASA...37....7S}).

No detection of the autocorrelation of \hi\ using single dish mode has been claimed but, using single dish telescopes such as the \textit{Green Bank Telescope} (GBT), statistically significant detections have been made by cross-correlation of the IM signal from \hi\ with optical galaxies from the WiggleZ survey \citep{2013ApJ...763L..20M, 2013MNRAS.434L..46S}. Results from cross-correlating the 2dF Galaxy Survey and \hi\ maps from the \textit{Parkes} radio telescope are used to confirm the relation between the star forming properties of galaxies and its \hi\ mass \citep{2018MNRAS.476.3382A}. Single dish observations are capable of providing precise measurements of \hi\ properties beyond the local Universe, as demonstrated by the results from cross-correlating eBOSS galaxies with \textit{GBT} data \citep{2022MNRAS.510.3495W}. Cross-correlation using the Canadian Hydrogen Intensity Mapping Experiment (CHIME) interferometer has been claimed by stacking the IM signal using eBOSS optical galaxy catalogues \citep{2022arXiv220201242C}.

Although single dish observations are expected to be the primary method for IM at low redshifts, interferometry can play an important role as well in probing smaller scales. Accessing non-linear scales of clustering helps further extract cosmological information (e.g. \citealt{2019MNRAS.485.4060P,2020JCAP...05..005D}), and the scales typically smaller than the size of dark matter halos can yield information about astrophysics of \hi\ galaxies \citep{2019MNRAS.484.1007W,2021MNRAS.502.5259C,2021JCAP...05..068S}. The combination of cosmological and astrophysical information in the non-linear scales of \hi\ clustering  provides strong incentives to interferometric IM. 

Surveys using radio interferometers, such as the MeerKAT International GHz Tiered Extragalactic Exploration (MIGHTEE) survey \citep{2016mks..confE...6J}, can be used to measure the \hi\ power spectrum. For surveys like MIGHTEE, it is believed that \hi\ clustering can be measured with high precision across a range of redshifts \citep{2021MNRAS.505.2039P}.

\hi\ can also be used to probe the epoch of reionization (EoR) at $z\gtrsim6$. During this period, structures such as the first stars and galaxies form and emit high energy photons to ionise the \hi\ inside the intergalactic medium. The high-redshift and low-frequency range of EoR naturally call for antenna array interferometers \citep{1997ApJ...475..429M}, such as Precision Array for Probing the Epoch of Reionization (PAPER, \citealt{2014ApJ...788..106P}), Murchison Widefield Array (MWA, \citealt{2019ApJ...884....1B}), Low-Frequency Array (LOFAR, \citealt{2017ApJ...838...65P}), and Hydrogen Epoch of Reionization Array (HERA, \citealt{2017PASP..129d5001D}). These arrays have relatively wide field-of-view (FOV) to balance the need for deep observations and large survey volumes.

Despite probing for very different scales and underlying signals, interferometric \hi\ IM at low redshifts and \hi\ observations of the EoR have very similar challenges.  In the frequency range of the observation, the radio sky is dominated by the diffuse foreground emissions from our galaxy and the local Universe \citep{2002ApJ...564..576D}. The foreground radiation is smooth in frequency and can be up to several orders of magnitude higher than the \hi. The \hi\ signal, on the other hand, is discrete in frequency. Therefore, we can use the smoothness of foreground in frequency by Fourier transforming the observed visibilities along the frequency axis, commonly called the ``delay transform'' \citep{2004ApJ...615....7M,2012ApJ...753...81P,2012ApJ...756..165P}. It separates smooth, large frequency structure of foreground and small oscillating frequency structure of \hi. The smooth foreground mainly resides on the large frequency-scale modes, creating the ``foreground wedge'' \citep{2014PhRvD..90b3018L} and an observation window outside these modes. Measuring the \hi\ power spectrum outside the foreground wedge is, therefore, sometimes called foreground avoidance. 

Apart from avoiding the foreground wedge, one can also try to subtract the foreground by using its smoothness in frequency and further extract information (e.g. \citealt{2009ApJ...695..183B}). The specific approaches can be generally split into two types, the parametric approaches that use polynomial fits to extract foreground (e.g. \citealt{2005ApJ...625..575S, 2015MNRAS.447.1973B}) and non-parametric approaches that use statistical methods to separate foreground components. The most standard approach for component separation is Principle Component Analysis (PCA). More advanced methods can be built upon it, such as Fast Independent Component Analysis (FastICA, e.g. \citealt{2012MNRAS.423.2518C}), Generalized Morphological Component Analysis (GMCA, e.g. \citealt{2016MNRAS.458.2928C}), and Gaussian Process Regression (GPR, e.g. \citealt{2018MNRAS.478.3640M}). The success of a detection of \hi\ signal in visibility data also relies heavily on mitigation of various systematics through the calibration of the data \citep{2016MNRAS.461.3135B}. This requires a thorough understanding of the instrument (e.g. \citealt{2016ApJ...825....9T}) and the properties of the foreground (e.g. \citealt{2020ApJ...893..118N}).


To tackle the difficult problem of measuring \hi\ clustering in the post-ionization Universe at small angular scales, it is necessary to create robust simulations for low-redshift IM using radio interferometers and the data analysis pipeline for \hi\ power spectrum estimation. {In this paper, we present an end-to-end simulation pipeline for low-redshift interferometric IM that generates realistic foreground and \hi\ signal, simulates interferometric observations and applies robust foreground mitigation strategies which are crucial for future detections for \textit{MeerKAT} and SKAO.} The visibility data is used to calculate the brightness temperature power spectrum using a quadratic estimator formalism. We use the configurations of \textit{MeerKAT} telescope and observational specifications mimicking a typical pointing of MIGHTEE survey to present a case study. {A detailed comparison of the effects of different foreground mitigation methods is conducted, including a direct comparison between foreground removal in visibility data and in images.} The aim of this pipeline is to provide a detailed look into the topics of IM delay power spectrum that have been extensively discussed in the context of EoR, but have not yet thoroughly quantified for observations of the low-redshift Universe. It also allows more realistic simulations that will enable us to fully study the challenges of interferometric IM towards future detection.

The paper is organized as follows: In Section \ref{sec:hips}, the basics of delay power spectrum in analytical formalism is reviewed. The quadratic estimator formalism for converting the visibility power spectrum to the brightness temperature power spectrum is discussed in Section \ref{sec:estform}. We present the simulation of the sky signal input in Section \ref{sec:skysim}. Foreground mitigation in visibility space is discussed in Section \ref{sec:fgremove}. Power spectrum estimation using interferometric images is presented in Section \ref{sec:image}. {Comparison between different methods with MIGHTEE-like noise level is made in Section \ref{sec:result}}. We conclude our findings in Section \ref{sec:conclusion}. Throughout this paper, we assume the Lambda cold dark matter cosmology from \cite{2020A&A...641A...6P}.

\section{\althi\ clustering from visibility}
\label{sec:hips}
In this section, we derive the connection between the visibility data and the power spectrum of cosmological \hi. Note for simplicity we do not consider RSDs \citep{1987MNRAS.227....1K}, which, in the scales of our interest, are dominated by effects from peculiar velocities of \hi\ galaxies (see e.g. \citealt{2021arXiv210411171C}). The density of \hi\ clustering is typically expressed as the brightness temperature $T_{\rm \hi}(\bm{x})$, 
\begin{equation}
    T_{\rm \hi} (\bm{x}) = \frac{1}{\mathbb{V}} \sum_i C_{\rm \hi}(z^i) M_{\rm \hi}^i \delta_{\rm D}^3(\bm{x}-\bm{x}^i),
\label{eq:thi}
\end{equation}
where $\mathbb{V}$ is the survey volume, $M_{\rm \hi}^i$ is the \hi\ mass of each sources within the survey volume, $z^i$ is the redshift each source is at, $\delta_{\rm D}^3$ is the Dirac delta function in comoving space, and 
\begin{equation}
    C_{\rm \hi} (z)=\frac{3A_{12}h_{\rm P}c^3(1+z)^2}{32\pi m_{\rm H}k_{\rm B} \nu_{21}^2 H(z)}
\end{equation}
is the conversion factor from \hi\ density to brightness temperature with $h_{\rm P}$ the Planck constant, $k_{\rm B}$ the Boltzmann constant, $m_{\rm H}$ the mass of the hydrogen atom, $A_{12}$ the emission coefficient of the 21-cm line transmission, $\nu_{21}$ the rest frequency of the 21-cm emission and $H(z)$ the Hubble parameter at redshift $z$ \citep{2016MNRAS.458.3399W}.

The power spectrum is measured in $\bm{k}$ space, which is the Fourier pair of comoving space coordinate $\bm{x}$. The Fourier convention used for the brightness temperature field is 
\begin{equation}
\begin{split}
    \Tilde{T}_{\rm \hi}(\bm{k}) =& \int \frac{{\rm d}^3 x}{\mathbb{V}} T_{\rm \hi}(\bm{x})\, e^{-i\bm{k}\bm{x}},\\
    T_{\rm \hi}(\bm{x})=&\frac{\mathbb{V}}{(2\pi)^3} \int {\rm d}^3k\, \Tilde{T}_{\rm \hi}(\bm{k})\,e^{i\bm{k}\bm{x}},
\end{split}
\end{equation}
and the corresponding convention for Dirac $\delta_{\rm D}$-function is:
\begin{equation}
\begin{split}
    \int \frac{{\rm d}^3 x}{\mathbb{V}} \delta_{\rm D}^3({\bm{x}-\bm{x_0}}) f(\bm{x}) = f(\bm{x_0}),\\ \mathbb{V} \int \frac{{\rm d}^3k}{(2\pi)^3}\tilde{\delta}_{\rm D}^3(\bm{k}-\bm{k}_0) \tilde{f}(\bm{k}) = \tilde{f}(\bm{k_0})
\end{split}
\end{equation}
for arbitrary function $f(\bm{x})$ in comoving space and $\tilde{f}(\bm{k})$ in Fourier space.

Assuming homogeneity, the two-point correlation function can be written as 
\begin{equation}
    \xi(\bm{s}) = \langle T_{\rm \hi}(\bm{x}) T_{\rm \hi}(\bm{x}+\bm{s}) \rangle = \int \frac{{\rm d}^3x}{\mathbb{V}} T_{\rm \hi}(\bm{x}) T_{\rm \hi}(\bm{x}+\bm{s}).
\end{equation}

The brightness temperature power spectrum is the Fourier transform of the two-point correlation function
\begin{equation}
    P_{21}(\bm{k}) = \int {\rm d^3}s\, \xi(\bm{s}) e^{-i\bm{k}\bm{s}} = \int {\rm d^3}s \frac{{\rm d}^3x}{\mathbb{V}} T_{\rm \hi}(\bm{x}) T_{\rm \hi}(\bm{x}+\bm{s})e^{-i\bm{k}\bm{s}},
\end{equation}
which can be similarly written in Fourier space as
\begin{equation}
\label{eq:pk}
\begin{split}
\mathbb{V}\langle \tilde{T}(\bm{k}) \tilde{T}^*(\bm{k}') \rangle = \tilde{\delta}_{\rm D}^3 (\bm{k}-\bm{k}')P_{21}(\bm{k}).
\end{split}
\end{equation}

The distribution of the brightness temperature can be observed using radio interferometers by measuring the radio sky through the correlations of signals across different pairs of antennas. At any given time, each pair measures the sky signal using the difference between the received signal phases determined by the position vector between the pair, i.e. the baseline. The length of the baseline in the units of the observing wavelength, $\{u,v,w\}$, corresponds to the scale of fluctuations the baseline is measuring. For a set of baselines with $u$-$v$ coordinates $\{\bm{b_\alpha}\} = \{(u_\alpha,v_\alpha,w_\alpha)\}c/f$, where $c$ is the speed of light and $f$ is the observing frequency, the visibility $V(u,v,w,f)$ generated on each baseline is related to the sky intensity distribution $I(l,m,f)$ \citep{2016era..book.....C}
\begin{equation}
\begin{split}
    V(u,v,w,f) = \int &\frac{{\rm d}l\,{\rm d}m}{\sqrt{1-l^2-m^2}} I(l,m,f)A(l,m,f)\\&\times{\rm exp}\big[-2\pi i \big(lu+mv+(1-n)w\big)\big],
\end{split}
\end{equation}
where $l,m$ are the sky coordinates on the celestial sphere, $n= \sqrt{1-l^2-m^2}$ and $A(l,m,f)$ is the beam response.

For wide field-of-view (FOV) instruments used in EoR observation, the curved sky poses a challenge in power spectrum estimation (e.g. \citealt{2015ApJ...807L..28T}). Here, we focus on IM in the low-redshift Universe, which is typically done using small-FOV dish arrays such as \textit{MeerKAT} and SKAO-mid. For \textit{MeerKAT} L-band receivers which is simulated in this paper, the beam width is $\sim 1\,{\rm deg}$ \citep{2021MNRAS.502.2970A}. Combined with the fact that beam properties of dish telescopes are relatively well-understood and yield more desirable features of foreground contamination \citep{2015ApJ...804...14T}, we expect applying simple flat-sky approximation is good enough for IM at low redshifts. Using the flat-sky approximation and Fourier transforming the visibility along the frequency axis we have
\begin{equation}
\begin{split}
    \Tilde{V}(u,v,\eta) \approx \int &{\rm d}l\,{\rm d}m\,{\rm d}f\, I(l,m,f)A(l,m,f)\,\\
    &\times{\rm exp}\big[-2\pi i \big(lu+mv+f\eta\big)\big],
\end{split}
\end{equation}
where $\eta$ is the Fourier pair of frequency $f$. 

In the flat-sky approximation, we can write down the transformation between the sky and comoving space
\begin{equation}
    l = \frac{r_x}{D_{\rm c}}, m=\frac{r_y}{D_{\rm c}}, f = \frac{f_{21}}{1+z}
\end{equation}
where we use $r_{x,y}$ to denote transverse scales. Note that the comoving distance $D_{\rm c}$ is the scale along the line-of-sight direction which we also denote as $r_z$.

The integral can be written as 
\begin{equation}
\begin{split}
\tilde{V} =& \int {\rm d}^3r \frac{-2k_{\rm B}f_{21}H(z)}{D_{\rm c}^2(z)\,(1+z)^2c\lambda^2} [AT](\bm{r})\,{\rm exp}\big[-2\pi i \big(lu+mv+f\eta\big)\big]\\
=& \frac{-2k_{\rm B}}{\lambda_{21}^3}\int {\rm d}^3r \frac{H(z)}{D_{\rm c}^2(z)}[AT](\bm{r})\,{\rm exp}\big[-2\pi i \big(lu+mv+f\eta\big)\big],
\end{split}
\end{equation}
where we use the Jacobian determinant to perform coordinate transformation and write $[AT](\bm{r}) = A(\bm{r})T(\bm{r})$ as the product of beam response and brightness temperature.

The delay power spectrum $P_{\rm d} = \langle |\tilde{V}(u,v,\eta)|^2\rangle $ can be written as
\begin{equation}
\begin{split}
&P_{\rm d} = \Big(\frac{2k_{\rm B}}{\lambda_{21}^3}\Big)^2\int {\rm d}^3r\, {\rm d}^3s \frac{H(z_r)}{D_{\rm c}^2(z_r)}\frac{H(z_{r+s})}{D_{\rm c}^2(z_{r+s})} \langle [AT]({\bm{r}})[AT]({\bm{r}+\bm{s}}) \rangle\\
&\times {\rm exp}\bigg[ -2\pi i \bigg( u\Big[\frac{r_x+s_x}{D_{\rm c}(z_{r+s})}-\frac{r_x}{D_{\rm c}(z_r)}\Big]+v\Big[\frac{r_y+s_y}{D_{\rm c}(z_{r+s})}-\frac{r_y}{D_{\rm c}(z_r)}\Big]
\\&+\eta\Big[\frac{f_{21}}{1+z_{r+s}}-\frac{f_{21}}{1+z_r}\Big] \bigg) \bigg].
\end{split}
\end{equation}
From the above equation one can see that the matching between the delay power spectrum and the \hi\ brightness temperature power spectrum is not exact. By assuming a narrow redshift range of integration, we can effectively use one central redshift $z_0$, so that
\begin{equation}
\frac{f_{21}}{1+z_{r+s}}-\frac{f_{21}}{1+z_r} \approx-\frac{f_{21}\Delta z_s}{(1+z_0)^2}
\approx -\frac{f_{21}s_zH(z_0)}{(1+z_0)^2c}.
\end{equation}

To make the equations more compact, we denote $X=D_{\rm c}(z_0)$ and $Y = \lambda_{21}(1+z_0)^2/H(z_0)$ with $H(z_0)$ being the Hubble parameter at $z_0$. This results in 
\begin{equation}
    \bm{k}_\perp = \frac{2\pi \bm{u}}{X},\; k_\parallel = -\frac{2\pi \eta}{Y},
\label{eq:kconvert}
\end{equation}
where $\bm{k}_\perp$ is the scales on the angular plane, $\bm{u}=\{u,v\}$ is the visibility space coordinates, and $k_\parallel$ is the scale along the line-of-sight. With this notation, we can write
\begin{equation}
\begin{split}
P_{\rm d} =& \Big(\frac{2k_B}{\lambda^2}\Big)^2\frac{1}{X^4Y^2}\int {\rm d}^3r\, {\rm d}^3s \,\langle[AT]({\bm{r}})[AT]({\bm{r}+\bm{s}})\rangle\\
&\times  {\rm exp}\bigg[-i\bigg(\frac{2\pi u}{D_{\rm c}(z_0)}s_x+\frac{2\pi v}{D_{\rm c}(z_0)}s_y-\frac{2\pi f_{21}H(z_0)\eta}{(1+z_0)^2c}s_z \bigg)\bigg]\\
=&\Big(\frac{2k_B}{\lambda^2}\Big)^2\frac{\mathbb{V}^4}{X^4Y^2}\int \frac{{\rm d}^3k'}{(2\pi)^3}\frac{{\rm d}^3k''}{(2\pi)^3} \tilde{A}(\bm{k}-\bm{k}')\tilde{A}^{*}(\bm{k}-\bm{k}'')\\
&\times\langle\tilde{T}(\bm{k}')\tilde{T}^{*}(\bm{k}'')\rangle,
\end{split}
\label{eq:psdelay}
\end{equation}
where $\bm{k}=\{\bm{k}_\perp,k_\parallel\}$ is the three-dimensional (3D) wave vector and $\tilde{A}$ is the Fourier transform of the beam response.

Following Eq. (\ref{eq:pk}), we can further simplify the previous equation to
\begin{equation}
\begin{split}
P_{\rm d} (u,v,\eta)
= \Big(\frac{2k_B}{\lambda^2}\Big)^2\frac{\mathbb{V}^2}{X^4Y^2}\int \frac{{\rm d}^3k'}{(2\pi)^3}|\tilde{A}(\bm{k}-\bm{k}')|^2P_{\rm T}(\bm{k}').
\end{split}
\label{eq:beamps}
\end{equation}

The above equation not only imposes flat-sky approximation, but also requires the evolution along the light-cone to be negligible. This assumption is reasonable for the simulation used in this paper. We leave the treatment to wide frequency ranges for future work.


\section{quadratic estimator of temperature power spectrum}
\label{sec:estform}
{In this section, we derive the implemented power spectrum estimator, with foreground mitigation and minimisation of measurement uncertainties}. In order to relate the foreground-dominated observations to the brightness temperature power spectrum of \hi, the sky signal measured in visibility space will be processed through a data analysis framework. Such a framework should allow estimations of the brightness temperature power spectrum of \hi\ clustering through appropriate means of statistics to mitigate foreground. These types of power spectrum frameworks using interferometric data have been extensively studied in the context of cosmic microwave background (e.g. \citealt{1997PhRvD..55.5895T,2003ApJ...591..575M}) and, more recently, EoR (e.g. \citealt{2011PhRvD..83j3006L,2014MNRAS.445.4351C,2019MNRAS.483.2207M}) and high-redshift IM (e.g. \citealt{2018MNRAS.473..261S,2021MNRAS.500.4398C}). See \cite{2020PASP..132f2001L} for a review. Here, we focus on interferometric low-redshift IM, by constructing a quadratic estimator which explicitly includes the operation of foreground mitigation and deconvolves the mode mixing introduced by the primary beam attenuation.

The visibility data consist of $N_{\rm bl}\times N_{\rm ch}\times N_{\rm steps}$ elements, where $N_{\rm bl}$ is the number of baselines, $N_{\rm ch}$ is the number of frequency channels and $N_{\rm steps}$ is the number of time-steps in the observations. By averaging the visibility data into $u$-$v$ grids, a data vector $\bm{V}$ with a length of $N_{\rm ch} \times N_{\bm{u}}$ can be constructed, where $N_{\rm ch}$ is the number of frequency channels and $N_{\bm{u}}$ is the number of grids on the $u$-$v$ plane. Note that $\bm{V}$ is a column vector that loops over both frequency channel and $u$-$v$ grids. Its elements can be written as
\begin{equation}
    \mathbf{V}^{j} = V(u_{ j}, v_{j}, f_{ j}),
\end{equation}
with $V(u_{j}, v_{j}, f_{ j})$ being the visibility data at the specific $u$-$v$ coordinate and frequency the $j^{\rm th}$ gridded baseline corresponds to. Similarly, we can also define a delay-transformed data vector
\begin{equation}
    \tilde{\mathbf{V}}^k = \tilde{V}(u_k,v_k,\eta_k),
\end{equation}
where $\eta$ is the Fourier-inverse of the observing frequencies. 

The visibility power spectrum of the 21cm emission can be discretized into the summation of the bandpower
\begin{equation}
    P_{\rm d}(\bm{u},\eta) = \sum_\alpha \chi_\alpha(\bm{u},\eta) P_{\rm d}(|\bm{u}|_\alpha,\eta_\alpha) = \sum_\alpha \chi_\alpha p^{\rm d}_\alpha,
\end{equation}
with $\chi^\alpha$ being the selection function, returning 1 if the baseline falls into the $\alpha^{\rm th}$ bin and 0 if not. We define $p^{\rm d}_\alpha \equiv P_{\rm d}(|\bm{u}|_\alpha,\eta_\alpha)$. 

Similarly the temperature bandpower is defined as
\begin{equation}
    P_{\rm T}(\bm{k}) = \sum_\beta \chi_\beta(\bm{k}) P_{\rm T}(\bm{k}_\beta) = \sum_\beta \chi_\beta(\bm{k}) p^{\rm T}_{\beta}.
\label{eq:tbandps}
\end{equation}

The estimator of the bandpower $\hat{p}^{\rm d}_\alpha$ can be constructed as 
\begin{equation}
    \hat{p}^{\rm d}_\alpha = \mathbf{V}^\dagger \mathbf{E}^{\rm d}_\alpha \mathbf{V} - \hat{b}^{\rm d}_\alpha.
\end{equation}
Here, $\mathbf{V}$ is the gridded visibility data vector, $\hat{b}^{\rm d}_\alpha$ is the correction term for bias, and $\mathbf{E}^{\rm d}_\alpha$ is the power spectrum estimation matrix which we will derive explicitly. 

When estimating $\hat{p}^{\rm d}_\alpha$, one can choose $\mathbf{V}$ to include all the visibility data. In our case, $\mathbf{E}^{\rm d}_\alpha$ is a block matrix of size $N_{\bm{u}}\times N_{\bm{u}}$, with each element being a $N_{\rm ch}\times N_{\rm ch}$ matrix
\[ \setlength{\fboxsep}{0pt}
\mathbf{E} = \left(\begin{array}{@{}c@{}l@{}c@{\mkern-5mu}c@{\,}rc*{2}{@{\;}c}@{}}%
\noalign{\vskip 1.5ex}
  \fbox{\,$\begin{matrix}
  \mathbf{E}_{11}^{1} & \hdots & \mathbf{E}_{1N_{\rm ch}}^{1} \\
  \vdots & \ddots & \vdots \\
  \mathbf{E}_{N_{\rm ch}}^{1} & \hdots & \mathbf{E}_{N_{\rm ch}N_{\rm ch}}^{1}
  \end{matrix}$\,}
 & \ \hdots  & \hdots  \\
  \vdots  &  \ \ddots  & \vdots \\
   \hdots & \hdots &  \fbox{\,$\begin{matrix}
  \mathbf{E}_{11}^{N_{\bm{u}}} & \hdots & \mathbf{E}_{1N_{\rm ch}}^{N_{\bm{u}}} \\
  \vdots & \ddots & \vdots \\
  \mathbf{E}_{N_{\rm ch}}^{N_{\bm{u}}} & \hdots & \mathbf{E}_{N_{\rm ch}N_{\rm ch}}^{N_{\bm{u}}}
  \end{matrix}$\,}\ \ 
\end{array}\right)
\]

The resulting $\mathbf{E}^{\rm d}_\alpha$ is computationally difficult to deal with due to its large size. Note that if different $u$-$v$ grids do not correlate, the off-diagonal blocks will be empty. The block elements of $\mathbf{E}^{\rm d}_\alpha$ will then only operate on the $N_{\rm ch}$ data points in the corresponding $u$-$v$ grid. Thus, we assume the off-diagonal part is negligible and only construct the matrix for one $u$-$v$ grid at a time, using an input data vector of size $N_{\rm ch}$ (the number of frequency channels). The beam mixes different $u$-$v$ modes, whose correlation is present in the non-block-diagonal part of $\mathbf{E}^{\rm d}_\alpha$. However, this approximation provides massive speed-up for computational efficiency. We leave the full treatment of $\mathbf{E}^{\rm d}_\alpha$ to future work. Under this assumption, the covariance matrix of the data vector can be written as \citep{2015PhRvD..91l3011D}
\begin{equation}
    C_{\bm{u}\bm{u}'ff'} \approx \delta_{\bm{u}\bm{u}'}^{\rm K} \hat{C}_{ff'}(|\bm{k}_\perp|),
\label{eq:empdata}
\end{equation}
where $\delta^{\rm K}$ is the Kronecker delta and $\hat{C}_{ff'}(|\bm{k}_\perp|)$ is the estimated data covariance in each annulus $|\bm{u}|$ bin, which is calculated from averaging the visibility covariance across the baselines that fall into the particular $|\bm{u}|$ bin.

For the data vector of one $u$-$v$ grid, the conversion from frequency to the delay time domain can be written as
\begin{equation}
    \tilde{\mathbf{V}}^k = \delta f \sum_{j} e^{-2\pi i \eta_k f_j} {\mathbf{V}}^j = \sum_j \mathbf{F}^k_{\ \, j} {\mathbf{V}}^j,
\end{equation}
where $\mathbf{F}^k_{\ \, j}$ is the discrete Fourier transform (DFT) kernel and $\delta f$ is the channel bandwidth.

The covariance matrix of the data vector of one $u$-$v$ grid can be written as
\begin{equation}
    \mathbf{C} \equiv \langle \mathbf{V} \mathbf{V}^\dagger \rangle = \mathbf{C}_{\rm fg} + \mathbf{N} + \sum_\alpha p^{\rm d}_\alpha \mathbf{C}_{,\alpha},
\end{equation}
with $\mathbf{C}_{\rm fg}$ being the covariance matrix of the radio foreground in the frequency domain, $\mathbf{N}$ the noise covariance matrix and $\sum_\alpha p^{\rm d}_\alpha \mathbf{C}_{,\alpha}$ the signal cross-correlation\footnote{The signal cross-correlation is a linear combination of $p^{\rm d}_\alpha$. Therefore, the coefficient of the expansion $\mathbf{C}_{,\alpha} = \partial\, \mathbf{C}/\partial p^{\rm d}_\alpha$.} which we decompose into some bandpower of visibility data $p^{\rm d}_\alpha$.

The elements of the signal covariance matrix can be written as 
\begin{equation}
\begin{split}
    & \big(\mathbf{C}_{\rm s}\big)_{ij} =  \langle V_{\rm \hi}(u,v,f_i)V^*_{\rm \hi}(u,v,f_j) \rangle \\ 
    & = \int {\rm d}\eta_1  {\rm d}\eta_2 \,{\rm exp}\big[2\pi i (f_i \eta_1 - f_j \eta_2) \big] \langle \tilde{V}_i \tilde{V}^*_j \rangle \\ 
    & = \int {\rm d}\eta_1  {\rm d}\eta_2 \,{\rm exp}\big[2\pi i (f_i \eta_1 - f_j \eta_2) \big] \tilde{\delta}_D(\eta_1-\eta_2) P_{\rm d} \\ 
    & =  \frac{1}{(N_{\rm ch}\delta f)^2} \sum_{\alpha} {\rm exp}\big[2\pi i (f_i-f_j)\eta_\alpha \big] p^{\rm d}_{\alpha }.
\end{split}    
\end{equation}
Therefore, we can write
\begin{equation}
    \big(\mathbf{C}_{,\alpha}\big)_{ij} =  \frac{1}{(N_{\rm ch}\delta f)^2} {\rm exp}\big[2\pi i (f_i-f_j)\eta_\alpha \big].
\end{equation}

If we aim to simply calculate the delay power of the visibility data instead of the temperature power and assume no foreground and noise contamination, the power spectrum estimation can be written as
\begin{equation}
    \hat{p}^{\rm d}_\alpha|_{\rm no\, fg,n} = \frac{\sum_{i} \chi_\alpha(u_{i},v_{ i},\eta_{i}) |\tilde{\mathbf{V}}^{i}|^2}{\sum_{i}\chi_\alpha(u_{i},v_{i},\eta_{i})} = \sum_i (\mathbf{w}_{\alpha})_{i} |\tilde{\mathbf{V}}^i|^2,
\label{eq:2dvisps}
\end{equation}
where $i$ loops over all Fourier transformed $u$-$v$ grids, $\chi_\alpha$ is the selection function and we express the normalized selection function as $\mathbf{w}_\alpha$ to make the expression more compact. With the above expression we can write 
\begin{equation}
    \hat{p}^{\rm d}_\alpha|_{\rm no\, fg,n} = \mathbf{V}^\dagger \mathbf{F}^\dagger {\rm diag}(\mathbf{w}_\alpha)\mathbf{F} \mathbf{V}.
\label{eq:pdanofgn}
\end{equation}
Here, ${\rm diag}(\mathbf{w}_\alpha)$ is a diagonal matrix with the $i^{\rm th}$ diagonal element being $(\mathbf{w}_\alpha)_i$. 

In the presence of thermal noise and foreground contamination, we can include some linear operation $\mathbf{R}$ in the estimator in order to mitigate the foreground. This operation $\mathbf{R}$ may include foreground removal, inverse covariance weighting, frequency tapering etc. Taking this operator into consideration we can rewrite the power spectrum estimation matrix $\mathbf{E}^{\rm d}_\alpha$ as
\begin{equation}
    \mathbf{E}^{\rm d}_\alpha =  {\rm diag}(\mathbf{S}_{\alpha}) \mathbf{R}^{\dagger} \mathbf{F}^\dagger {\rm diag}(\mathbf{w}_\alpha)\mathbf{F}\, \mathbf{R}.
\label{eq:eda}
\end{equation}
Here, $\mathbf{S}_\alpha$ is a normalisation vector which we can solve for by taking the expectation value of the power spectrum estimator 
\begin{equation}
    \langle \hat{p}^{\rm d}_\alpha \rangle = \sum_\beta {\rm tr}\big[ \mathbf{C}_{, \beta} \mathbf{E}_\alpha \big]p^{\rm d}_\beta + {\rm tr}\big[\big(\mathbf{N} + \mathbf{C}_{\rm fg}\big) \mathbf{E}^{\rm d}_\alpha \big] - \hat{b}^{\rm d}_\alpha.
\end{equation}
To solve for $\mathbf{S}_\alpha$, we impose
\begin{equation}
    \sum_\beta {\rm tr}\big[ \mathbf{C}_{, \beta} \mathbf{E}_\alpha \big] = 1,
\end{equation}
and the covariance of the estimation is
\begin{equation}
    \Sigma_{\alpha \alpha '} = \langle \hat{p}^{\rm d}_\alpha \hat{p}^{\rm d}_{\alpha '} \rangle -  \langle \hat{p}^{\rm d}_\alpha \rangle\langle \hat{p}^{\rm d}_{\alpha'} \rangle = 2 {\rm tr}\big[ \mathbf{C} \,\mathbf{E}^{\rm d}_\alpha \mathbf{C}\, \mathbf{E}^{\rm d}_{\alpha '}\big].
\label{eq:sigmaalpha}
\end{equation}
Note that here we choose $\mathbf{S}_\alpha$ as a vector, or effectively as a diagonal normalization matrix as discussed in \cite{1997MNRAS.289..285H} and \cite{1997PhRvD..55.5895T}. This leads to the correlation of the variance of different bandpowers, as the above matrix $\Sigma_{\alpha \alpha '}$ will have non-diagonal components. To decorrelate the variance, one should choose the normalization matrix to be $\mathbf{F}^{-1/2}$ where $\mathbf{F}$ is the bandpower Fisher matrix \citep{1998tsra.conf..270T,2000MNRAS.312..285H}. As mentioned earlier though, we calculate the bandpower one $u$-$v$ grid at a time for computational efficiency. Thus, despite using a normalization method that results in correlation between variances of different bandpowers, the non-diagonal part of $\Sigma_{\alpha \alpha '}$ is not included in our calculation. See \cite{2014PhRvD..89b3002D} for a discussion on overcoming real world obstacles such as large data volume and error properties.

In the ideal case where the foreground and noise covariance is known, one should always choose $\hat{b}^{\rm d}_\alpha = {\rm tr}\big[\big(\mathbf{N} + \mathbf{C}_{\rm fg}\big) \mathbf{E}^{\rm d}_\alpha \big]$ and use inverse covariance weighting $\mathbf{R} = \mathbf{C}^{-1}$, so that we have an unbiased estimation of the power spectrum with minimum uncertainty (i.e. the optimal estimator). In reality though, we only have a guess for the true covariance of the foreground and noise, and therefore the estimator can be written as
\begin{equation}
    \hat{p}^{\rm d}_\alpha = \mathbf{V}^\dagger \mathbf{E}^{\rm d}_\alpha \mathbf{V} - {\rm tr}\big[\big(\mathbf{N}^{\rm psd} + \mathbf{C}_{\rm fg}^{\rm psd}\big) \mathbf{E}^{\rm d}_\alpha \big],
\label{eq:estpdalpha}
\end{equation}
with a pseudo covariance as our best guess of the true foreground and noise covariance.

If no extra data processing is applied, $\mathbf{R}$ is simply the identity matrix. Frequency tapering can be applied by choosing $\mathbf{R}$ to be a diagonal matrix with the diagonal components being the frequency window (see e.g. \citealt{2021arXiv210802263T}). For any foreground mitigation strategy $\mathbf{R}$ that perfectly removes the foreground components in the data vector, the resulting estimator should revert to the optimal case. Thus, we can decompose the operation matrix into
\begin{equation}
    \mathbf{R}_{\rm fg} = (\mathbf{C} - \mathbf{C}_{\rm fg}^{\rm psd})^{-1} \mathbf{A}.
\end{equation}
Here, $\mathbf{A}$ is the component separation matrix that should remove the contribution of the foregrounds. Since we believe the resulting data vector after the operation of $\mathbf{A}$ is free of foreground, a subsequent inverse covariance matrix is applied. See \cite{2021MNRAS.501.1463K} for an example of this using GPR. 

Finally, after obtaining the estimation of the delay power spectrum on each $u$-$v$ grid, we apply annulus bins in $u$-$v$ space to $\hat{p}^{\rm d}_\alpha$. The temperature power spectrum can be estimated with the annulus-binned delay power spectrum via
\begin{equation}
    \hat{p}^{\rm T}_\beta = \sum_\alpha \big(\mathcal{M}\big)^{-1}_{\beta\alpha} \hat{p}^{\rm d}_\alpha.
\label{eq:ptbpda}
\end{equation}
Here, $\mathcal{M}$ is a conversion matrix derived from Eq. (\ref{eq:beamps}) (see Appendix \ref{sec:apdxps} for detailed derivation)
\begin{equation}
\begin{split}
    \big(\mathcal{M}\big)_{\alpha\beta}  = & \Big(\frac{2k_{\rm B}}{\lambda^2}\Big)^2\frac{N_{\rm ch}\,\delta f}{Y\sum_i \chi_\alpha(\bm{k}_i)} \sum_i  \int \frac{{\rm d^2}\bm{k}_\perp}{(2\pi)^2} \chi_\alpha(\bm{k}_i) \\&\times \Big|\tilde{A}_\perp \Big(\bm{k}^i_\perp-\bm{k}_\perp\Big)\Big|^2\,\chi_\beta(\bm{k}_\perp,k_\parallel^i),
\end{split}
\label{eq:mab}
\end{equation}
where $i$ loops over all Fourier transformed $u$-$v$ grids and $\tilde{A}_\perp$ is the Fourier transformed beam response in the transverse plane defined in Eq. (\ref{eq:fbeam2d}).
The variance of the estimation can also be propagated assuming each bandpower is an independent measurement of the clustering.

If we choose the same number of bins for $|\bm{u}|_\alpha$ and $|\bm{k}_\perp|_\beta$, we can calculate the above square matrix and estimate the cylindrical temperature power spectrum $\big[\hat{P}_{21}^\gamma\big]_\beta$ via a matrix inversion. The resulting cylindrical power spectrum can be further averaged into 1D $\{k_i\}$ bins
\begin{equation}
    \hat{p}^{\rm 1d}_i = \frac{\sum_\beta w_i(\bm{k}_\perp^\beta,k_\parallel^\beta) \hat{p}^{\rm T}_\beta}{\sum_\beta w_i(\bm{k}_\perp^\beta,k_\parallel^\beta)},
\label{eq:1dtps}
\end{equation}
where $w_i$ is a combination of the selection function and wedge criteria; $w_i$ returns 1 if $(\bm{k}_\perp^\beta,k_\parallel^\beta)$ falls into the bin and is not avoided by the foreground wedge criteria.

\section{Simulation of the Radio Sky}
\label{sec:skysim}
In this section, we describe the simulations of the sky signal and the experimental set-up we use to generate visibility data consistent with \textit{MeerKAT} observations. The \textit{MeerKAT} array consists of 64 dish telescopes. It observes the sky in the \textit{L} band and the \textit{UHF} band. In this paper, we are focusing on the \textit{L}-band observations. The \textit{L}-band receivers have 4096 frequency channels and observe the sky with a time resolution of 8 s. Also, they have a frequency resolution of 208.984 kHz. Following \cite{2021MNRAS.505.2039P}, we choose 220 frequency channels centred at 1115.14 MHz and set the pointing centre at the Cosmic Evolution Survey (COSMOS) field \citep{2007ApJS..172....1S} at RA=150.12 deg and Dec=2.21 deg with an 11.2-hour tracking. The frequency range of the sub-band we choose covers a narrow redshift bin $z\sim0.25-0.3$. {The size of the input sky image is $3.7 \times 3.7 \,{\rm deg}^2$, much larger than the \textit{MeerKAT} dish FoV to ensure completeness}.

\subsection{Simulation of Galactic Foregrounds}
\label{subsec:sync}

The biggest challenge for simulating interferometric observations is the extreme angular resolution, which is typically $\sim$arcsec. Note that, we are only interested in the cosmological clustering and can therefore ignore the scales corresponding to the longest baselines. {For $k\gtrsim 20 {\rm Mpc}^{-1}$, the point source assumption of \hi\ sources breaks down and the information within these small scales is beyond the interests of cosmology.} Therefore, we simulate \textsc{healpix} \citep{2005ApJ...622..759G, Zonca2019} $\rm N_{side}=8192$ maps, which corresponds to a pixel size of (0.43 arcmin$)^2$ and clustering scales of $k\sim 40 {\rm Mpc}^{-1}$ at $z\sim0.25-0.3$.

At frequencies around 1GHz, the dominant component of smooth Galactic foregrounds is the synchrotron radiation. The template we use to generate the signal is the ``Haslam map'' at 408MHz \citep{1981A&A...100..209H,1982A&AS...47....1H}. We use the de-sourced, de-striped version of the map described in \cite{2015MNRAS.451.4311R}. The pixel size of the original map is (6.87 arcmin$)^2$ corresponding to \textsc{healpix} $\rm N_{side} = 512$. To reduce the pixel size of the template, we follow the method in \cite{2015MNRAS.451.4311R} and generate Gaussian random structure to fill in the small scales.

\begin{figure}
    \centering
    \includegraphics[width=\linewidth]{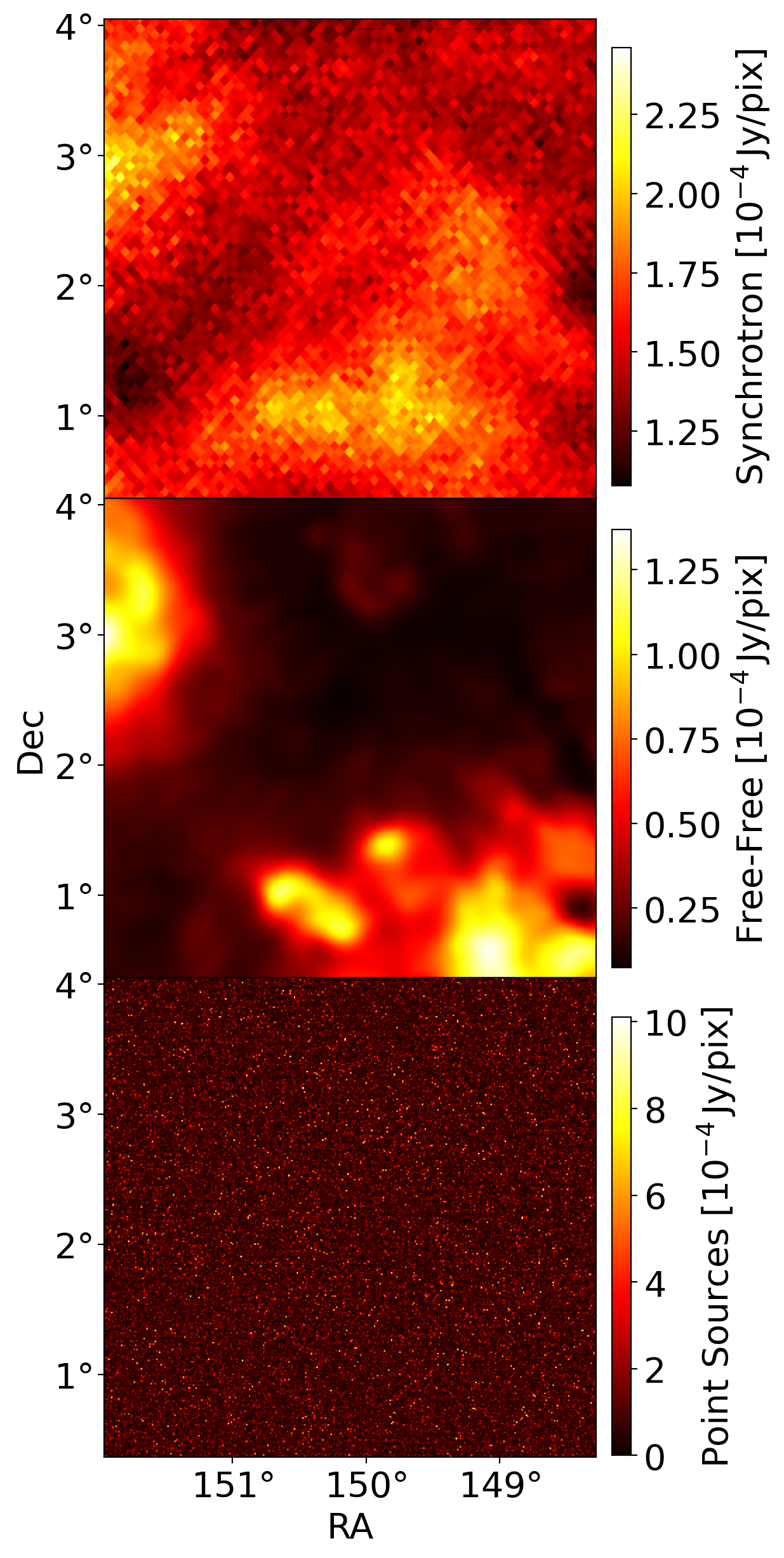}
    \caption{Top panel: the simulated synchrotron radiation around the pointing centre at RA=150.12 deg and Dec=2.21 deg, originally generated at 408MHz and extrapolated to 1.4 GHz. Middle panel: the simulated free-free emission around the pointing centre at 1.4 GHz. Bottom panel: the simulated extragalactic point sources around the pointing centre at 1.4 GHz. {Values above $10^{-3}$ Jy per pixel is set to $10^{-3}$ for better presentation}. {The pixel size of the images is set to $0.01 \times 0.01 \,{\rm deg}^2$}. {The size of the input sky image is $3.7 \times 3.7 \,{\rm deg}^2$}.}
    \label{fig:skysim}
\end{figure}

The spectral index of the emission can be extrapolated from observations of synchrotron-dominated radio sky at different wavelengths (e.g. \citealt{2021MNRAS.tmp.1362S}). Here, we use the 1.4GHz and 2.3GHz maps from the Global Sky Model \citep{2017MNRAS.464.3486Z}, which are based on observations of \cite{2001A&A...376..861R} and \cite{1998MNRAS.297..977J}. The resulting input sky image at 1.4 GHz is presented at the top panel of Figure \ref{fig:skysim}. Note that upgrading the resoultion using \textsc{healpix} creates numerical artifacts at small scales seen in the map. As we show in Section \ref{subsec:avoid}, the synchrotron component has a trivial impact on power spectrum on small scales.

Apart from the synchrotron radiation, the free-free emission is another important component of the smooth foregrounds (e.g. \citealt{2020MNRAS.496.1232L}). It is believed to be well-approximated by a Gaussian distribution (e.g. \citealt{2014MNRAS.444.3183A}). Alternatively, one can also use existing ${\rm H}_\alpha$ templates (e.g. \citealt{2018MNRAS.473.4242O}). We use the ${\rm H}_\alpha$ template of \cite{2003ApJS..146..407F} and the conversion factor from unit Rayleigh to brightness temperature as in \cite{2003MNRAS.341..369D} assuming a constant spectral index extrapolated from 2.326 GHz and 1.420 GHz. We upgrade the pixel size of the map from \textsc{healpix} $\rm N_{side} = 1024$ to \textsc{healpix} $\rm N_{side} = 8192$. The input sky image of the free-free emission at 1.4 GHz is shown in the middle panel of Figure \ref{fig:skysim}.

\subsection{Simulation of Extragalactic Foreground}
\label{subsec:pointsource}
The discrete extragalactic radio sources dominate the power spectrum of the total sky signal and are the biggest source of foreground contamination on small scales. Therefore, a realistic modelling of the discrete radio sources is crucial to the accurate simulation of the foregrounds (e.g. \citealt{2009MNRAS.394.1575L}). Three approaches have been used: Gaussian realizations from a given angular power spectrum (e.g. \citealt{2005ApJ...625..575S}), a point source catalogue for the particular patch of the sky of interest (e.g. \citealt{2021MNRAS.505.2039P}), or Poisson realizations using flux count statistics (e.g. \citealt{2013MNRAS.434.1239B}). We simulate the sky in a stochastic fashion and choose the flux count approach.

Radio galaxies are the most important targets for observations in the radio band, and the source counts are well studied by many surveys such as the NVSS survey \citep{1998AJ....115.1693C}, the VLA-Deep Field \citep{2003A&A...403..857B}, and more recently the \textit{MeerKAT} DEEP2 field \citep{2020ApJ...888...61M} and the ongoing VLASS survey \citep{2020PASP..132c5001L}. We follow the source count statistics at 1.4GHz described in \cite{2021ApJ...909..193M}, which uses NVSS and DEEP2 to account for the bright and faint end of the distribution respectively.

Two assumptions are made for the simulation of extragalactic sources. First, they are treated as point sources. Although large radio interferometers such as \textit{MeerKAT} can resolve most of the sources observed, the arcsecond angular scales are not of cosmological interest. Second, we assume that point sources above the flux density of $10^{-3}$Jy can be efficiently removed or avoided. This is based on the fact that some of the fields that are being investigated, such as COSMOS and the DEEP2 field, bright radio sources are avoided or modelled as demonstrated in \cite{2021MNRAS.505.2039P}. For other fields with more bright sources, we can expect the source ``peeling'' to be efficient above this flux cut given the sensitivity of the MIGHTEE survey \citep{2016mks..confE...6J}. {Relaxing the flux cut can lead to the foreground power saturating all $k_\parallel$ scales of our interests.} We leave a more careful treatment to the effects of imperfect peeling for future work. 

The all-sky distribution of point sources is generated with the following steps:
\begin{itemize}
    \item We grid the flux density range ($10^{-7}$ to $10^{-3}$ Jy) with 20
    narrow, logarithmic bins. In each bin, a total number of sources is calculated based on the source count statistics. We find that using more bins has little impact on the results.
    \item For each bin, a random sub-sample of pixels is selected and each is assigned a source with the average flux density of the bin. Note this is different from simulations of single dish observations due to the angular resolution of the observations. For lower resolution, in a flux bin, each pixel will have multiple sources on average. For each pixel a source number will be Poisson sampled. Here, the number of pixels will be larger than the number of sources, and therefore only a sample of pixels is uniformly selected.
    \item Following \cite{2021ApJ...909..193M}, for each pixel a Gaussian random spectral index with an average of -0.7 and standard deviation of 0.2 is assigned.
\end{itemize}
We present the simulated extragalactic foreground in the bottom panel of Fig. \ref{fig:skysim}.

Note that, the extragalactic point sources also cluster (e.g. \citealt{2003A&A...405...53O,2018MNRAS.474.4133H,2020A&A...643A.100S}). It is straightforward to generate Gaussian fluctuations based on input angular power spectrum to account for the clustering component. It scales approximately as $w(\theta) \propto \theta^{-0.8}$ \citep{1974ApJ...189L..51P} and is therefore negligible on small scales of our interest.

\subsection{Simulations of \althi\ Signal}
\label{subsec:hisim}
In this subsection, we describe the simulations of \hi\ used in this paper. The \hi\ signal is intrinsically different to the foregrounds, since the foregrounds are smooth in frequency while the \hi\ signal from a particular frequency corresponds to specific cosmological redshift. As a consequence of this, simulating all-sky maps of \hi\ at different frequencies are computationally expensive and difficult. Instead, we use 3D simulations of \hi\ in cubic boxes of cosmological volumes. While more sophisticated simulations of \hi\ can be found (e.g. \citealt{2017MNRAS.464.4204C,2018ApJ...866..135V}), {for our purpose} we use halo-model-based \citep{2002PhR...372....1C} log-normal simulations introduced in \cite{2019MNRAS.484.1007W} using \textsc{powerbox} \citep{2018JOSS....3..850M}. This formalism allows us to efficiently generate many realizations to test our pipeline. The simulation involves the following steps:
\begin{itemize}
    \item Assuming the halo mass function of \cite{2008ApJ...688..709T} and the halo bias of \cite{2010ApJ...724..878T}, we calculate the halo auto power spectrum using \textsc{halomod} \citep{2021A&C....3600487M} for a fixed redshift at the centre of our redshift bin $z\sim 0.27$. All steps before light-cone construction assume this fixed redshift. The power spectrum is then used as an input to generate log-normal discrete samples of halo centres using \textsc{powerbox} in a $\rm 250^3\,Mpc^3$ box with $800^3$ resolution, corresponding to $k_{\rm min} = 0.025\,{\rm Mpc^{-1}}$ and $k_{\rm max} = 10.0\,{\rm Mpc^{-1}}$.
    \item Each halo is then randomly assigned a halo mass based on the halo mass function of \cite{2008ApJ...688..709T} using \textsc{hmf} \citep{2013A&C.....3...23M}. 
    \item Assuming the galaxy halo occupation distribution (HOD) of \cite{2005ApJ...633..791Z}, a number of galaxies is assigned to each halo. By Bernoulli sampling with $p=\langle N_{\rm cen}^{\rm g}\rangle$, we determine whether each halo has a central galaxy. For halos with central galaxies, the number of satellite galaxies is determined by a Poisson distribution with mean $\langle N_{\rm sat}^{\rm g}\rangle$.
    \item The positions of central galaxies are assigned to the halo centres. For satellite galaxies, we assume the positions follow the probability distribution of a NFW profile \citep{1996ApJ...462..563N} and assign a random distance to the halo centre following that distribution. The angular positions of the galaxies with regard to the halo centres are then uniformly sampled and combined with the distance we can assign physical coordinates to the satellite galaxies.
    \item Each galaxy is assigned a \hi\ mass following log-normal distributions. The central and satellite \hi\ mass of each halo is calculated base on the \hi\ HOD $\langle M_{\rm cen,sat}^{\rm \hi} \rangle$ following \cite{2020MNRAS.493.5434S}. The mean of the distribution is set to be $M_{\rm field} = \langle M_{\rm cen,sat}^{\rm \hi} \rangle/\langle N_{\rm cen,sat}^{\rm g} \rangle$ and a standard deviation of $0.25M_{\rm field}$.
\end{itemize}

The above steps generate a catalogue of \hi\ galaxies with their positions and \hi\ mass in the comoving space. To map it onto the sky to construct the light-cone, we perform the following steps:
\begin{itemize}
    \item The centre of the simulation box is assumed to be the pointing centre and at the centre of the frequency range. Using that, we can assign a position vector to the pointing centre $X_{\rm cen}(\sin\theta\cos\phi, \sin\theta\sin\phi, \cos\phi)^{\rm T}$ relative to the observer, where $X_{\rm cen}$ is the comoving distance of the central redshift and $(\theta, \phi)$ is the sky coordinate of the pointing centre. 
    \item The position vector of each \hi\ galaxy can be solved for. The modulus of each position vector is the comoving distance for each source. The comoving distances can be conversely used to assign redshifts and subsequently frequency channels to \hi\ galaxies.
    \item The i$^{\rm th}$ galaxy is assigned a flux density following Eq. (\ref{eq:mass2flux}) as discussed in Appendix \ref{sec:hiflux}.
\end{itemize}

The resulting \hi\ simulation can be passed to the visibility simulation described in the next subsection.

\subsection{Simulation of Instrument}
\label{subsec:simvis}
In this subsection, we outline the simulation of visibility data using the sky input including foregrounds and \hi. We use \textsc{oskar} \citep{5613289} to generate visibility data. \textsc{oskar} is a \textsc{c++}-based simulation tool for radio interferometers, supporting GPU-accelerated computation for efficiency. It takes in a sky model, observation strategy, and the telescope array specifications, including station placement and primary beam, to simulate visibility data as a measurement set file. \footnote{\url{https://casadocs.readthedocs.io/en/latest/notebooks/casa-fundamentals.html}}

\begin{figure}
    \centering
    \includegraphics[width=\linewidth]{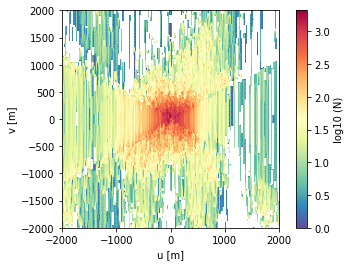}
    \caption{The $u$-$v$ coverage of simulated baselines. The $u$-$v$ plane is cut to only show the scales of our interest corresponding to $k \lesssim 20\,{\rm Mpc^{-1}}$, and gridded with a pixel size of (20m$)^2$ for presentation. The colormap denotes the number of baselines in each pixel. }
    \label{fig:uvcoverage}
\end{figure}

{We simulate observations by} the 64-dish \textit{MeerKAT} telescope array. We assume the beam of each dish is Gaussian with a full-width-half-maximum (FWHM) of 57.5 arcmin at 1.5GHz \citep{2020ApJ...888...61M}. {Bright foreground sources in sidelobes can have a non-trivial effect on foreground cleaning. In interferometric observations, this can be resolved by the secondary and position-dependent calibration steps (see e.g. Section 2 of \citealt{2022MNRAS.509.2150H}) and we leave the treatment of beam sidelobes for future work.} The pointing centre is set to be RA=150.12 deg and Dec=2.21 deg at the COSMOS field with an 11.2-hour tracking. The \textit{MeerKAT} telescope has a time resolution of 8 seconds, generating $\sim 10^7$ instantaneous baselines in one tracking. For computational efficiency, we choose the time resolution to be $\delta t=40$ seconds for the simulation and have verified that there is no visible difference compared to a full time-resolution simulation. The simulated $u$-$v$ coverage {of the 11.2-hour tracking} is presented in Figure \ref{fig:uvcoverage}.

We generate thermal noise per baseline following the radiometer equation \citep{2016era..book.....C}
\begin{equation}
    \sigma_{\rm N} = \frac{2k_{\rm B}T_{\rm sys}}{A_{\rm e}\sqrt{\delta f \delta t}},
\label{eq:sigmaN}
\end{equation}
where $T_{\rm sys}$ is the temperature of the receiver system, $A_{\rm e}$ is the effective aperture of the dish, $\delta f = 208.984$kHz is the channel bandwidth, and $\delta t = 40$s is the time resolution as mentioned above. We use $A_{\rm e}/T_{\rm sys} = 6.22 {\rm m^2\,K^{-1}}$\footnote{\url{https://www.sarao.ac.za/science/meerkat/about-meerkat/}} and generate random Gaussian noise for the complex visibility.


{The noise level of a single pointing for a small range of 220 channels at $z\sim 0.25-0.3$ is quite {high}, {with the amplitude of the noise covariance comparable to the foregrounds}. For the rest of the study, we consider the following two thermal noise scenarios. {In order to }isolate the effects of foregrounds from the thermal noise, we simulate visibility data with noise level scaled down by a factor of 40.
{From now on, this simulation is referred to as the ``low noise level'' case.} Note that we do not use the results of the low noise level case for realistic forecasts, but to showcase the effects of different foreground mitigation strategies in Section \ref{sec:fgremove}.}

{For the second noise scenario, we aim to show the robustness of the methods {for the realistic thermal noise level matching the corresponding sub-band of the MIGHTEE survey while simulating  observations of one field for simplicity.}
The MIGHTEE survey consists of 52 pointings with a total observational time of $\sim$1920 hours \citep{2016mks..confE...6J}. To match the noise level of our simulation of 11.2 hours to the entire survey, we first scale the thermal noise level down {dividing by} a factor of $\sqrt{(1920/52)/11.2}$, matching the total integration time for a single pointing. In real observations, the scaling of the thermal noise will be achieved through coherently averaging across different nights of visibility data (see e.g. \citealt{2020MNRAS.493.1662M}). We then further reduce the thermal noise level {dividing by} a factor of $\sqrt[4]{52}$, so that the thermal noise power spectrum matches the incoherent averaging of all 52 pointings. {From now on, the simulation with MIGHTEE-like noise level for $z \sim 0.25-0.3$ is referred to as the ``high noise level'' case.}}

\section{Foreground Mitigation in Visibility Space}
\label{sec:fgremove}
In this section, we examine the effects of three different foreground mitigation strategies in visibility space on the power spectrum measurement. First, we validate our power spectrum estimator using \hi\ only simulations in Section \ref{subsec:valid} . We confirm the wedge structure of the foreground power in our simulation in Section \ref{subsec:wedge}. We then apply foreground avoidance in Section \ref{subsec:avoid}. We further explore foreground subtraction by polynomial fitting the foreground covariance in Section \ref{subsec:subtraction}. Component separation for the visibility data using Principle Component Analysis (PCA) is presented in Section \ref{subsec:pca}. As mentioned in Section \ref{subsec:simvis}, the results shown here are for the low noise level case to isolate the effects of foregrounds from the thermal noise.

\subsection{Validation of Power Spectrum Estimator}
\label{subsec:valid}

\begin{figure}
    \centering
    \includegraphics[width=\linewidth]{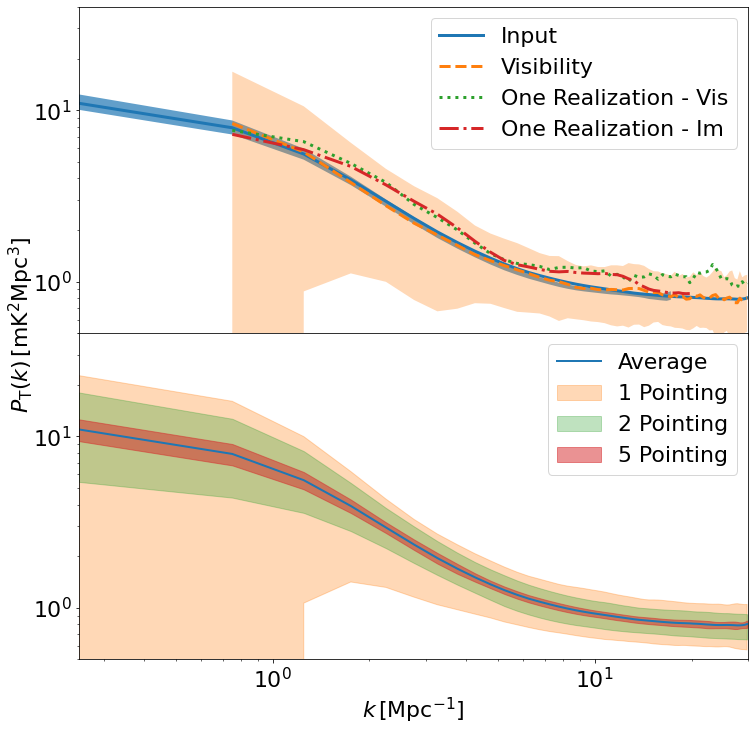}
    \caption{Top panel: The brightness temperature power spectrum of input \hi\ signal (``Input''), comparing to the results using the power spectrum estimator described in Section \ref{sec:estform}, averaged over 20 different \hi\ only realizations (``Visibility''). The shaded blue region shows the standard deviation of the input power spectrum. The shaded orange region shows the standard deviation of the output power spectrum. Note that for the power spectrum from simulated visibility, we cut the first $|\bm{u}|$ bin and consequently first $k$ bin for reasons discussed in Section \ref{subsec:avoid}. Therefore, the orange region starts at larger $k$ than the input. The green, dotted line shows the result of one specific realization using the power spectrum estimator. The red, dash-dotted line shows the input \hi\ power spectrum from the simulated lightcone within the telescope FoV from the same realization, corresponding to the power spectrum of the input image within the FoV ("One Realizaiton - Im"). Bottom panel: The variance of \hi\ power spectrum due to the limited survey volume for 1, 2, and 5 pointings as shown in the shaded regions comparing to the average \hi\ power spectrum of the entire box.}
    \label{fig:hionly}
\end{figure}

In this subsection, we validate the power spectrum estimator described in Section \ref{sec:estform}. We simulate visibility data with only \hi\ input, and pass the output data to the estimator. We choose to grid the $u$-$v$ plane with $\Delta u = \Delta v = 10$ from 0 to 6000. Here, we apply gridding for computational efficiency, reducing the number of baselines from $N_{\rm bl} \sim 2\times 10^6$ to $N_{\rm bl} \sim 3\times 10^4$. The results from gridded visibility will be sub-optimal \citep{2014PhRvD..90b3018L}, but in the case of noise covariance being dominant, the effects on the uncertainties are small.

{The power spectrum is computed in cylindrical space with bandpowers further averaged into $\{|\bm{u}|\}$ bins}. We choose the edges of the $u$-$v$ annulus bins to be $\{|\bm{u}|_\alpha\} = [0,100,200,...,6000]$ and $\{|\bm{k}_\perp|_\beta\} = 2\pi\{|\bm{u}|_\alpha\}/X$ for the cylindrical power spectrum used in Eq. (\ref{eq:ptbpda}) and Eq. (\ref{eq:1dtps}). {From now on, we refer to $\{|\bm{u}|_\alpha\} = [0,100,200,...,6000]$ as the ``annulus $\{|\bm{u}|\}$ bins'' used in our simulation.} For the one-dimensional (1D) power spectrum, we choose the edges of the $k$ bins to have $\Delta k = 0.5\,{\rm Mpc^{-1}}$ from 0.5 ${\rm Mpc^{-1}}$ to 20 ${\rm Mpc^{-1}}$. The delay power spectrum is calculated using Eq. (\ref{eq:pdanofgn}) and converted to the temperature power spectrum using Eq. (\ref{eq:ptbpda}) and Eq. (\ref{eq:1dtps}).

{We simulate 20 realisations with different dark matter halo and \hi\ content} and show the mean and the standard deviation of the power spectrum results in the top panel of Figure \ref{fig:hionly}. The blue line shows the mean of input \hi\ boxes averaged over the realizations (``Input''). The blue shaded region represents the standard deviation of the input power spectrum. The averaged output of the power spectrum estimator is shown as the orange dashed line. As one can see, the results from the estimator agree tightly with the input. The standard deviation of the output power spectrum is shown as the orange shaded region. The large variance is due to the fact that the input \hi\ box is much larger than the telescope FoV. At scales of 1-halo correlation and shot noise, the variance of the power spectrum is large for the small volume of one pointing. The number density of massive \hi\ galaxies in the telescope FoV fluctuates from one point to another. To illustrate this variance, we calculate the variance of the input power spectrum using bootstrapping by dividing the input \hi\ box into sub-boxes matching the survey volume of 1, 2, and 5 pointings and present the results in the bottom panel of Figure \ref{fig:hionly}. The variance from the volume of 1 pointing agrees well with the variance of the output. The variance decreases as the survey volume goes up. For the MIGHTEE survey with $\sim 50$ pointings, the effect of this variance will be negligible.

We have shown that the output power spectrum for one realization and one pointing will not correspond to the input simulation box due to small survey volume, but agrees very well after being averaged over multiple pointings. One may think that instead of the entire \hi\ box, the output power spectrum is the power spectrum of the part of the box within the telescope FoV. We further clarify that, the output \hi\ power spectrum from one realization does not correspond exactly to ``the input image'', i.e. the \hi\ signal within the telescope primary-beam FoV as well. The \hi\ sources outside the telescope FoV, though being heavily attenuated, still contribute to the visibility data and, therefore, the output from our estimator differs from ``the input image''. In the top panel of Figure \ref{fig:hionly}, we show, for one realization, the output power spectrum (``One Realization - Vis'') against the power spectrum of the \hi\ signal within the telescope FoV (``One Realization - Im'').

As discussed in Section \ref{subsec:simvis}, we aim to forecast for the MIGHTEE survey using one pointing. We choose the specific \hi\ realization shown in Figure \ref{fig:hionly}, with the foreground signal and thermal noise added, to present our results for the rest of this section.

\subsection{Foreground wedge}
\label{subsec:wedge}

\begin{figure*}
    \centering
    \includegraphics[width=\linewidth]{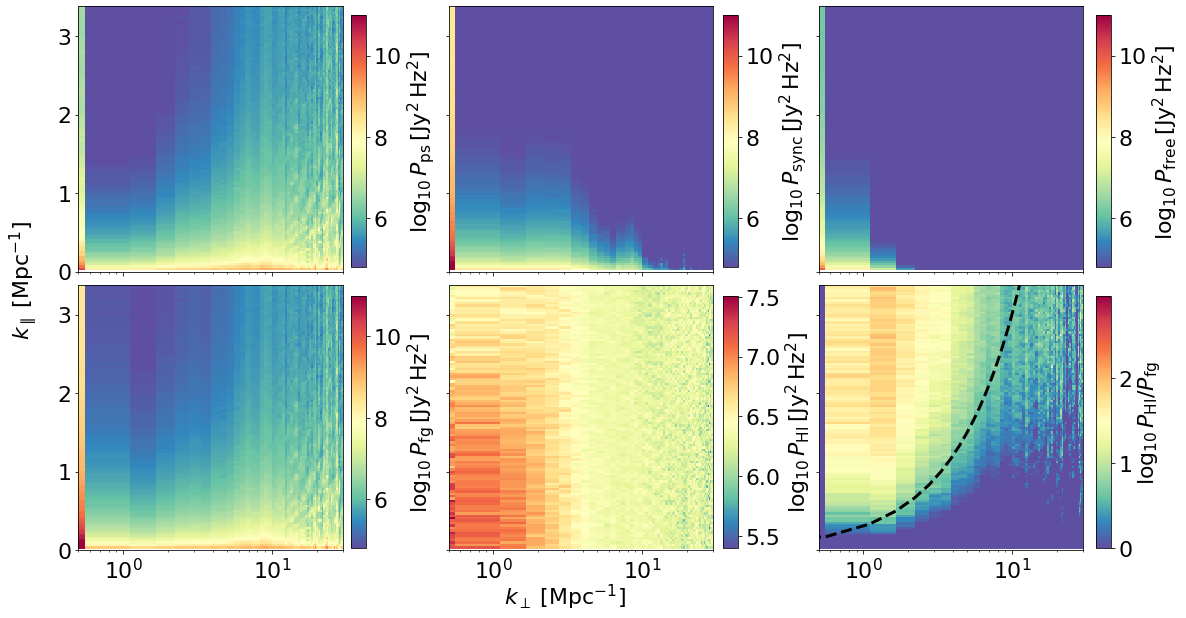}
    \caption{The cylindrical delay power spectrum for different components of the simulation. The delay time $\gamma$ and the radius of annulus $u$-$v$ bins $|\bm{u}|$ have been converted to $k$-space coordinates following Eq. (\ref{eq:kconvert}). All delay power spectra are in Jy$^2$Hz$^2$. Top-left panel: the delay power spectrum of the extragalactic point sources. Top-centre panel: the synchrontron radiation. Top-right panel: the free-free emission. Lower-left panel: the total foreground emission. Lower-centre panel: the \hi\ signal. Lower-right panel: the ratio of the \hi\ power spectrum over the total foreground power spectrum. Values below 1 are masked for clearer view of the foreground wedge. The black dashed line shows a schematic illustration of the foreground wedge. Note that this line is simply for illustration and the actual criteria used to avoid foreground in cylindrical $k$-space are discussed in Section \ref{subsec:avoid}. The color bars for top three panels and the bottom left panel are kept the same for direct comparison.}
    \label{fig:fgwindow}
\end{figure*}

In this subsection, we present the cylindrical delay power spectrum $P_{\rm d}(\bm{k}_{\perp},k_\parallel)$ for each component of the simulation. $P_{\rm d}(\bm{k}_{\perp},k_\parallel)$ contains information on the clustering of \hi\ on both the transverse plane and along the frequency direction. Since \hi\ is a cosmological signal, there is no preferred direction for the \hi\ power spectrum except for the finger-of-god (FoG) effect. The foregrounds, on the other hand, are smooth in frequency. Therefore, the foreground power resides mostly in low $k_\parallel$ modes and decreases sharply as $k_\parallel$ increases. It is important to use simulations to understand which $(\bm{k}_\perp, k_\parallel)$ modes are contaminated by the foregrounds in order to extract information on \hi.

We use the simulated visibility data of different components without the thermal noise to visualize the ``foreground wedge'' \citep{2014PhRvD..90b3019L}. In this subsection, we calculate the delay power spectrum of different components using Eq. (\ref{eq:estpdalpha}), with uniform weighting $\mathbf{R} = \mathbf{I}$  to calculate $\mathbf{E}^{\rm d}_{\alpha}$ in Eq. (\ref{eq:eda}). We do not include frequency tapering, because as shown later the foreground contamination in our case is severe, and we find that the tapering does not have significant effects on containing the foreground wedge. We also do not perform inverse covariance weighting for this subsection, the reasons for which will be discussed in Section \ref{subsec:avoid}. The cylindrical power spectra shown throughout this paper are the outputs of Eq. (\ref{eq:estpdalpha}), with no $k_\parallel$-filter applied as we only filter out small $k_\parallel$ scales when converting to 1D power as in Eq. (\ref{eq:1dtps}).

We present the cylindrical power spectrum for each component of the signal in Figure \ref{fig:fgwindow}. The top panels, from left to right, show the delay power spectra of point source, synchrontron and free-free emissions respectively. The sum of these three components is shown in the lower left panel. The foreground power is significantly larger in low $k_\parallel$ modes than in high $k_\parallel$ modes. The dominant component of the foregrounds is extragalactic radio sources, especially for large $|\bm{k}_\perp|>1\, {\rm Mpc}^{-1}$ where it is at least 2 orders of magnitude larger than synchrontron and the free-free emission. The point source power spectrum increases and leaks more into high $k_\parallel$ modes when considering smaller angular scales, while synchrontron and the free-free emission behave in the opposite way. On short baselines corresponding to $|\bm{u}|<100$ and $|\bm{k}_\perp|\lesssim 0.5 \, {\rm Mpc}^{-1}$, the delay power spectrum of foreground is exceptionally large comparing to other $|\bm{k}_\perp|$ bins. This is for several reasons. First, at the particular pointing we are simulating which is close to the Galactic plane, the synchrontron radiation is bright at large scales; for $|\bm{k}_\perp|\lesssim 0.5 \, {\rm Mpc}^{-1}$, the synchrontron radiation is at least an order of magnitude larger than other components. Second, the first $u$-$v$ bin, namely $|\bm{u}|$ from 0 to 100, contains contributions from the shortest baselines $|\bm{u}|\sim 10$ to 100, covering an order of magnitude in angular scale. The rapidly decreasing power spectrum at these scales means the overall contribution to the first $|\bm{u}|$ bin is overestimated. Finally, the simulation of a limited patch of the sky introduces a windowing effect, leaking the power of the monopole into low $|\bm{k}_\perp|$ modes. When we convert the delay power spectrum to the temperature power spectrum, the conversion matrix mixes angular modes that leads to further overestimation for nearby $|\bm{u}|$ bins. Therefore, we discard the first $|\bm{u}|$ bin and start from $|\bm{u}|>100$. 

The bottom-centre panel of Figure \ref{fig:fgwindow} shows the delay power spectrum of \hi. As expected, the \hi\ power spectrum only depends on the 1D wavenumber $k = \sqrt{|\bm{k}_\perp|^2+k_\parallel^2}$ as we do not include redshift space distortions. Comparing to the foreground power spectrum, \hi\ can be orders of magnitude larger for high $k_\parallel$ modes. We show the ratio between \hi\ and total foreground power in the bottom right panel. The ``wedge'' where foreground contamination is most severe is at low $k_\parallel$ scales and gets larger at longer baselines where the power spectrum of point sources increases. It leaves out a potential observation window for \hi\ power spectrum. At a given $|\bm{k}_\perp|$, we can filter out low $k_\parallel$ modes and only take the measurement of power spectrum at higher $k_\parallel$. This method of excluding foreground contaminated modes is called foreground avoidance, which we further explore in the next subsection.

\subsection{Foreground Avoidance}
\label{subsec:avoid}
In this subsection, we include the contributions from foregrounds and thermal noise and use a foreground avoidance strategy to measure the \hi\ power spectrum from our simulation. The estimator described in Section \ref{sec:estform} applies foreground avoidance in Eq (\ref{eq:1dtps}). Namely, it includes $w_i(\bm{k}_\perp,k_\parallel)$ to encode the wedge criteria (defined later), which has the impact that when $k_\parallel$ is too small, $w_i$ is zero and the foreground contaminated modes are filtered out. We can then calculate the resulting 1D \hi\ power spectrum.

To estimate the temperature power spectrum from visibility data we follow Eq. (\ref{eq:estpdalpha}). Since we do not analytically model the foreground covariance and simply avoid contaminated modes, $\mathbf{N}^{\rm psd}+\mathbf{C}_{\rm fg}^{\rm psd}$ in Eq. (\ref{eq:estpdalpha}) is the noise covariance {for a single $u$-$v$ grid {point} $p$} in our simulation
\begin{equation}
    \mathbf{N}_{ij}^{p} = \delta_{ij} \sigma^2_{\rm N}/N_i^p,
\label{eq:noisecov}
\end{equation}
where $N_i$ is the number of baselines for the $i^{\rm th}$ frequency channel {in the $p^{\rm th}$ $u$-$v$ grid}. Throughout this paper, the data covariance $\mathbf{C}$ is taken to be the empirical data covariance from visibility data in each annulus $|\bm{u}|$ bin used to estimate the uncertainty following Eq. (\ref{eq:sigmaalpha}). {The empirical data covariance $\mathbf{C}$ is calculated from the visibility data following Eq. (\ref{eq:empdata}) as discussed in Section \ref{sec:estform}.}
{Note that here $\hat{C}_{ff'}(|\bm{k}_\perp|)$ is calculated from the visibility data before gridding. We then subtract the noise variance $\sigma^2_{\rm N}$ and add the noise covariance of the gridded data following Eq. (\ref{eq:noisecov}). Throughout this paper, $\sigma^2_{\rm N}$ is treated as a known quantity and, therefore, the noise covariance is also known.}

Real observations are likely to be noise dominated with the exact position of the foreground wedge also unknown. Assuming that we are working blind, we start with the generic criteria to be used in Eq. (\ref{eq:1dtps}). The selection function $w_i(\bm{k}_\perp,k_\parallel)$ returns 1 only if $(\bm{k}_\perp,k_\parallel)$ falls into the 1D $k$ bin and satisfies
\begin{equation}
    k_\parallel > c_k\,k_\perp,
\end{equation}
where $c_k$ is a coefficient describing the position of the wedge.

\begin{figure}
    \centering
    \includegraphics[width=\linewidth]{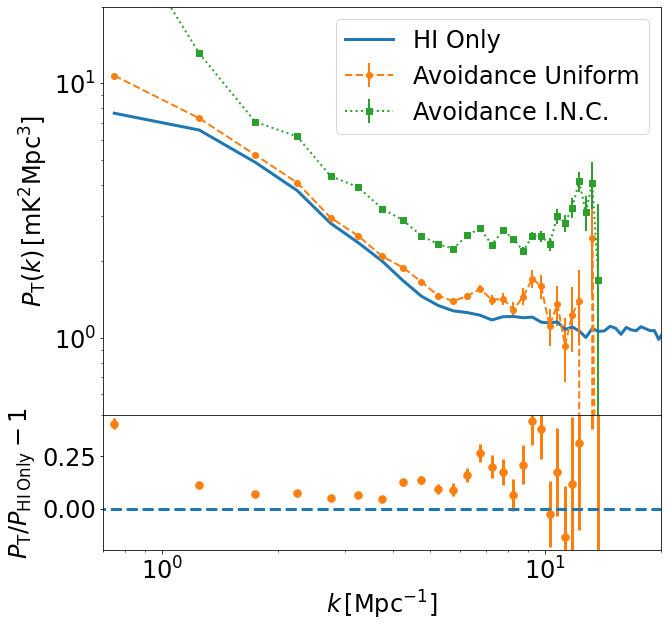}
    \caption{Top panel: the 1D brightness temperature power spectrum from \hi\ only visibility data (``HI only''), low thermal noise simulation using foreground avoidance method described in Section \ref{subsec:avoid} (``Avoidance Uniform''), and the same avoidance method but with inverse noise covariance weighting (``Avoidance I.N.C.''). Bottom panel: the fractional difference between the ``HI only'' and the ``Avoidance Uniform'' results. }
    \label{fig:fgavoid}
\end{figure}

To determine the value of $c_k$, we start with $c_k = 0$ and increase $c_k$ iteratively with a step size of 0.05. We find that for $c_k = 0.2$ and $c_k=0.25$, the difference in power spectrum estimation is within 5\%. The convergence suggests that the foreground power is largely avoided. The \hi\ power spectrum estimated using foreground avoidance with $c_k = 0.25$ (''Avoidance Uniform'') is shown in Figure \ref{fig:fgavoid}. {Note that, the standard ``horizon criteria'' (see Eq. (13) of \citealt{2014PhRvD..90b3018L}), assuming a maximum angular extent $\sin \theta_0 = 1$, corresponds to $c_k \sim 0.232$, suggesting that the foreground contamination is likely to be {the dominant constraint} for MIGHTEE observations. }

{As seen by the fact that the estimated power is always higher compared to the input, the foreground power contamination is present for all $k$ modes and results in an overestimation of the power spectrum by around 10\%. For wavenumbers $k\sim 1\,{\rm Mpc^{-1}}$, the contamination is more severe because at larger angular scales synchrotron and free-free emissions have a large effect on the delay power spectrum as shown in Figure \ref{fig:fgwindow}. The contamination is also more severe for $k>5\,{\rm Mpc^{-1}}$ modes, since at higher $k_\perp$ the power spectrum of point sources is much larger and leaks more into the window. }

\begin{figure}
    \centering
    \includegraphics[width=\linewidth]{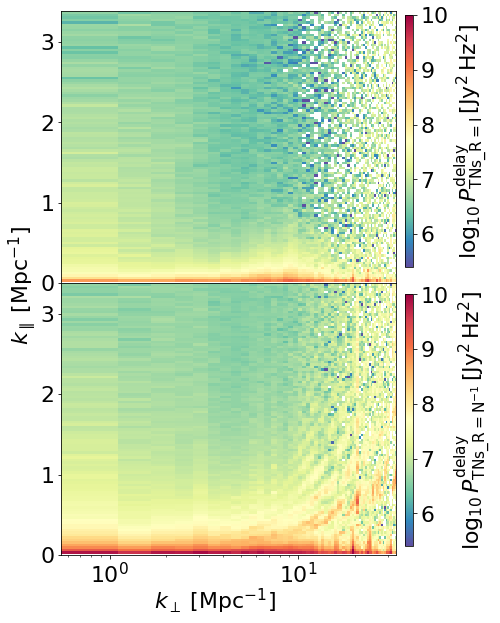}
    \caption{Top panel: the cylindrical delay power spectrum for the uniform weighting case, with noise covariance subtracted for the low noise level case ($P^{\rm delay}_{\rm TNs\_R=I}$). Bottom panel: the cylindrical delay power spectrum for the inverse noise covariance weighting case, with noise covariance subtracted for the low noise level case as described in Section \ref{subsec:avoid} ($P^{\rm delay}_{\rm TNs\_R=N^{-1}}$). The empty (white) regions at high $|\bm{k}_\perp|$ in the figure are for negative power due to thermal noise covariance subtraction. The red and orange regions ($P^{\rm delay}_{\rm TNs}>10^9 \, {\rm Jy^2Hz^2}$) in the bottom panel is much larger than ones in the upper panel, suggesting much more severe foreground leakage into higher $k_\parallel$ modes.}
    \label{fig:dps2dlowtn}
\end{figure}

{The noise level of a $u$-$v$ grid point varies across frequencies due to the change in observing wavelengths and therefore applying inverse covariance weighting gives uneven weights across frequency channels, mixing different $k_\parallel$ modes in the estimator.} The mixture of different $k_\parallel$ modes can lead to a spillover of the foreground power into the observational window (see also \citealt{2018ApJ...868...26C}). To illustrate this, in Figure \ref{fig:fgavoid}, we also show the power spectrum estimation made by choosing {inverse noise covariance weighting $\mathbf{R} = \mathbf{N}^{-1}$} (''Avoidance I.N.C.''), keeping everything else the same as the {uniform weighting $\mathbf{R} = \mathbf{I}$} case. 

Note that here we do not choose the inverse of the total data covariance, since in this low thermal noise case foregrounds contribute a substantial fraction of the total covariance. In realistic observations, on the other hand, the noise covariance is expected to dominate. Therefore, we choose $\mathbf{R}$ in Eq. (\ref{eq:eda}) to be the inverse noise covariance to illustrate the mode-mixing. As one can see from Figure \ref{fig:fgavoid}, when applying $\mathbf{R} = \mathbf{N}^{-1}$, the power spectrum is significantly overestimated. {Our claim that the foreground leakage into higher $k_\parallel$ modes is responsible can be verified by investigating the cylindrical delay power spectrum, as we show in Figure \ref{fig:dps2dlowtn}. For $\mathbf{R} = \mathbf{I}$ case in the top panel, the foreground wedge is most visible at $k_\parallel \lesssim 0.3 \, {\rm Mpc^{-1}}$ as the red and orange regions, where it is 2-4 orders of magnitude larger than the \hi\ signal. For $\mathbf{R} = \mathbf{N}^{-1}$, on the other hand, the red and orange regions can be seen at $k_\parallel \sim 0.5 \, {\rm Mpc^{-1}}$. In short, if foreground contamination is severe, additional weighting further mixes different $k_\parallel$ modes, further contaminating high $k_\parallel$ modes which leads to the overestimation in the \hi\ power spectrum.}

\subsection{Subtracting Foreground using Fitted Foreground Covariance}
\label{subsec:subtraction}
In this subsection, we explore polynomial fitting of the foreground covariance in visibility space, utilizing the fact that the foreground emission is smooth in frequency. {The polynomial fitting of the covariance is similar to} polynomial fitting of the signal along the line-of-sight direction in the image domain (e.g. \citealt{2009ApJ...695..183B}) or in $u$-$v$ space (e.g. \citealt{2010MNRAS.405.2492H}). The main difference is that we perform the fitting on the empirical data covariance instead of directly on the visibility data. If the subtraction is perfect, i.e. the resulting fit is the foreground covariance, there should be no increase in the measurement errors. 

{We split the visibility data into the same $\{|\bm{u}|_\alpha\}$ bins for the power spectrum estimation defined in Section \ref{subsec:valid} and calculate the empirical data covariance using Eq. (\ref{eq:empdata}).} Note that for most $|\bm{u}|$ bins, the dominant foreground component is the point sources which are believed to be approximately Poisson distributed and assumed to be so in our simulation. In the limit of a narrow frequency range, the covariance matrix of the Poisson foreground should be real-valued (see derivations in Section 4 of \citealt{2017ApJ...845....7M}). We verify that this is indeed the case for our visibility data, where the real part of the each element in the covariance matrix is at least an order of magnitude larger than the imaginary part.

\begin{figure}
    \centering
    \includegraphics[width=\linewidth]{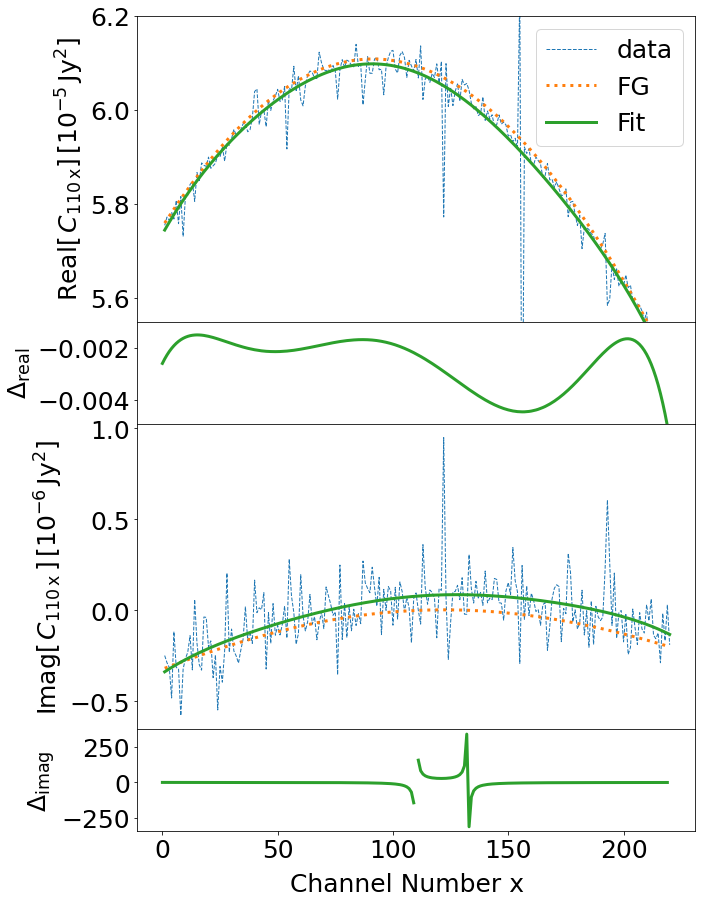}
    \caption{Top panel: the real part of the $110^{\rm th}$ row (the centre frequency row) of the empirical data covariance matrix in the $6^{\rm th}$ $|\bm{u}|$ bin corresponding to $k_\perp \sim 3.5 {\rm Mpc^{-1}}$ (``data''). The same row in the same bin of the covariance matrix for foreground component only visibility is shown for comparison (``FG''). The extracted smooth component using $6^{\rm th}$ order polynomial fits is also shown (``Fit''). Below it is the fractional difference between the real part of the fitted covariance and the actual foreground covariance on the $110^{\rm th}$ row in the $6^{\rm th}$ $|\bm{u}|$ bin. Bottom panel: the same with the top panel, except for the imaginary part of the covariance matrix.}
    \label{fig:covfitting}
\end{figure}

The resulting data covariance can then be processed to extract the foreground by applying a $k_\parallel$ filter to remove the rapidly oscillating \hi\ component. We do this by applying polynomial fitting to each row of the covariance matrix, aiming to extract the smoothly varying part. We find that $6^{\rm th}$ order fitting respectively for real and imaginary part of the empirical covariance matrix is sufficient, as showcased in Figure \ref{fig:covfitting}.  

\begin{figure}
    \centering
    \includegraphics[width=\linewidth]{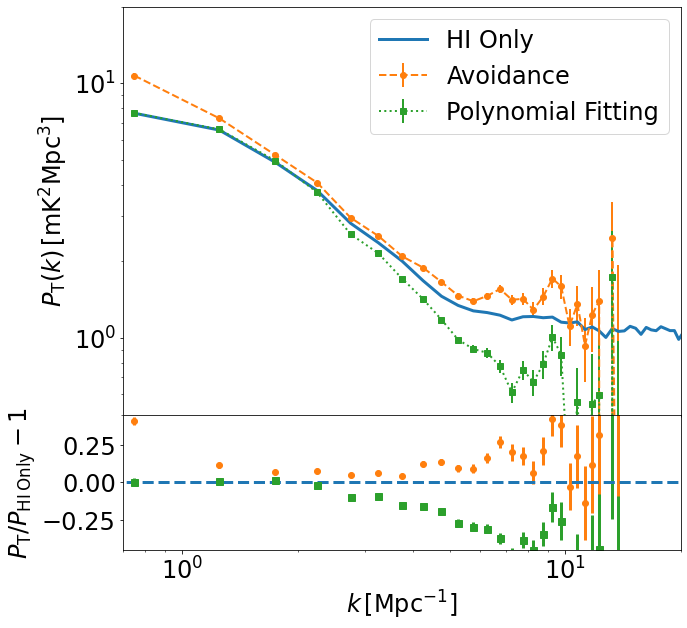}
    \caption{The 1D brightness temperature power spectrum from the \hi\ only visibility data (``HI Only''), the low thermal noise simulation using foreground avoidance method (``Avoidance''), and the foreground subtraction method using polynomial fitting of the empirical data covariance (``Polynomial Fitting''). Shown also is the fractional difference between the ``HI Only'', ``Avoidance'' and ``Polynomial Fitting'' results. While the avoidance under subtract the foregrounds, the polynomial fitting over subtracts on small scales.}
    \label{fig:fgsub}
\end{figure}

\begin{figure}
    \centering
    \includegraphics[width=\linewidth]{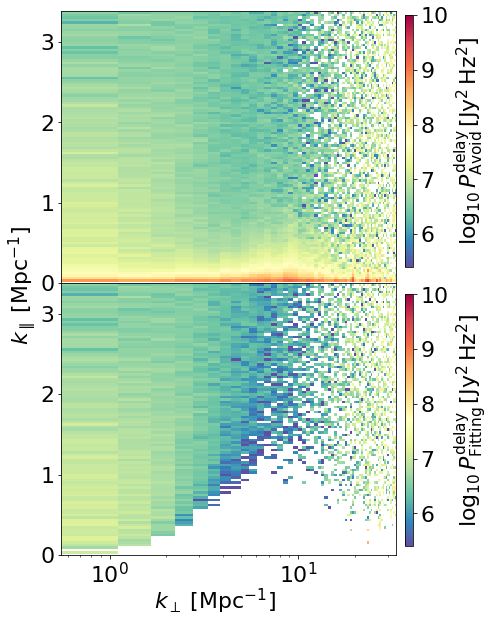}
    \caption{Top panel: the cylindrical delay power spectrum with noise covariance subtracted using foreground avoidance method ($P^{\rm delay}_{\rm Avoid}$). Bottom panel: the cylindrical delay power spectrum with noise covariance subtracted using polynomial fitting ($P^{\rm delay}_{\rm Fitting}$). The empty (white) regions indicate overcleaning which results in negative power.}
    \label{fig:dpssub}
\end{figure}

We take the fitting results as $\mathbf{C}_{\rm fg}^{\rm psd}$ and use Eq. (\ref{eq:noisecov}) to calculate the noise covariance to subtract noise and foreground covariances via Eq. (\ref{eq:estpdalpha}). We choose {uniform weighting $\mathbf{R} = \mathbf{I}$} with the same $c_k = 0.25$ as in Section \ref{subsec:avoid} and present our result in Figure \ref{fig:fgsub}. As shown in the figure, comparing to the direct avoidance, subtracting the fitted foreground covariance corrects the amplitude at large $k\lesssim 1\, {\rm Mpc^{-1}}$ scales. However, it overcleans on $k\gtrsim 2\,{\rm Mpc^{-1}}$ which results in a $> 10\%$ signal loss. The overcleaning is due to the fact that, despite in most cases the fitted covariance matches the foreground covariance up to per cent level as shown in the top panel of Figure \ref{fig:covfitting}, it still wrongly subtracts \hi\ features. This can be seen in the bottom panel of Figure \ref{fig:covfitting} where for the imaginary part of the covariance, foreground is relatively small. Comparing the fitted covariance with the actual foreground covariance, we can see that some additional structure over a large frequency range is mistaken to be a contribution from the foregrounds. This results in overcleaning, typically in relatively low $k_\parallel$ modes. 

We also show the cylindrical delay power spectrum to further verify this in Figure \ref{fig:dpssub}. Comparing to direct avoidance in the upper panel, the subtraction overcleans signal at lower $k_\parallel$ which results in lower and even negative power as shown in the bottom panel of Figure \ref{fig:dpssub}.
 
\subsection{Foreground Removal with Principal Component Analysis}
\label{subsec:pca}
In this section, we examine component separation in visibility data to mitigate foreground contamination. From Section \ref{subsec:avoid}, we see that to exclude the foreground power leakage into the observation window, a foreground removal method is required. We found that direct subtraction is likely to result in signal loss on small scales despite correcting the amplitude of the power spectrum well at large scales. It highlights the need for component separation techniques. 

Here, we provide a case study of the most standard technique of PCA. PCA is proven to be very robust and works similarly well comparing to methods such as fastICA (see e.g. \citealt{2021MNRAS.504..208C}). For sky maps/images, maps in different frequency channels are mean-centred and then an empirical frequency-frequency covariance matrix can be constructed for data analysis (e.g. \citealt{2015MNRAS.454.3240B}). For visibility data on the other hand, each baseline corresponds to a different Fourier mode and can not be processed in the same way. As discussed in Section \ref{subsec:avoid}, we follow \cite{2015PhRvD..91l3011D} to calculate empirical covariance using Eq. (\ref{eq:empdata}) in annulus $|\bm{u}|$ bins and perform PCA in each bin independently. 

\begin{figure}
    \centering
    \includegraphics[width=\linewidth]{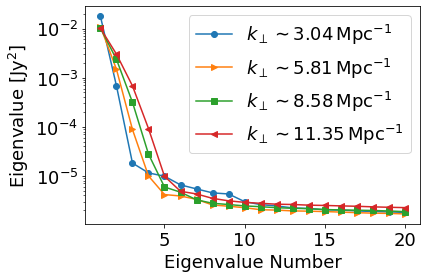}
    \caption{The first 20 eigenvalues, ranked from the biggest to the smallest, of the empirical data covariance matrices for the $k_\perp \sim 3.04\,{\rm Mpc^{-1}}$, $\sim 5.81\,{\rm Mpc^{-1}}$, $\sim 8.58\,{\rm Mpc^{-1}}$ and $ \sim 11.35\,{\rm Mpc^{-1}}$ bins. Note the significant drop-off in the first few bins and then the plateau beyond the fifth eigenvalue.}
    \label{fig:eigval}
\end{figure}

The first 20 eigenvalues of the empirical covariance matrices in four different bins are presented in Figure \ref{fig:eigval}. The largest component is typically an order of magnitude higher than the second eigenvalue, and the eigenvalues hit the plateau at about the fifth eigenvalue, suggesting there is a mixture of \hi\ and foregrounds in these modes. This is also supported by the fact that for higher $k_\perp$ the eigenvalues hit the plateau later, due to the more severe foreground contamination at higher $k_\perp$ we saw in Figure \ref{fig:fgwindow}. {We find that choosing $N_{\rm fg}=5$ removes foreground at small $k_\parallel$ modes but overcleans the \hi\ over a wide range of scales, something that we would like to avoid. We instead find $N_{\rm fg}=2$ suits our purpose best. It still leaves a surpressed foreground wedge, which we can avoid by applying a ``loose wedge criterion''.}

\begin{figure}
    \centering
    \includegraphics[width=\linewidth]{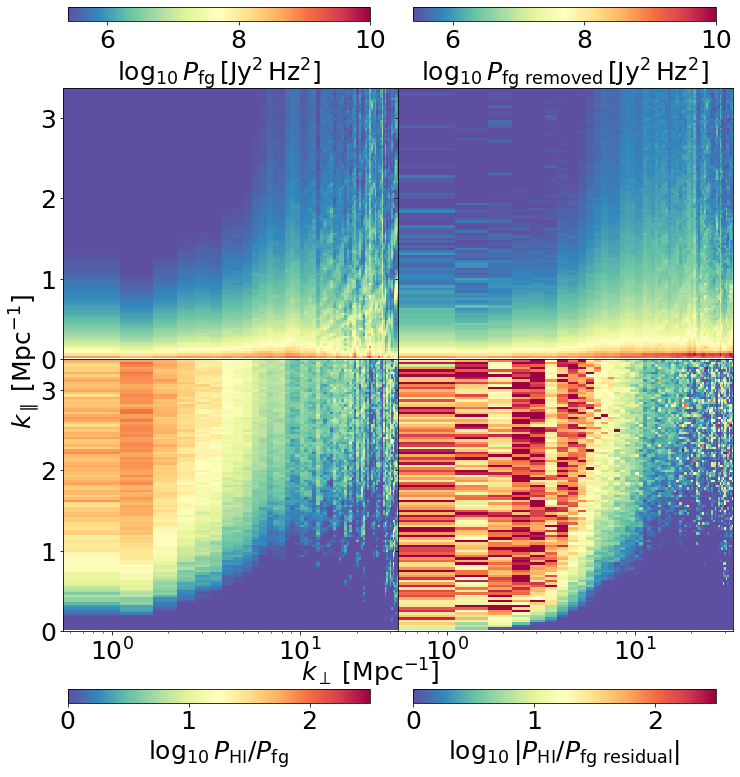}
    \caption{Top-left panel: the cylindrical delay power spectrum of total foreground in Jy$^2$Hz$^2$. Top-right panel: the cylindrical delay power spectrum of the removed component by PCA for low thermal noise simulation. {The color bar is kept the same to the upper left panel for direct comparison.} Bottom-left panel: the ratio between the cylindrical delay power spectrum of the \hi\ component of visibility data over the power spectrum of the foregrounds. Bottom-right panel: the absolute value of ratio between the cylindrical delay power spectrum of the \hi\ component of visibility data over the power spectrum of the residual of the foregrounds. Values below 1 are set to be 1 for better presentation in bottom-left and bottom-right panels.}
    \label{fig:pcawindow}
\end{figure}

PCA returns the source mixing matrix $\hat{\mathbf{A}}$ which extracts the foreground components. Note that, following the framework of Section \ref{sec:estform}, we should include $\hat{\mathbf{A}}$ in $\mathbf{R}$ to construct our estimator. However, the component separation here is performed on the full visibility data to ensure sufficient cleaning, whereas for power spectrum estimation is applied to the gridded visibilities for computational efficiency. Therefore, we do not include $\hat{\mathbf{A}}$ in $\mathbf{R}$. Here, the foreground is subtracted before gridding and a new gridded data vector $\mathbf{V}'$ with foreground components removed is used, with the assumption that the operations of gridding and foreground subtraction are commutable.

We present the cylindrical delay power spectrum with {uniform weighting} $\mathbf{R} = \mathbf{I}$ in Figure \ref{fig:pcawindow}. For the $N_{\rm fg} =2$ case we can see that the reconstructed foreground matches the clustering of the actual foreground quite well by comparing the upper panels of Figure \ref{fig:pcawindow}. {PCA overcleans the foregrounds at low $k_\perp$, high $k_\parallel$ modes as seen by comparing the top-left and top-right panels of Figure \ref{fig:pcawindow}}. Slight overcleaning can be acceptable, as long as the residual does not leak too much negative power into the observation window. The amplitude of the negative residual power is orders of magnitude smaller than \hi, as shown in the lower right panel of Figure \ref{fig:pcawindow}. Comparing the bottom two panels, we see that PCA widens the observation window at lower $k_\perp$ by efficiently cleaning the foreground power, while not having visible improvements on higher $k_\perp$ modes. Comparing to foreground avoidance, a loose selection criterion can be applied for small $k_\perp$. We find that the power spectrum converges on all scales when $c_k = 0.04$ for $k_\perp<5\,{\rm Mpc^{-1}}$ and $c_k = 0.3$ for $k_\perp>5\,{\rm Mpc^{-1}}$. 

\begin{figure}
    \centering
    \includegraphics[width=\linewidth]{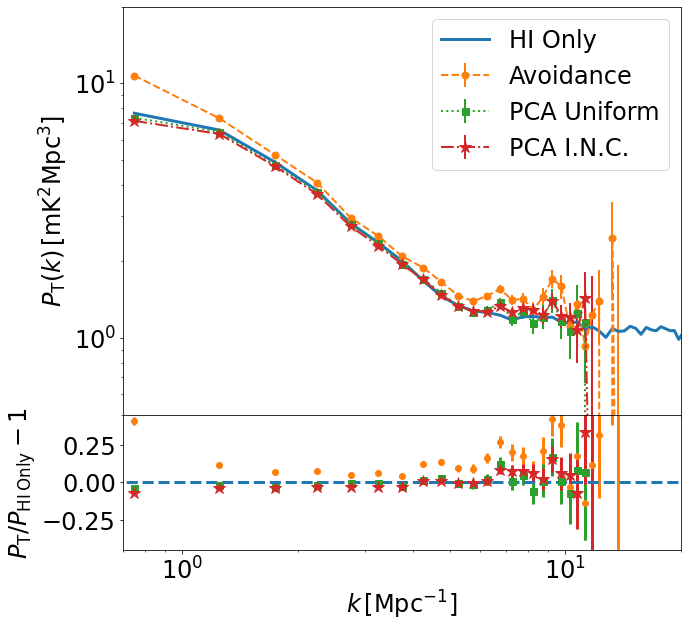}
    \caption{The 1D brightness temperature power spectrum from the \hi\ only visibility data (``HI Only''), simulation with low thermal noise level using foreground avoidance method described in Section \ref{subsec:avoid} (``Avoidance''), simulation with low thermal noise level using PCA foreground removal with uniform weighting (``PCA Uniform'') and with inverse noise covariance weighting (``PCA I.N.C.''). In addition, we have provided the fractional difference between the estimated 1D brightness temperature power spectrum and the \hi\ only simulation for low thermal noise case. }
    \label{fig:pcalowtn}
\end{figure}

\begin{figure}
    \centering
    \includegraphics[width=\linewidth]{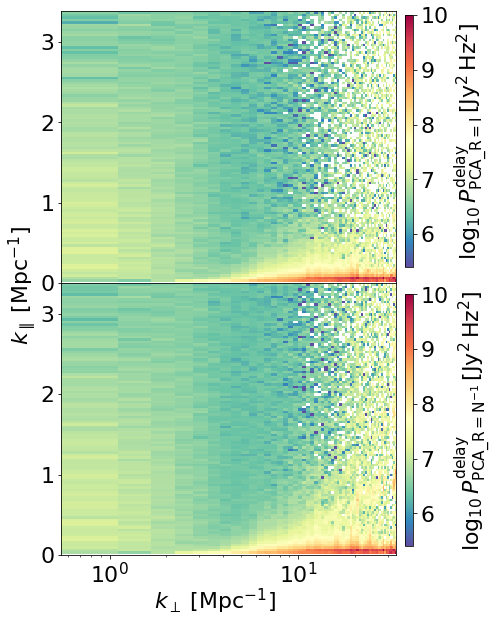}
    \caption{Top panel: the cylindrical delay power spectrum with noise covariance subtracted using PCA and uniform weighting described in Section \ref{subsec:pca} for the low thermal noise simulation ($P^{\rm delay}_{\rm PCA\_ R=I}$). Bottom panel: the same with the top panel but with noise covariance subtracted using PCA and inverse noise covariance weighting for the low thermal noise simulation ($P^{\rm delay}_{\rm PCA\_ R=N^{-1}}$).}
    \label{fig:dpspcalowtn}
\end{figure}

We show the results for the 1D temperature power spectrum in Figure \ref{fig:pcalowtn}. Foreground removal using PCA accounts for the leakage of the foreground power into the observation window and returns a result which agrees relatively tightly with the \hi\ only case. Note that for $k>5\,{\rm Mpc^{-1}}$, simple avoidance results in a 10\% level overestimation due to foreground contamination, whereas for PCA the power spectrum estimation is relatively accurate up to $k\sim 10\,{\rm Mpc^{-1}}$. Moreover, as discussed in Section \ref{subsec:avoid}, the foreground contamination prevents us from using the inverse covariance weighting as it will further mixes different $k_\parallel$ modes resulting in more contamination. For the PCA case where the foregrounds are sufficiently removed, we can re-apply inverse covariance weighting. We show the results with $\mathbf{R}=\mathbf{N}^{-1}$ in Figure \ref{fig:pcalowtn} as well. {The power spectrum for PCA method with inverse covariance weighting matches the \hi\ only case and produces smaller uncertainties.} We also show the cylindrical delay power spectrum in Figure \ref{fig:dpspcalowtn} {to verify there is no visible foreground leakage into higher $k_\parallel$. Comparing the top panel for uniform weighting and the bottom panel for inverse noise covariance weighting, the mode mixing is most visible at higher $k_\perp$ outside the observation window, whereas the difference at lower $k_\perp$ is negligible.}

\section{Foreground Mitigation in Image Space}
\label{sec:image}
In this section, we temporarily depart from visibility-only approach and investigate {foreground cleaning by applying PCA directly on the image} and perform power spectrum estimation in image space similar to \cite{2021MNRAS.500.2264H} in order to compare with methods in visibility space. We use \textsc{CASA} \textit{tclean} \citep{1974A&AS...15..417H,2007ASPC..376..127M} to produce the images. The pixel size of the images is set to $25.37 \times 25.37\, {\rm arcsecond}^2$, corresponding to $k\sim 45\, {\rm Mpc^{-1}}$ with $192\times192$ standard grids to match the size of the primary beam. {Only dirty images are generated, since we find that iterative cleaning takes out part of the \hi\ signal.}

{The dirty images are in the units of Jy per point spread function (PSF). We calculate the area of PSF around its centre to convert intensity to temperature unit. Starting with the centre and neighbouring pixels, we iteratively expand the integration area until the PSF area decreases, suggesting that the effects of sidelobe structures are starting to dominate. We then rescale the images from Jy per PSF to Jy per pixel using the PSF area calculated. The primary beam effect is then removed by dividing by the beam attentuation term. The processed dirty image is then converted to temperature unit and used to calculate the temperature power spectrum.} {We do not deconvolve the shape of the PSF, which leaves a scale-dependant attenuation effect in the power spectrum. Therefore, when investigating the effects of PCA, we compare the power spectrum results with the dirty images of \hi\ and thermal noise simulation. We leave the full treatment of the PSF for future work.}

We aim to present the imaging approach as a qualitative comparison to the visibility approach and do not derive in detail how to subtract thermal noise covariance and estimate uncertainties. We simulate a thermal noise only image and use it to directly subtract the thermal noise power spectrum. The uncertainties are estimated using the sampling variance in $k$ bins. 

\begin{figure}
    \centering
    \includegraphics[width=0.8\linewidth]{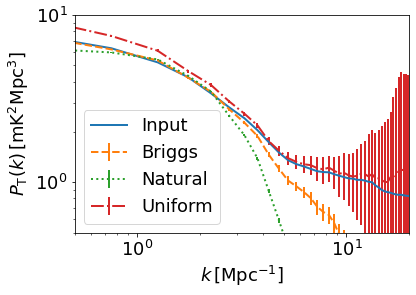}
    \caption{The power spectrum of input \hi\ corresponding to the image size used in Section \ref{sec:image} (``Input''), comparing to the power spectrum of images using \hi\ visibilities produced by natural weighting (``Natural''), uniform weighting (``Uniform'') and Briggs weighting with robustness parameter equal to 0.5 (``Briggs'').}
    \label{fig:imhitn}
\end{figure}

{We test our imaging and power spectrum estimation pipeline} with \hi\ and thermal noise only visibility data and compare it with the input image within the telescope FoV. We find that the weighting of baselines can have a major effect on the resulting power spectrum as shown in Figure \ref{fig:imhitn}. When uniform weighting in imaging is used, the narrow PSF results in overestimation of the power spectrum amplitude at small $k$. On the other hand, natural weighting achieves maximum sensitivity but underestimates the power spectrum due to the large PSF confusing different sources. Using Briggs weighting \citep{1995AAS...18711202B} and testing different robust parameters, we find that when the robustness parameter is set to 0.5, the resulting power spectrum achieves a balance between accuracy and sensitivity. {The power spectrum of the image matches the input from $k_\parallel \sim 0.5\,{\rm Mpc^{-1}}$ up to $k_\parallel \sim 5\,{\rm Mpc^{-1}}$, but suffers signal loss from the PSF convolution at smaller scales.} 

\begin{figure}
    \centering
    \includegraphics[width=\linewidth]{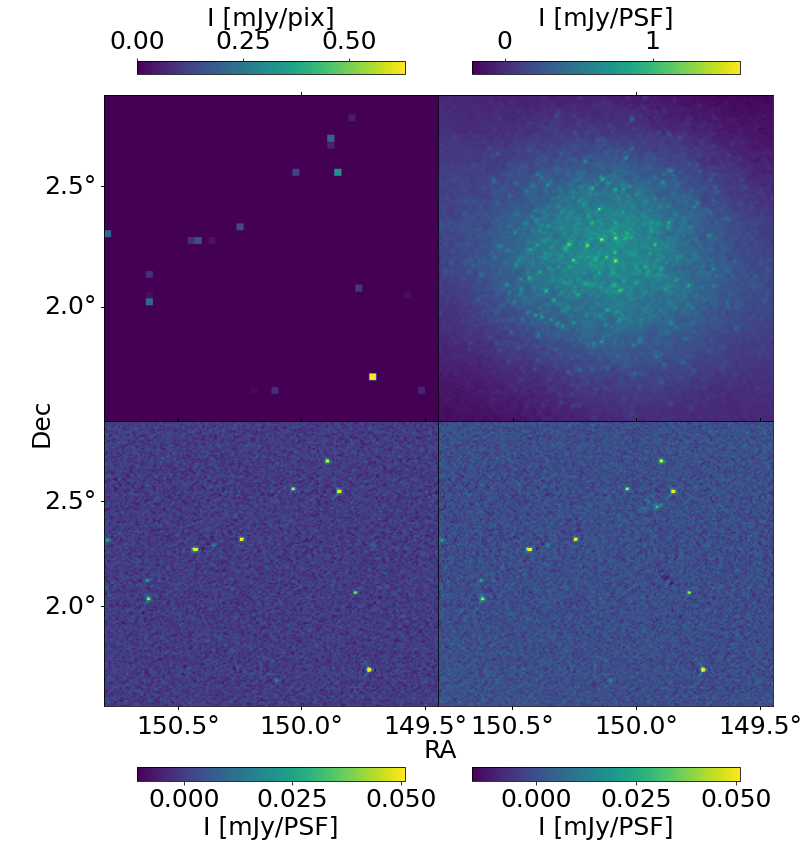}
    \caption{Top-left panel: the input image of \hi\ signal in the 20$^{\rm th}$ frequency bin around 1.096 GHz in Jy per pixel. {The pixel size of the image is set to $(100\ {\rm arcseconds})^2$ to show clearly the position of the \hi\ sources}. Top-right panel: the output image of the simulation in the same frequency bin, with \hi, foreground and thermal noise. Bottom-left panel: the output image for \hi\ and thermal noise only visibility simulation image without the foreground. Bottom-right panel: the image shown in the top right panel after PCA cleaning. The image after PCA cleaning (bottom-right panel) matches the \hi\ and thermal noise only simulation well (bottom-left panel). {In bottom panels, pixels brighter than 0.05 mJy per PSF are set to 0.05 mJy per PSF for better presentation.}  }
    \label{fig:impca}
\end{figure}

We generate image cubes for the simulated visibilities. The image cube is then mean-centered and subsequently a PCA is performed. {We find that in image space, the mixture of \hi\ signal with the thermal noise and foreground is more severe due to the convolution of the PSF and loss of information by gridding. More modes need to be removed for image space PCA comparing to the visibility space PCA in order to subtract the foreground at small $k_\parallel$ and widen the observation window.} We find removing a total number of 7 PCA modes gives the best result. {We find that choosing more modes leads to overcleaning in the high $k_\parallel$ modes while choosing less modes does not sufficiently remove the foregrounds at low $k_\parallel$.} We show the results for the $10^{\rm th}$ frequency bin in Figure \ref{fig:impca}. Comparing the lower panels, one can see PCA removes most of the foreground signal. The undercleaning of some foreground structure leaves residual sources, which are smooth in frequency. The cleaned images are then converted from Jy per PSF to Jy per pixel and corrected for primary beam attenuation.

\begin{figure}
    \centering
    \includegraphics[width=\linewidth]{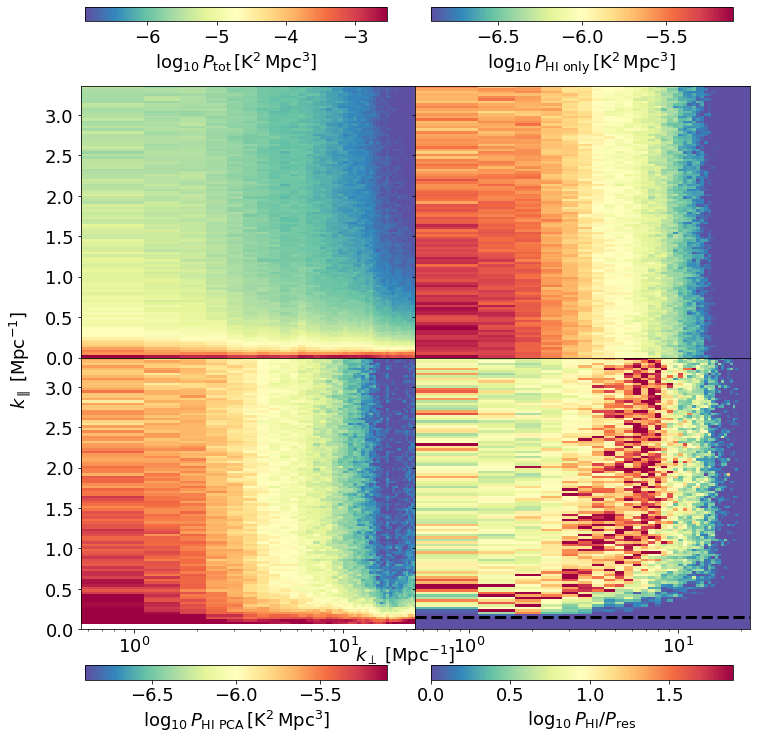}
    \caption{Upper Left Panel: The cylindrical temperature power spectrum of the images of the total intensity with thermal noise subtracted. Values under $10^{-7}\,{\rm K^2Mpc^3}$ are masked for better presentation. Upper Right Panel: The cylindrical temperature power spectrum of the images of the \hi\ signal. Values under $10^{-7}\,{\rm K^2Mpc^3}$ are masked. Lower Left Panel: The cylindrical temperature power spectrum of the images of total intensity after PCA cleaning with thermal noise subtracted. Values under $10^{-7}\,{\rm K^2Mpc^3}$ are masked. Lower Right Panel: The ratio between \hi\ and the residual foreground power spectrum. Values under 1 are set to 1 for better presentation.}
    \label{fig:ptim2d}
\end{figure}

\begin{figure}
    \centering
    \includegraphics[width=0.8\linewidth]{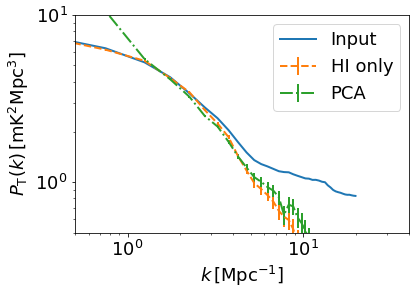}
    \caption{The power spectrum of input \hi\ corresponding to the image size used in Section \ref{sec:image} (''Input''), comparing to the power spectrum of images using \hi\ visibilities (''\hi\ only'') and power spectrum of total intensity images of simulation with low thermal noise level after PCA cleaning (''PCA'').}
    \label{fig:ptpca}
\end{figure}

We verify the frequency smoothness of residual foreground structure by calculating the cylindrical temperature power spectrum presented in Figure \ref{fig:ptim2d}. Comparing the power spectrum of images of \hi\ and images of total intensity after PCA cleaning, the foreground residual mainly resides in low $k_\parallel$ ranges. We find that the observation window probes the region $k_\parallel >0.2\,{\rm Mpc^{-1}}$ by checking the foreground wedge. We verify that raising this threshold does not have a significant impact on the resulting power spectrum. 

The 1D temperature power spectrum result is presented in Figure \ref{fig:ptpca}. Compared to the \hi\ only power spectrum, the PCA results overestimate the large $k\lesssim 1\,{\rm Mpc^{-1}}$ scales. This suggests that large scale information is lost due to the excess gridding and imaging, as we do not see such effects in Section \ref{subsec:pca} when we apply PCA to the visibility data. Furthermore, scales $k\gtrsim 5\,{\rm Mpc^{-1}}$ are not recovered, due to the effects of weighting and PSF discussed previously.

In conclusion, we find that power spectrum estimation in image space is sensitive to the choice of baseline weighting and subsequent deconvolution of PSF, which requires more careful treatment. Applying component separation to the image cube recovers \hi\ clustering at scales of roughly $1\,{\rm Mpc^{-1}}$ to $5\,{\rm Mpc^{-1}}$. {We compare it to PCA in the visibility data in the next section.}

\section{Comparing Foreground Mitigation Methods with MIGHTEE-like noise level}
\label{sec:result}
{In this section, we present a direct comparison of the foreground mitigation methods for power spectrum estimation using our simulation with realistic thermal noise level. As mentioned in Section \ref{subsec:simvis}, we generate MIGHTEE-like thermal noise consistent with the noise level of the entire MIGHTEE survey of 52 pointings and 1920 hours. We then apply the foreground mitigation methods investigated in the previous sections.} {The edges of the 1D k-bins are set to be logarithmically distributed from 0.1 to 30 ${\rm Mpc^{-1}}$ with 14 bins due to the lower signal-to-noise ratio.}

\begin{figure}
    \centering
    \includegraphics[width=\linewidth]{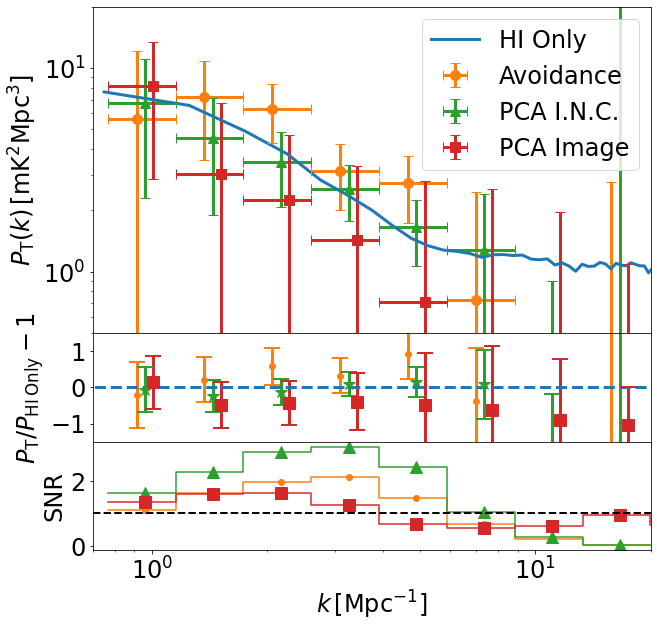}
    \caption{The 1-D brightness temperature power spectrum from the \hi\ only visibility data (``HI Only''), simulation with MIGHTEE-like noise using foreground avoidance method described in Section \ref{subsec:avoid} (``Avoidance''), simulation using PCA in visibility space with inverse noise covariance weighting (``PCA I.N.C.''), and simulation using PCA in image space (``PCA Image''). In addition, the fractional difference between the estimated 1D brightness temperature power spectrum and the \hi\ only simulation is presented below. The centres of the k-bins for ``Avoidance'' and ``PCA Image'' are misplaced by 5 \% for better comparison. {The signal-to-noise ratios of the power spectrum measurements are shown at the bottom.}}
    \label{fig:mighteeps}
\end{figure}

Based on the previous discussion, we adopt the foreground avoidance described in Section \ref{subsec:avoid}, foreground removal using PCA in visibility with $\mathbf{R} = \mathbf{N}^{-1}$ described in Section \ref{subsec:pca}, and in image space described in Section \ref{sec:image}. The resulting projected power spectrum recovery is shown in Figure \ref{fig:mighteeps}. We find that for direct avoidance method, the foreground wedge criterion remains the same at $c_k = 0.25$. For PCA the same $n_{\rm fg}=2$ number of modes are removed and we find that power spectrum result converges when $c_k = 0.1$ for $k_\perp<5\,{\rm Mpc^{-1}}$ and $c_k = 0.25$ for $k_\perp>5\,{\rm Mpc^{-1}}$, tighter than the low thermal noise case in Section \ref{subsec:pca}. For the image output, we apply the exact same process in Section \ref{sec:image} with $n_{\rm fg}=7$ number of modes removed. Overall, the methods used are robust to a substantially increased noise level. 

Comparing the results from avoidance and the PCA method, we find that PCA in visibility space provides less biased estimation of the \hi\ power spectrum with projected uncertainties roughly 30\% smaller than direct avoidance. This is due to the larger observation window and the enabling of the inverse covariance weighting as discussed in Section \ref{sec:fgremove}. {Overall, PCA in visibility space gives measurements of the \hi\ power spectrum with signal-to-noise ratio $\lesssim 3$ from $k \sim 0.5\,{\rm Mpc^{-1}}$ up to $k \sim 5\,{\rm Mpc^{-1}}$ with $\Delta k \sim 0.5\,{\rm Mpc^{-1}}$.}

Comparing the projected power spectrum using the image space PCA method in Figure \ref{fig:mighteeps} and in Section \ref{sec:image}, there is further signal loss in the presence of a higher thermal noise level. {Overall, PCA in image space results in an overestimation of the \hi\ power spectrum at $k < 1.0\,{\rm Mpc^{-1}}$. Compared to the visibility space PCA case, the measurement errors in the image space PCA case are $\sim 50\%$ larger and the signal-to-noise ratio is $\lesssim 2$ from $k \sim 0.5\,{\rm Mpc^{-1}}$ up to $k \sim 5\,{\rm Mpc^{-1}}$.}

{The fractional differences of the projected power spectrum measurements and the input as shown in Figure \ref{fig:mighteeps} provides a direct comparison of the methods used. The visibility space PCA method has least bias on all scales $0.5\,{\rm Mpc^{-1}}\lesssim k\lesssim 10\,{\rm Mpc^{-1}}$ and the true signal is within the 1$\sigma$ uncertainty. Foreground avoidance in visibility space and PCA in image space have larger bias in power spectrum estimation. We find that the projected uncertainties for PCA in image space are 50\% to 2 times larger than PCA in visibility space. This is due to three reasons. First, information is lost due to $u$-$v$ gridding and limited image size which does not use all the baselines. Second, we weight the baselines with Briggs weighting robustness equal to 0.5 for reasons discussed in Section \ref{sec:image}, which is far from optimal sensitivity using natural weighting. Finally, the mixture of \hi\ and foregrounds in principal components is more severe in image space due to the convolution of PSF, making foreground cleaning less effective. We also emphasize that the uncertainties for images are estimated using sampling variance in $k$ space {while the ones for visibilities are estimated using the quadratic estimator discussed in Section \ref{sec:estform}}, which may not be a fair comparison (see e.g. \citealt{2021ApJS..255...26T}).}

\section{Conclusion}
\label{sec:conclusion}
In this paper, we present an end-to-end pipeline of realistic signal modelling, observation simulation and data analysis for low-redshift \hi\ IM using radio interferometers. We have built a powerful simulation tool capable of generating input signal of \hi, foregrounds and thermal noise for a given field of observation. The generated sky signal is used to simulate visibility data for any given observation strategy. We have developed a quadratic estimator for \hi\ power spectrum estimation using visibility data. Multiple foreground mitigation strategies are examined.

We generate simulations of visibility data mimicking a typical tracking of \textit{MeerKAT} for 220 frequency channels at $z\sim 0.25-0.30$, consistent with existing observations of deep fields such as COSMOS and DEEP2 to validate our estimation pipeline. By calculating the cylindrical power spectrum and comparing the contribution of different components, we find:
\begin{itemize}
    \item The Galactic foreground signal, including the synchrotron radiation and the free-free emssion, mainly affects angular scales larger than $\sim 0.1$ degree. Its power drops significantly at smaller angular scales and is much smaller than the \hi\ signal. 
    \item Extragalactic radio sources dominate the foreground signal on smaller angular scales and increase its leakage into higher $k_\parallel$ modes. Comparing to the \hi\ signal, the foreground contamination is severe and leaves a limited observation window for detection even for a deep field with sufficient source peeling.
    \item For scales probed by interferometry at low redshifts, the foreground can be well described by the covariance. This is due to the fact that at the scales of our interest, it is dominated by contributions from Poisson point sources. 
    \item For observations of one field using small-FOV arrays such as \textit{MeerKAT}, the limited survey volume induces large variance in the \hi\ signal. We find that the variance due to survey volume becomes trivial for volumes larger than that of 5 MIGHTEE-like fields for a narrow frequency range, $V \sim 5\times 20\times20\times 200\,{\rm Mpc^3}$.
\end{itemize}

The foreground contamination calls for careful treatment of foreground mitigation methods. We compare, in detail, different ways to mitigate the impact of foregrounds using visibility data and conclude:
\begin{itemize}
    \item When bright point sources are sufficiently removed, there exists an observation window at large $k_\parallel$ in which we can directly estimate the \hi\ power spectrum. It provides biased estimation of the power spectrum, with an overestimation of 10\%. It is due to the leakage of the foregrounds into large $k_\parallel$ that can not be completely excluded. Furthermore, data analysis techniques such as inverse covariance weighting become difficult as different $k_\parallel$ modes further mix under a non-uniform weighting, leading to more foreground contamination.
    
    \item Fitting the empirical covariance to extract out the smooth part for different annulus $|\bm{u}|$ bins, we find that subtracting the foreground covariance corrects for the overestimation at large $k<1\, {\rm Mpc^{-1}}$ scales. It overcleans the foreground and results in signal loss at smaller scales. Extracting the smooth structure of data covariance falsely includes structure of \hi\ covariance over large frequency ranges, which leads to overcleaning at small $k_\parallel$.
    
    \item By binning the visibility in annulus $|\bm{u}|$ bins and performing PCA in each bin, we find that the foreground contamination is reduced. The observation window widens due to improved cleaning of low $k_\parallel$ modes and this works better with short baselines up to $k\sim 5\, {\rm Mpc^{-1}}$. It allows accurate estimation of \hi\ power spectrum from $k\sim 0.5 \, {\rm Mpc^{-1}}$ to $k\sim 10 \, {\rm Mpc^{-1}}$. The reduced contamination allows inverse covariance weighting with little extra spillover of the foregrounds. Comparing to direct avoidance, it does not have the overestimation bias and has uncertainties that are $\sim 30$\% smaller.
    \item We find that foreground mitigation in visibility space is robust to high levels of thermal noise {consistent with the noise level of MIGHTEE for all 52 pointings and a redshift bin of $z\sim 0.25-0.3$}. PCA in visibility space gives projected measurement of the power spectrum with signal-to-noise ratio $\sim$3 up to $k\sim 5\, {\rm Mpc^{-1}}$ and possible detection up to $k\sim 10\, {\rm Mpc^{-1}}$. {Due to RFI contaminations, data blocks from the MIGHTEE survey are usually divided into narrow frequency sub-bands. Therefore, our results will likely apply to any sub-band from observations in the L-band.}
\end{itemize}
Our findings suggest the feasibility of using the visibility power spectrum and foreground extraction to measure the clustering of \hi. At the angular scales of our interest $\sim$arcmin, the foreground components are largely stochastic and therefore can be removed using covariance-based methods such as PCA. Using surveys such as MIGHTEE, interferometric \hi\ IM will be able to map the evolution of \hi\ clustering at inner halo scales.

We have investigated the important question of choosing visibility space or image space to measure the \hi\ power spectrum by directly comparing these two qualitatively. We generate image cubes of our visibility data to investigate power spectrum estimation in image space. We find:
\begin{itemize}
    \item The power spectrum of the images relies heavily on the choice of baseline weighting, with uniform weighting overestimating the power spectrum and natural weighting causing signal loss due to the large PSF.
    \item We find that using Briggs weighting and perform PCA for the image cube, we can recover the \hi\ power spectrum from $k\sim 1 \, {\rm Mpc^{-1}}$ to $k\sim 5 \, {\rm Mpc^{-1}}$. The component separation is less robust in the presence of large thermal noise.
    \item Due to the sub-optimal weighting and information loss from visibility to image space, power spectrum estimation in image space performs relatively poorly comparing to the visibility approach, resulting in a more biased estimation with larger error bars. Comparing to PCA in visibility space, PCA in image space leads to reconstructed uncertainties that are at least 50\% larger.
\end{itemize}

Overall, we find that for future surveys such as MIGHTEE, \hi\ IM using visibility data is capable of measuring the \hi\ power spectrum across a wide range of scales with high accuracy in narrow redshift bins, probing the redshift evolution of \hi\ inside dark matter halos. We provide a proof of concept study for data analysis using interferometric IM. {Results from our pipeline strongly favour component separation and power spectrum estimation directly in visibility data without imaging.} The pipeline can further include realistic beam model, polarization leakage, calibration error and more. It will enable more studies into this topic towards future detection and constraints of cosmic \hi\ in the near future.

\section*{Acknowledgements}

We thank the anonymous referee for useful comments which improved our manuscript. We thank Steven Cunnington for discussions. LW is a UK Research and Innovation Future Leaders Fellow [grant MR/V026437/1]. Apart from aforementioned packages, this work also uses \textsc{pytorch} \citep{NEURIPS2019_9015}, \textsc{numpy} \citep{2020Natur.585..357H}, \textsc{scipy} \citep{2020NatMe..17..261V}, \textsc{astropy} \citep{2018AJ....156..123A}, \textsc{camb} \citep{2000ApJ...538..473L}, \textsc{numba} \citep{10.1145/2833157.2833162} and \textsc{matplotlib} \citep{Hunter:2007}. For the purpose of open access, the author has applied a Creative Commons Attribution (CC BY) licence to any Author Accepted Manuscript version arising.

\section*{Data Availability}
Data underlying this paper will be shared on reasonable request to the corresponding author.



\bibliographystyle{mnras}
\bibliography{example} 

\begin{thebibliography}{}
\makeatletter
\relax
\def\mn@urlcharsother{\let\do\@makeother \do\$\do\&\do\#\do\^\do\_\do\%\do\~}
\def\mn@doi{\begingroup\mn@urlcharsother \@ifnextchar [ {\mn@doi@}
  {\mn@doi@[]}}
\def\mn@doi@[#1]#2{\def\@tempa{#1}\ifx\@tempa\@empty \href
  {http://dx.doi.org/#2} {doi:#2}\else \href {http://dx.doi.org/#2} {#1}\fi
  \endgroup}
\def\mn@eprint#1#2{\mn@eprint@#1:#2::\@nil}
\def\mn@eprint@arXiv#1{\href {http://arxiv.org/abs/#1} {{\tt arXiv:#1}}}
\def\mn@eprint@dblp#1{\href {http://dblp.uni-trier.de/rec/bibtex/#1.xml}
  {dblp:#1}}
\def\mn@eprint@#1:#2:#3:#4\@nil{\def\@tempa {#1}\def\@tempb {#2}\def\@tempc
  {#3}\ifx \@tempc \@empty \let \@tempc \@tempb \let \@tempb \@tempa \fi \ifx
  \@tempb \@empty \def\@tempb {arXiv}\fi \@ifundefined
  {mn@eprint@\@tempb}{\@tempb:\@tempc}{\expandafter \expandafter \csname
  mn@eprint@\@tempb\endcsname \expandafter{\@tempc}}}

\bibitem[\protect\citeauthoryear{{Alam} et~al.,}{{Alam}
  et~al.}{2021}]{2021PhRvD.103h3533A}
{Alam} S.,  et~al., 2021, \mn@doi [\prd] {10.1103/PhysRevD.103.083533}, \href
  {https://ui.adsabs.harvard.edu/abs/2021PhRvD.103h3533A} {103, 083533}

\bibitem[\protect\citeauthoryear{{Alonso}, {Ferreira}  \& {Santos}}{{Alonso}
  et~al.}{2014}]{2014MNRAS.444.3183A}
{Alonso} D.,  {Ferreira} P.~G.,   {Santos} M.~G.,  2014, \mn@doi [\mnras]
  {10.1093/mnras/stu1666}, \href
  {https://ui.adsabs.harvard.edu/abs/2014MNRAS.444.3183A} {444, 3183}

\bibitem[\protect\citeauthoryear{{Anderson} et~al.,}{{Anderson}
  et~al.}{2018}]{2018MNRAS.476.3382A}
{Anderson} C.~J.,  et~al., 2018, \mn@doi [\mnras] {10.1093/mnras/sty346}, \href
  {https://ui.adsabs.harvard.edu/abs/2018MNRAS.476.3382A} {476, 3382}

\bibitem[\protect\citeauthoryear{{Asad} et~al.,}{{Asad}
  et~al.}{2021}]{2021MNRAS.502.2970A}
{Asad} K.~M.~B.,  et~al., 2021, \mn@doi [\mnras] {10.1093/mnras/stab104}, \href
  {https://ui.adsabs.harvard.edu/abs/2021MNRAS.502.2970A} {502, 2970}

\bibitem[\protect\citeauthoryear{{Astropy Collaboration} et~al.,}{{Astropy
  Collaboration} et~al.}{2018}]{2018AJ....156..123A}
{Astropy Collaboration} et~al., 2018, \mn@doi [\aj] {10.3847/1538-3881/aabc4f},
  \href {https://ui.adsabs.harvard.edu/abs/2018AJ....156..123A} {156, 123}

\bibitem[\protect\citeauthoryear{{Barry}, {Hazelton}, {Sullivan}, {Morales}  \&
  {Pober}}{{Barry} et~al.}{2016}]{2016MNRAS.461.3135B}
{Barry} N.,  {Hazelton} B.,  {Sullivan} I.,  {Morales} M.~F.,   {Pober} J.~C.,
  2016, \mn@doi [\mnras] {10.1093/mnras/stw1380}, \href
  {https://ui.adsabs.harvard.edu/abs/2016MNRAS.461.3135B} {461, 3135}

\bibitem[\protect\citeauthoryear{{Barry} et~al.,}{{Barry}
  et~al.}{2019}]{2019ApJ...884....1B}
{Barry} N.,  et~al., 2019, \mn@doi [\apj] {10.3847/1538-4357/ab40a8}, \href
  {https://ui.adsabs.harvard.edu/abs/2019ApJ...884....1B} {884, 1}

\bibitem[\protect\citeauthoryear{{Battye}, {Davies}  \& {Weller}}{{Battye}
  et~al.}{2004}]{2004MNRAS.355.1339B}
{Battye} R.~A.,  {Davies} R.~D.,   {Weller} J.,  2004, \mn@doi [\mnras]
  {10.1111/j.1365-2966.2004.08416.x}, \href
  {https://ui.adsabs.harvard.edu/abs/2004MNRAS.355.1339B} {355, 1339}

\bibitem[\protect\citeauthoryear{{Battye}, {Browne}, {Dickinson}, {Heron},
  {Maffei}  \& {Pourtsidou}}{{Battye} et~al.}{2013}]{2013MNRAS.434.1239B}
{Battye} R.~A.,  {Browne} I.~W.~A.,  {Dickinson} C.,  {Heron} G.,  {Maffei} B.,
    {Pourtsidou} A.,  2013, \mn@doi [\mnras] {10.1093/mnras/stt1082}, \href
  {https://ui.adsabs.harvard.edu/abs/2013MNRAS.434.1239B} {434, 1239}

\bibitem[\protect\citeauthoryear{{Bigot-Sazy} et~al.,}{{Bigot-Sazy}
  et~al.}{2015}]{2015MNRAS.454.3240B}
{Bigot-Sazy} M.~A.,  et~al., 2015, \mn@doi [\mnras] {10.1093/mnras/stv2153},
  \href {https://ui.adsabs.harvard.edu/abs/2015MNRAS.454.3240B} {454, 3240}

\bibitem[\protect\citeauthoryear{{Bonaldi} \& {Brown}}{{Bonaldi} \&
  {Brown}}{2015}]{2015MNRAS.447.1973B}
{Bonaldi} A.,  {Brown} M.~L.,  2015, \mn@doi [\mnras] {10.1093/mnras/stu2601},
  \href {https://ui.adsabs.harvard.edu/abs/2015MNRAS.447.1973B} {447, 1973}

\bibitem[\protect\citeauthoryear{{Bondi} et~al.,}{{Bondi}
  et~al.}{2003}]{2003A&A...403..857B}
{Bondi} M.,  et~al., 2003, \mn@doi [\aap] {10.1051/0004-6361:20030382}, \href
  {https://ui.adsabs.harvard.edu/abs/2003A&A...403..857B} {403, 857}

\bibitem[\protect\citeauthoryear{{Bowman}, {Morales}  \& {Hewitt}}{{Bowman}
  et~al.}{2009}]{2009ApJ...695..183B}
{Bowman} J.~D.,  {Morales} M.~F.,   {Hewitt} J.~N.,  2009, \mn@doi [\apj]
  {10.1088/0004-637X/695/1/183}, \href
  {https://ui.adsabs.harvard.edu/abs/2009ApJ...695..183B} {695, 183}

\bibitem[\protect\citeauthoryear{{Briggs}}{{Briggs}}{1995}]{1995AAS...18711202B}
{Briggs} D.~S.,  1995, in American Astronomical Society Meeting Abstracts. p.
  112.02

\bibitem[\protect\citeauthoryear{{Bull}, {Ferreira}, {Patel}  \&
  {Santos}}{{Bull} et~al.}{2015}]{2015ApJ...803...21B}
{Bull} P.,  {Ferreira} P.~G.,  {Patel} P.,   {Santos} M.~G.,  2015, \mn@doi
  [\apj] {10.1088/0004-637X/803/1/21}, \href
  {https://ui.adsabs.harvard.edu/abs/2015ApJ...803...21B} {803, 21}

\bibitem[\protect\citeauthoryear{{CHIME Collaboration} et~al.,}{{CHIME
  Collaboration} et~al.}{2022}]{2022arXiv220201242C}
{CHIME Collaboration} et~al., 2022, arXiv e-prints, \href
  {https://ui.adsabs.harvard.edu/abs/2022arXiv220201242C} {p. arXiv:2202.01242}

\bibitem[\protect\citeauthoryear{{Chang}, {Pen}, {Peterson}  \&
  {McDonald}}{{Chang} et~al.}{2008}]{2008PhRvL.100i1303C}
{Chang} T.-C.,  {Pen} U.-L.,  {Peterson} J.~B.,   {McDonald} P.,  2008, \mn@doi
  [\prl] {10.1103/PhysRevLett.100.091303}, \href
  {https://ui.adsabs.harvard.edu/abs/2008PhRvL.100i1303C} {100, 091303}

\bibitem[\protect\citeauthoryear{{Chapman} et~al.,}{{Chapman}
  et~al.}{2012}]{2012MNRAS.423.2518C}
{Chapman} E.,  et~al., 2012, \mn@doi [\mnras]
  {10.1111/j.1365-2966.2012.21065.x}, \href
  {https://ui.adsabs.harvard.edu/abs/2012MNRAS.423.2518C} {423, 2518}

\bibitem[\protect\citeauthoryear{{Chapman}, {Zaroubi}, {Abdalla}, {Dulwich},
  {Jeli{\'c}}  \& {Mort}}{{Chapman} et~al.}{2016}]{2016MNRAS.458.2928C}
{Chapman} E.,  {Zaroubi} S.,  {Abdalla} F.~B.,  {Dulwich} F.,  {Jeli{\'c}} V.,
   {Mort} B.,  2016, \mn@doi [\mnras] {10.1093/mnras/stw161}, \href
  {https://ui.adsabs.harvard.edu/abs/2016MNRAS.458.2928C} {458, 2928}

\bibitem[\protect\citeauthoryear{{Chatterjee}, {Bharadwaj}  \&
  {Marthi}}{{Chatterjee} et~al.}{2021}]{2021MNRAS.500.4398C}
{Chatterjee} S.,  {Bharadwaj} S.,   {Marthi} V.~R.,  2021, \mn@doi [\mnras]
  {10.1093/mnras/staa3348}, \href
  {https://ui.adsabs.harvard.edu/abs/2021MNRAS.500.4398C} {500, 4398}

\bibitem[\protect\citeauthoryear{{Chen}, {Wolz}, {Spinelli}  \&
  {Murray}}{{Chen} et~al.}{2021}]{2021MNRAS.502.5259C}
{Chen} Z.,  {Wolz} L.,  {Spinelli} M.,   {Murray} S.~G.,  2021, \mn@doi
  [\mnras] {10.1093/mnras/stab386}, \href
  {https://ui.adsabs.harvard.edu/abs/2021MNRAS.502.5259C} {502, 5259}

\bibitem[\protect\citeauthoryear{{Cheng} et~al.,}{{Cheng}
  et~al.}{2018}]{2018ApJ...868...26C}
{Cheng} C.,  et~al., 2018, \mn@doi [\apj] {10.3847/1538-4357/aae833}, \href
  {https://ui.adsabs.harvard.edu/abs/2018ApJ...868...26C} {868, 26}

\bibitem[\protect\citeauthoryear{{Choudhuri}, {Bharadwaj}, {Ghosh}  \&
  {Ali}}{{Choudhuri} et~al.}{2014}]{2014MNRAS.445.4351C}
{Choudhuri} S.,  {Bharadwaj} S.,  {Ghosh} A.,   {Ali} S.~S.,  2014, \mn@doi
  [\mnras] {10.1093/mnras/stu2027}, \href
  {https://ui.adsabs.harvard.edu/abs/2014MNRAS.445.4351C} {445, 4351}

\bibitem[\protect\citeauthoryear{{Chung} et~al.,}{{Chung}
  et~al.}{2021}]{2021arXiv210411171C}
{Chung} D.~T.,  et~al., 2021, arXiv e-prints, \href
  {https://ui.adsabs.harvard.edu/abs/2021arXiv210411171C} {p. arXiv:2104.11171}

\bibitem[\protect\citeauthoryear{{Condon} \& {Ransom}}{{Condon} \&
  {Ransom}}{2016}]{2016era..book.....C}
{Condon} J.~J.,  {Ransom} S.~M.,  2016, {Essential Radio Astronomy}

\bibitem[\protect\citeauthoryear{{Condon}, {Cotton}, {Greisen}, {Yin},
  {Perley}, {Taylor}  \& {Broderick}}{{Condon}
  et~al.}{1998}]{1998AJ....115.1693C}
{Condon} J.~J.,  {Cotton} W.~D.,  {Greisen} E.~W.,  {Yin} Q.~F.,  {Perley}
  R.~A.,  {Taylor} G.~B.,   {Broderick} J.~J.,  1998, \mn@doi [\aj]
  {10.1086/300337}, \href
  {https://ui.adsabs.harvard.edu/abs/1998AJ....115.1693C} {115, 1693}

\bibitem[\protect\citeauthoryear{{Cooray} \& {Sheth}}{{Cooray} \&
  {Sheth}}{2002}]{2002PhR...372....1C}
{Cooray} A.,  {Sheth} R.,  2002, \mn@doi [\physrep]
  {10.1016/S0370-1573(02)00276-4}, \href
  {https://ui.adsabs.harvard.edu/abs/2002PhR...372....1C} {372, 1}

\bibitem[\protect\citeauthoryear{{Cornwell}, {Golap}  \&
  {Bhatnagar}}{{Cornwell} et~al.}{2008}]{2008ISTSP...2..647C}
{Cornwell} T.~J.,  {Golap} K.,   {Bhatnagar} S.,  2008, \mn@doi [IEEE Journal
  of Selected Topics in Signal Processing] {10.1109/JSTSP.2008.2005290}, \href
  {https://ui.adsabs.harvard.edu/abs/2008ISTSP...2..647C} {2, 647}

\bibitem[\protect\citeauthoryear{{Crain} et~al.,}{{Crain}
  et~al.}{2017}]{2017MNRAS.464.4204C}
{Crain} R.~A.,  et~al., 2017, \mn@doi [\mnras] {10.1093/mnras/stw2586}, \href
  {https://ui.adsabs.harvard.edu/abs/2017MNRAS.464.4204C} {464, 4204}

\bibitem[\protect\citeauthoryear{{Cunnington}, {Irfan}, {Carucci}, {Pourtsidou}
   \& {Bobin}}{{Cunnington} et~al.}{2021}]{2021MNRAS.504..208C}
{Cunnington} S.,  {Irfan} M.~O.,  {Carucci} I.~P.,  {Pourtsidou} A.,   {Bobin}
  J.,  2021, \mn@doi [\mnras] {10.1093/mnras/stab856}, \href
  {https://ui.adsabs.harvard.edu/abs/2021MNRAS.504..208C} {504, 208}

\bibitem[\protect\citeauthoryear{{DeBoer} et~al.,}{{DeBoer}
  et~al.}{2017}]{2017PASP..129d5001D}
{DeBoer} D.~R.,  et~al., 2017, \mn@doi [\pasp]
  {10.1088/1538-3873/129/974/045001}, \href
  {https://ui.adsabs.harvard.edu/abs/2017PASP..129d5001D} {129, 045001}

\bibitem[\protect\citeauthoryear{{Di Matteo}, {Perna}, {Abel}  \& {Rees}}{{Di
  Matteo} et~al.}{2002}]{2002ApJ...564..576D}
{Di Matteo} T.,  {Perna} R.,  {Abel} T.,   {Rees} M.~J.,  2002, \mn@doi [\apj]
  {10.1086/324293}, \href
  {https://ui.adsabs.harvard.edu/abs/2002ApJ...564..576D} {564, 576}

\bibitem[\protect\citeauthoryear{{Dickinson}, {Davies}  \& {Davis}}{{Dickinson}
  et~al.}{2003}]{2003MNRAS.341..369D}
{Dickinson} C.,  {Davies} R.~D.,   {Davis} R.~J.,  2003, \mn@doi [\mnras]
  {10.1046/j.1365-8711.2003.06439.x}, \href
  {https://ui.adsabs.harvard.edu/abs/2003MNRAS.341..369D} {341, 369}

\bibitem[\protect\citeauthoryear{{Dillon} et~al.,}{{Dillon}
  et~al.}{2014}]{2014PhRvD..89b3002D}
{Dillon} J.~S.,  et~al., 2014, \mn@doi [\prd] {10.1103/PhysRevD.89.023002},
  \href {https://ui.adsabs.harvard.edu/abs/2014PhRvD..89b3002D} {89, 023002}

\bibitem[\protect\citeauthoryear{{Dillon} et~al.,}{{Dillon}
  et~al.}{2015}]{2015PhRvD..91l3011D}
{Dillon} J.~S.,  et~al., 2015, \mn@doi [\prd] {10.1103/PhysRevD.91.123011},
  \href {https://ui.adsabs.harvard.edu/abs/2015PhRvD..91l3011D} {91, 123011}

\bibitem[\protect\citeauthoryear{{Dodelson}}{{Dodelson}}{2003}]{2003moco.book.....D}
{Dodelson} S.,  2003, {Modern cosmology}

\bibitem[\protect\citeauthoryear{{Eisenstein} \& {Hu}}{{Eisenstein} \&
  {Hu}}{1998}]{1998ApJ...496..605E}
{Eisenstein} D.~J.,  {Hu} W.,  1998, \mn@doi [\apj] {10.1086/305424}, \href
  {https://ui.adsabs.harvard.edu/abs/1998ApJ...496..605E} {496, 605}

\bibitem[\protect\citeauthoryear{{Finkbeiner}}{{Finkbeiner}}{2003}]{2003ApJS..146..407F}
{Finkbeiner} D.~P.,  2003, \mn@doi [\apjs] {10.1086/374411}, \href
  {https://ui.adsabs.harvard.edu/abs/2003ApJS..146..407F} {146, 407}

\bibitem[\protect\citeauthoryear{{G{\'o}rski}, {Hivon}, {Banday}, {Wandelt},
  {Hansen}, {Reinecke}  \& {Bartelmann}}{{G{\'o}rski}
  et~al.}{2005}]{2005ApJ...622..759G}
{G{\'o}rski} K.~M.,  {Hivon} E.,  {Banday} A.~J.,  {Wandelt} B.~D.,  {Hansen}
  F.~K.,  {Reinecke} M.,   {Bartelmann} M.,  2005, \mn@doi [\apj]
  {10.1086/427976}, \href {http://adsabs.harvard.edu/abs/2005ApJ...622..759G}
  {622, 759}

\bibitem[\protect\citeauthoryear{{Hale}, {Jarvis}, {Delvecchio}, {Hatfield},
  {Novak}, {Smol{\v{c}}i{\'c}}  \& {Zamorani}}{{Hale}
  et~al.}{2018}]{2018MNRAS.474.4133H}
{Hale} C.~L.,  {Jarvis} M.~J.,  {Delvecchio} I.,  {Hatfield} P.~W.,  {Novak}
  M.,  {Smol{\v{c}}i{\'c}} V.,   {Zamorani} G.,  2018, \mn@doi [\mnras]
  {10.1093/mnras/stx2954}, \href
  {https://ui.adsabs.harvard.edu/abs/2018MNRAS.474.4133H} {474, 4133}

\bibitem[\protect\citeauthoryear{{Hamilton}}{{Hamilton}}{1997}]{1997MNRAS.289..285H}
{Hamilton} A.~J.~S.,  1997, \mn@doi [\mnras] {10.1093/mnras/289.2.285}, \href
  {https://ui.adsabs.harvard.edu/abs/1997MNRAS.289..285H} {289, 285}

\bibitem[\protect\citeauthoryear{{Hamilton} \& {Tegmark}}{{Hamilton} \&
  {Tegmark}}{2000}]{2000MNRAS.312..285H}
{Hamilton} A.~J.~S.,  {Tegmark} M.,  2000, \mn@doi [\mnras]
  {10.1046/j.1365-8711.2000.03074.x}, \href
  {https://ui.adsabs.harvard.edu/abs/2000MNRAS.312..285H} {312, 285}

\bibitem[\protect\citeauthoryear{{Harker} et~al.,}{{Harker}
  et~al.}{2010}]{2010MNRAS.405.2492H}
{Harker} G.,  et~al., 2010, \mn@doi [\mnras]
  {10.1111/j.1365-2966.2010.16628.x}, \href
  {https://ui.adsabs.harvard.edu/abs/2010MNRAS.405.2492H} {405, 2492}

\bibitem[\protect\citeauthoryear{{Harris} et~al.,}{{Harris}
  et~al.}{2020}]{2020Natur.585..357H}
{Harris} C.~R.,  et~al., 2020, \mn@doi [\nat] {10.1038/s41586-020-2649-2},
  \href {https://ui.adsabs.harvard.edu/abs/2020Natur.585..357H} {585, 357}

\bibitem[\protect\citeauthoryear{{Haslam}, {Klein}, {Salter}, {Stoffel},
  {Wilson}, {Cleary}, {Cooke}  \& {Thomasson}}{{Haslam}
  et~al.}{1981}]{1981A&A...100..209H}
{Haslam} C.~G.~T.,  {Klein} U.,  {Salter} C.~J.,  {Stoffel} H.,  {Wilson}
  W.~E.,  {Cleary} M.~N.,  {Cooke} D.~J.,   {Thomasson} P.,  1981, \aap, \href
  {https://ui.adsabs.harvard.edu/abs/1981A&A...100..209H} {100, 209}

\bibitem[\protect\citeauthoryear{{Haslam}, {Salter}, {Stoffel}  \&
  {Wilson}}{{Haslam} et~al.}{1982}]{1982A&AS...47....1H}
{Haslam} C.~G.~T.,  {Salter} C.~J.,  {Stoffel} H.,   {Wilson} W.~E.,  1982,
  \aaps, \href {https://ui.adsabs.harvard.edu/abs/1982A&AS...47....1H} {47, 1}

\bibitem[\protect\citeauthoryear{{Heywood} et~al.,}{{Heywood}
  et~al.}{2022}]{2022MNRAS.509.2150H}
{Heywood} I.,  et~al., 2022, \mn@doi [\mnras] {10.1093/mnras/stab3021}, \href
  {https://ui.adsabs.harvard.edu/abs/2022MNRAS.509.2150H} {509, 2150}

\bibitem[\protect\citeauthoryear{{H{\"o}gbom}}{{H{\"o}gbom}}{1974}]{1974A&AS...15..417H}
{H{\"o}gbom} J.~A.,  1974, \aaps, \href
  {https://ui.adsabs.harvard.edu/abs/1974A&AS...15..417H} {15, 417}

\bibitem[\protect\citeauthoryear{{Hothi} et~al.,}{{Hothi}
  et~al.}{2021}]{2021MNRAS.500.2264H}
{Hothi} I.,  et~al., 2021, \mn@doi [\mnras] {10.1093/mnras/staa3446}, \href
  {https://ui.adsabs.harvard.edu/abs/2021MNRAS.500.2264H} {500, 2264}

\bibitem[\protect\citeauthoryear{{Hu}, {Wang}, {Wu}, {Wang}, {Zhang}  \&
  {Chen}}{{Hu} et~al.}{2020}]{2020MNRAS.493.5854H}
{Hu} W.,  {Wang} X.,  {Wu} F.,  {Wang} Y.,  {Zhang} P.,   {Chen} X.,  2020,
  \mn@doi [\mnras] {10.1093/mnras/staa650}, \href
  {https://ui.adsabs.harvard.edu/abs/2020MNRAS.493.5854H} {493, 5854}

\bibitem[\protect\citeauthoryear{Hunter}{Hunter}{2007}]{Hunter:2007}
Hunter J.~D.,  2007, \mn@doi [Computing in Science \& Engineering]
  {10.1109/MCSE.2007.55}, 9, 90

\bibitem[\protect\citeauthoryear{{Jarvis} et~al.,}{{Jarvis}
  et~al.}{2016}]{2016mks..confE...6J}
{Jarvis} M.,  et~al., 2016, in MeerKAT Science: On the Pathway to the SKA. p.~6
  (\mn@eprint {arXiv} {1709.01901})

\bibitem[\protect\citeauthoryear{{Jonas}, {Baart}  \& {Nicolson}}{{Jonas}
  et~al.}{1998}]{1998MNRAS.297..977J}
{Jonas} J.~L.,  {Baart} E.~E.,   {Nicolson} G.~D.,  1998, \mn@doi [\mnras]
  {10.1046/j.1365-8711.1998.01367.x}, \href
  {https://ui.adsabs.harvard.edu/abs/1998MNRAS.297..977J} {297, 977}

\bibitem[\protect\citeauthoryear{{Kaiser}}{{Kaiser}}{1987}]{1987MNRAS.227....1K}
{Kaiser} N.,  1987, \mn@doi [\mnras] {10.1093/mnras/227.1.1}, \href
  {https://ui.adsabs.harvard.edu/abs/1987MNRAS.227....1K} {227, 1}

\bibitem[\protect\citeauthoryear{{Kern} \& {Liu}}{{Kern} \&
  {Liu}}{2021}]{2021MNRAS.501.1463K}
{Kern} N.~S.,  {Liu} A.,  2021, \mn@doi [\mnras] {10.1093/mnras/staa3736},
  \href {https://ui.adsabs.harvard.edu/abs/2021MNRAS.501.1463K} {501, 1463}

\bibitem[\protect\citeauthoryear{{Kovetz} et~al.,}{{Kovetz}
  et~al.}{2017}]{2017arXiv170909066K}
{Kovetz} E.~D.,  et~al., 2017, arXiv e-prints, \href
  {https://ui.adsabs.harvard.edu/abs/2017arXiv170909066K} {p. arXiv:1709.09066}

\bibitem[\protect\citeauthoryear{{Lacy} et~al.,}{{Lacy}
  et~al.}{2020}]{2020PASP..132c5001L}
{Lacy} M.,  et~al., 2020, \mn@doi [\pasp] {10.1088/1538-3873/ab63eb}, \href
  {https://ui.adsabs.harvard.edu/abs/2020PASP..132c5001L} {132, 035001}

\bibitem[\protect\citeauthoryear{Lam, Pitrou  \& Seibert}{Lam
  et~al.}{2015}]{10.1145/2833157.2833162}
Lam S.~K.,  Pitrou A.,   Seibert S.,  2015, in Proceedings of the Second
  Workshop on the LLVM Compiler Infrastructure in HPC. LLVM '15.
Association for Computing Machinery, New York, NY, USA,
  \mn@doi{10.1145/2833157.2833162}, \url
  {https://doi.org/10.1145/2833157.2833162}

\bibitem[\protect\citeauthoryear{{Lewis}, {Challinor}  \& {Lasenby}}{{Lewis}
  et~al.}{2000}]{2000ApJ...538..473L}
{Lewis} A.,  {Challinor} A.,   {Lasenby} A.,  2000, \mn@doi [\apj]
  {10.1086/309179}, \href
  {https://ui.adsabs.harvard.edu/abs/2000ApJ...538..473L} {538, 473}

\bibitem[\protect\citeauthoryear{{Lian}, {Xu}, {Zhu}  \& {Hu}}{{Lian}
  et~al.}{2020}]{2020MNRAS.496.1232L}
{Lian} X.,  {Xu} H.,  {Zhu} Z.,   {Hu} D.,  2020, \mn@doi [\mnras]
  {10.1093/mnras/staa1179}, \href
  {https://ui.adsabs.harvard.edu/abs/2020MNRAS.496.1232L} {496, 1232}

\bibitem[\protect\citeauthoryear{{Liu} \& {Shaw}}{{Liu} \&
  {Shaw}}{2020}]{2020PASP..132f2001L}
{Liu} A.,  {Shaw} J.~R.,  2020, \mn@doi [\pasp] {10.1088/1538-3873/ab5bfd},
  \href {https://ui.adsabs.harvard.edu/abs/2020PASP..132f2001L} {132, 062001}

\bibitem[\protect\citeauthoryear{{Liu} \& {Tegmark}}{{Liu} \&
  {Tegmark}}{2011}]{2011PhRvD..83j3006L}
{Liu} A.,  {Tegmark} M.,  2011, \mn@doi [\prd] {10.1103/PhysRevD.83.103006},
  \href {https://ui.adsabs.harvard.edu/abs/2011PhRvD..83j3006L} {83, 103006}

\bibitem[\protect\citeauthoryear{{Liu}, {Tegmark}  \& {Zaldarriaga}}{{Liu}
  et~al.}{2009}]{2009MNRAS.394.1575L}
{Liu} A.,  {Tegmark} M.,   {Zaldarriaga} M.,  2009, \mn@doi [\mnras]
  {10.1111/j.1365-2966.2009.14426.x}, \href
  {https://ui.adsabs.harvard.edu/abs/2009MNRAS.394.1575L} {394, 1575}

\bibitem[\protect\citeauthoryear{{Liu}, {Parsons}  \& {Trott}}{{Liu}
  et~al.}{2014a}]{2014PhRvD..90b3018L}
{Liu} A.,  {Parsons} A.~R.,   {Trott} C.~M.,  2014a, \mn@doi [\prd]
  {10.1103/PhysRevD.90.023018}, \href
  {https://ui.adsabs.harvard.edu/abs/2014PhRvD..90b3018L} {90, 023018}

\bibitem[\protect\citeauthoryear{{Liu}, {Parsons}  \& {Trott}}{{Liu}
  et~al.}{2014b}]{2014PhRvD..90b3019L}
{Liu} A.,  {Parsons} A.~R.,   {Trott} C.~M.,  2014b, \mn@doi [\prd]
  {10.1103/PhysRevD.90.023019}, \href
  {https://ui.adsabs.harvard.edu/abs/2014PhRvD..90b3019L} {90, 023019}

\bibitem[\protect\citeauthoryear{{Madau}, {Meiksin}  \& {Rees}}{{Madau}
  et~al.}{1997}]{1997ApJ...475..429M}
{Madau} P.,  {Meiksin} A.,   {Rees} M.~J.,  1997, \mn@doi [\apj]
  {10.1086/303549}, \href
  {https://ui.adsabs.harvard.edu/abs/1997ApJ...475..429M} {475, 429}

\bibitem[\protect\citeauthoryear{{Mao}, {Tegmark}, {McQuinn}, {Zaldarriaga}  \&
  {Zahn}}{{Mao} et~al.}{2008}]{2008PhRvD..78b3529M}
{Mao} Y.,  {Tegmark} M.,  {McQuinn} M.,  {Zaldarriaga} M.,   {Zahn} O.,  2008,
  \mn@doi [\prd] {10.1103/PhysRevD.78.023529}, \href
  {https://ui.adsabs.harvard.edu/abs/2008PhRvD..78b3529M} {78, 023529}

\bibitem[\protect\citeauthoryear{{Masui} et~al.,}{{Masui}
  et~al.}{2013}]{2013ApJ...763L..20M}
{Masui} K.~W.,  et~al., 2013, \mn@doi [\apjl] {10.1088/2041-8205/763/1/L20},
  \href {https://ui.adsabs.harvard.edu/abs/2013ApJ...763L..20M} {763, L20}

\bibitem[\protect\citeauthoryear{{Matthews}, {Condon}, {Cotton}  \&
  {Mauch}}{{Matthews} et~al.}{2021}]{2021ApJ...909..193M}
{Matthews} A.~M.,  {Condon} J.~J.,  {Cotton} W.~D.,   {Mauch} T.,  2021,
  \mn@doi [\apj] {10.3847/1538-4357/abdd37}, \href
  {https://ui.adsabs.harvard.edu/abs/2021ApJ...909..193M} {909, 193}

\bibitem[\protect\citeauthoryear{{Mauch} et~al.,}{{Mauch}
  et~al.}{2020}]{2020ApJ...888...61M}
{Mauch} T.,  et~al., 2020, \mn@doi [\apj] {10.3847/1538-4357/ab5d2d}, \href
  {https://ui.adsabs.harvard.edu/abs/2020ApJ...888...61M} {888, 61}

\bibitem[\protect\citeauthoryear{{McMullin}, {Waters}, {Schiebel}, {Young}  \&
  {Golap}}{{McMullin} et~al.}{2007}]{2007ASPC..376..127M}
{McMullin} J.~P.,  {Waters} B.,  {Schiebel} D.,  {Young} W.,   {Golap} K.,
  2007, in {Shaw} R.~A.,  {Hill} F.,   {Bell} D.~J.,  eds,  Astronomical
  Society of the Pacific Conference Series Vol. 376, Astronomical Data Analysis
  Software and Systems XVI. p.~127

\bibitem[\protect\citeauthoryear{{Mertens}, {Ghosh}  \& {Koopmans}}{{Mertens}
  et~al.}{2018}]{2018MNRAS.478.3640M}
{Mertens} F.~G.,  {Ghosh} A.,   {Koopmans} L.~V.~E.,  2018, \mn@doi [\mnras]
  {10.1093/mnras/sty1207}, \href
  {https://ui.adsabs.harvard.edu/abs/2018MNRAS.478.3640M} {478, 3640}

\bibitem[\protect\citeauthoryear{{Mertens} et~al.,}{{Mertens}
  et~al.}{2020}]{2020MNRAS.493.1662M}
{Mertens} F.~G.,  et~al., 2020, \mn@doi [\mnras] {10.1093/mnras/staa327}, \href
  {https://ui.adsabs.harvard.edu/abs/2020MNRAS.493.1662M} {493, 1662}

\bibitem[\protect\citeauthoryear{{Meyer}, {Robotham}, {Obreschkow},
  {Westmeier}, {Duffy}  \& {Staveley-Smith}}{{Meyer}
  et~al.}{2017}]{2017PASA...34...52M}
{Meyer} M.,  {Robotham} A.,  {Obreschkow} D.,  {Westmeier} T.,  {Duffy} A.~R.,
   {Staveley-Smith} L.,  2017, \mn@doi [\pasa] {10.1017/pasa.2017.31}, \href
  {https://ui.adsabs.harvard.edu/abs/2017PASA...34...52M} {34, 52}

\bibitem[\protect\citeauthoryear{{Morales} \& {Hewitt}}{{Morales} \&
  {Hewitt}}{2004}]{2004ApJ...615....7M}
{Morales} M.~F.,  {Hewitt} J.,  2004, \mn@doi [\apj] {10.1086/424437}, \href
  {https://ui.adsabs.harvard.edu/abs/2004ApJ...615....7M} {615, 7}

\bibitem[\protect\citeauthoryear{{Morales}, {Beardsley}, {Pober}, {Barry},
  {Hazelton}, {Jacobs}  \& {Sullivan}}{{Morales}
  et~al.}{2019}]{2019MNRAS.483.2207M}
{Morales} M.~F.,  {Beardsley} A.,  {Pober} J.,  {Barry} N.,  {Hazelton} B.,
  {Jacobs} D.,   {Sullivan} I.,  2019, \mn@doi [\mnras]
  {10.1093/mnras/sty2844}, \href
  {https://ui.adsabs.harvard.edu/abs/2019MNRAS.483.2207M} {483, 2207}

\bibitem[\protect\citeauthoryear{Mort, Dulwich, Salvini, Adami  \& Jones}{Mort
  et~al.}{2010}]{5613289}
Mort B.~J.,  Dulwich F.,  Salvini S.,  Adami K.~Z.,   Jones M.~E.,  2010, in
  2010 IEEE International Symposium on Phased Array Systems and Technology. pp
  690--694, \mn@doi{10.1109/ARRAY.2010.5613289}

\bibitem[\protect\citeauthoryear{{Murray}}{{Murray}}{2018}]{2018JOSS....3..850M}
{Murray} S.~G.,  2018, \mn@doi [The Journal of Open Source Software]
  {10.21105/joss.00850}, \href
  {https://ui.adsabs.harvard.edu/abs/2018JOSS....3..850M} {3, 850}

\bibitem[\protect\citeauthoryear{{Murray}, {Power}  \& {Robotham}}{{Murray}
  et~al.}{2013}]{2013A&C.....3...23M}
{Murray} S.~G.,  {Power} C.,   {Robotham} A.~S.~G.,  2013, \mn@doi [Astronomy
  and Computing] {10.1016/j.ascom.2013.11.001}, \href
  {https://ui.adsabs.harvard.edu/abs/2013A&C.....3...23M} {3, 23}

\bibitem[\protect\citeauthoryear{{Murray}, {Trott}  \& {Jordan}}{{Murray}
  et~al.}{2017}]{2017ApJ...845....7M}
{Murray} S.~G.,  {Trott} C.~M.,   {Jordan} C.~H.,  2017, \mn@doi [\apj]
  {10.3847/1538-4357/aa7d0a}, \href
  {https://ui.adsabs.harvard.edu/abs/2017ApJ...845....7M} {845, 7}

\bibitem[\protect\citeauthoryear{{Murray}, {Diemer}, {Chen}, {Neuhold},
  {Schnapp}, {Peruzzi}, {Blevins}  \& {Engelman}}{{Murray}
  et~al.}{2021}]{2021A&C....3600487M}
{Murray} S.~G.,  {Diemer} B.,  {Chen} Z.,  {Neuhold} A.~G.,  {Schnapp} M.~A.,
  {Peruzzi} T.,  {Blevins} D.,   {Engelman} T.,  2021, \mn@doi [Astronomy and
  Computing] {10.1016/j.ascom.2021.100487}, \href
  {https://ui.adsabs.harvard.edu/abs/2021A&C....3600487M} {36, 100487}

\bibitem[\protect\citeauthoryear{{Myers} et~al.,}{{Myers}
  et~al.}{2003}]{2003ApJ...591..575M}
{Myers} S.~T.,  et~al., 2003, \mn@doi [\apj] {10.1086/375509}, \href
  {https://ui.adsabs.harvard.edu/abs/2003ApJ...591..575M} {591, 575}

\bibitem[\protect\citeauthoryear{{Nasirudin}, {Murray}, {Trott}, {Greig},
  {Joseph}  \& {Power}}{{Nasirudin} et~al.}{2020}]{2020ApJ...893..118N}
{Nasirudin} A.,  {Murray} S.~G.,  {Trott} C.~M.,  {Greig} B.,  {Joseph} R.~C.,
   {Power} C.,  2020, \mn@doi [\apj] {10.3847/1538-4357/ab8003}, \href
  {https://ui.adsabs.harvard.edu/abs/2020ApJ...893..118N} {893, 118}

\bibitem[\protect\citeauthoryear{{Navarro}, {Frenk}  \& {White}}{{Navarro}
  et~al.}{1996}]{1996ApJ...462..563N}
{Navarro} J.~F.,  {Frenk} C.~S.,   {White} S. D.~M.,  1996, \mn@doi [\apj]
  {10.1086/177173}, \href
  {https://ui.adsabs.harvard.edu/abs/1996ApJ...462..563N} {462, 563}

\bibitem[\protect\citeauthoryear{{Olivari}, {Dickinson}, {Battye}, {Ma},
  {Costa}, {Remazeilles}  \& {Harper}}{{Olivari}
  et~al.}{2018}]{2018MNRAS.473.4242O}
{Olivari} L.~C.,  {Dickinson} C.,  {Battye} R.~A.,  {Ma} Y.~Z.,  {Costa} A.~A.,
   {Remazeilles} M.,   {Harper} S.,  2018, \mn@doi [\mnras]
  {10.1093/mnras/stx2621}, \href
  {https://ui.adsabs.harvard.edu/abs/2018MNRAS.473.4242O} {473, 4242}

\bibitem[\protect\citeauthoryear{{Overzier}, {R{\"o}ttgering}, {Rengelink}  \&
  {Wilman}}{{Overzier} et~al.}{2003}]{2003A&A...405...53O}
{Overzier} R.~A.,  {R{\"o}ttgering} H.~J.~A.,  {Rengelink} R.~B.,   {Wilman}
  R.~J.,  2003, \mn@doi [\aap] {10.1051/0004-6361:20030527}, \href
  {https://ui.adsabs.harvard.edu/abs/2003A&A...405...53O} {405, 53}

\bibitem[\protect\citeauthoryear{{Padmanabhan}, {Refregier}  \&
  {Amara}}{{Padmanabhan} et~al.}{2019}]{2019MNRAS.485.4060P}
{Padmanabhan} H.,  {Refregier} A.,   {Amara} A.,  2019, \mn@doi [\mnras]
  {10.1093/mnras/stz683}, \href
  {https://ui.adsabs.harvard.edu/abs/2019MNRAS.485.4060P} {485, 4060}

\bibitem[\protect\citeauthoryear{{Pandey} et~al.,}{{Pandey}
  et~al.}{2021}]{2021arXiv210513545P}
{Pandey} S.,  et~al., 2021, arXiv e-prints, \href
  {https://ui.adsabs.harvard.edu/abs/2021arXiv210513545P} {p. arXiv:2105.13545}

\bibitem[\protect\citeauthoryear{{Parsons}, {Pober}, {McQuinn}, {Jacobs}  \&
  {Aguirre}}{{Parsons} et~al.}{2012a}]{2012ApJ...753...81P}
{Parsons} A.,  {Pober} J.,  {McQuinn} M.,  {Jacobs} D.,   {Aguirre} J.,  2012a,
  \mn@doi [\apj] {10.1088/0004-637X/753/1/81}, \href
  {https://ui.adsabs.harvard.edu/abs/2012ApJ...753...81P} {753, 81}

\bibitem[\protect\citeauthoryear{{Parsons}, {Pober}, {Aguirre}, {Carilli},
  {Jacobs}  \& {Moore}}{{Parsons} et~al.}{2012b}]{2012ApJ...756..165P}
{Parsons} A.~R.,  {Pober} J.~C.,  {Aguirre} J.~E.,  {Carilli} C.~L.,  {Jacobs}
  D.~C.,   {Moore} D.~F.,  2012b, \mn@doi [\apj] {10.1088/0004-637X/756/2/165},
  \href {https://ui.adsabs.harvard.edu/abs/2012ApJ...756..165P} {756, 165}

\bibitem[\protect\citeauthoryear{{Parsons} et~al.,}{{Parsons}
  et~al.}{2014}]{2014ApJ...788..106P}
{Parsons} A.~R.,  et~al., 2014, \mn@doi [\apj] {10.1088/0004-637X/788/2/106},
  \href {https://ui.adsabs.harvard.edu/abs/2014ApJ...788..106P} {788, 106}

\bibitem[\protect\citeauthoryear{Paszke et~al.,}{Paszke
  et~al.}{2019}]{NEURIPS2019_9015}
Paszke A.,  et~al., 2019, in Wallach H.,  Larochelle H.,  Beygelzimer A.,
  d\textquotesingle Alch\'{e}-Buc F.,  Fox E.,   Garnett R.,  eds, , Advances
  in Neural Information Processing Systems 32.
Curran Associates, Inc., pp 8024--8035

\bibitem[\protect\citeauthoryear{{Patil} et~al.,}{{Patil}
  et~al.}{2017}]{2017ApJ...838...65P}
{Patil} A.~H.,  et~al., 2017, \mn@doi [\apj] {10.3847/1538-4357/aa63e7}, \href
  {https://ui.adsabs.harvard.edu/abs/2017ApJ...838...65P} {838, 65}

\bibitem[\protect\citeauthoryear{{Paul}, {Santos}, {Townsend}, {Jarvis},
  {Maddox}, {Collier}, {Frank}  \& {Taylor}}{{Paul}
  et~al.}{2021}]{2021MNRAS.505.2039P}
{Paul} S.,  {Santos} M.~G.,  {Townsend} J.,  {Jarvis} M.~J.,  {Maddox} N.,
  {Collier} J.~D.,  {Frank} B.~S.,   {Taylor} R.,  2021, \mn@doi [\mnras]
  {10.1093/mnras/stab1089}, \href
  {https://ui.adsabs.harvard.edu/abs/2021MNRAS.505.2039P} {505, 2039}

\bibitem[\protect\citeauthoryear{{Peebles}}{{Peebles}}{1974}]{1974ApJ...189L..51P}
{Peebles} P.~J.~E.,  1974, \mn@doi [\apjl] {10.1086/181462}, \href
  {https://ui.adsabs.harvard.edu/abs/1974ApJ...189L..51P} {189, L51}

\bibitem[\protect\citeauthoryear{{Peebles}}{{Peebles}}{1980}]{1980lssu.book.....P}
{Peebles} P.~J.~E.,  1980, {The large-scale structure of the universe}

\bibitem[\protect\citeauthoryear{{Planck Collaboration} et~al.,}{{Planck
  Collaboration} et~al.}{2020}]{2020A&A...641A...6P}
{Planck Collaboration} et~al., 2020, \mn@doi [\aap]
  {10.1051/0004-6361/201833910}, \href
  {https://ui.adsabs.harvard.edu/abs/2020A&A...641A...6P} {641, A6}

\bibitem[\protect\citeauthoryear{{Reich}, {Testori}  \& {Reich}}{{Reich}
  et~al.}{2001}]{2001A&A...376..861R}
{Reich} P.,  {Testori} J.~C.,   {Reich} W.,  2001, \mn@doi [\aap]
  {10.1051/0004-6361:20011000}, \href
  {https://ui.adsabs.harvard.edu/abs/2001A&A...376..861R} {376, 861}

\bibitem[\protect\citeauthoryear{{Remazeilles}, {Dickinson}, {Banday},
  {Bigot-Sazy}  \& {Ghosh}}{{Remazeilles} et~al.}{2015}]{2015MNRAS.451.4311R}
{Remazeilles} M.,  {Dickinson} C.,  {Banday} A.~J.,  {Bigot-Sazy} M.~A.,
  {Ghosh} T.,  2015, \mn@doi [\mnras] {10.1093/mnras/stv1274}, \href
  {https://ui.adsabs.harvard.edu/abs/2015MNRAS.451.4311R} {451, 4311}

\bibitem[\protect\citeauthoryear{{Santos}, {Cooray}  \& {Knox}}{{Santos}
  et~al.}{2005}]{2005ApJ...625..575S}
{Santos} M.~G.,  {Cooray} A.,   {Knox} L.,  2005, \mn@doi [\apj]
  {10.1086/429857}, \href
  {https://ui.adsabs.harvard.edu/abs/2005ApJ...625..575S} {625, 575}

\bibitem[\protect\citeauthoryear{{Santos} et~al.,}{{Santos}
  et~al.}{2017}]{2017arXiv170906099S}
{Santos} M.~G.,  et~al., 2017, arXiv e-prints, \href
  {https://ui.adsabs.harvard.edu/abs/2017arXiv170906099S} {p. arXiv:1709.06099}

\bibitem[\protect\citeauthoryear{{Sarkar}, {Bharadwaj}  \& {Marthi}}{{Sarkar}
  et~al.}{2018}]{2018MNRAS.473..261S}
{Sarkar} A.~K.,  {Bharadwaj} S.,   {Marthi} V.~R.,  2018, \mn@doi [\mnras]
  {10.1093/mnras/stx2344}, \href
  {https://ui.adsabs.harvard.edu/abs/2018MNRAS.473..261S} {473, 261}

\bibitem[\protect\citeauthoryear{{Schaan} \& {White}}{{Schaan} \&
  {White}}{2021}]{2021JCAP...05..068S}
{Schaan} E.,  {White} M.,  2021, \mn@doi [\jcap]
  {10.1088/1475-7516/2021/05/068}, \href
  {https://ui.adsabs.harvard.edu/abs/2021JCAP...05..068S} {2021, 068}

\bibitem[\protect\citeauthoryear{{Scoville} et~al.,}{{Scoville}
  et~al.}{2007}]{2007ApJS..172....1S}
{Scoville} N.,  et~al., 2007, \mn@doi [\apjs] {10.1086/516585}, \href
  {https://ui.adsabs.harvard.edu/abs/2007ApJS..172....1S} {172, 1}

\bibitem[\protect\citeauthoryear{{Siewert} et~al.,}{{Siewert}
  et~al.}{2020}]{2020A&A...643A.100S}
{Siewert} T.~M.,  et~al., 2020, \mn@doi [\aap] {10.1051/0004-6361/201936592},
  \href {https://ui.adsabs.harvard.edu/abs/2020A&A...643A.100S} {643, A100}

\bibitem[\protect\citeauthoryear{{Spinelli}, {Zoldan}, {De Lucia}, {Xie}  \&
  {Viel}}{{Spinelli} et~al.}{2020}]{2020MNRAS.493.5434S}
{Spinelli} M.,  {Zoldan} A.,  {De Lucia} G.,  {Xie} L.,   {Viel} M.,  2020,
  \mn@doi [\mnras] {10.1093/mnras/staa604}, \href
  {https://ui.adsabs.harvard.edu/abs/2020MNRAS.493.5434S} {493, 5434}

\bibitem[\protect\citeauthoryear{{Spinelli}, {Bernardi}, {Garsden},
  {Greenhill}, {Fialkov}, {Dowell}  \& {Price}}{{Spinelli}
  et~al.}{2021}]{2021MNRAS.tmp.1362S}
{Spinelli} M.,  {Bernardi} G.,  {Garsden} H.,  {Greenhill} L.~J.,  {Fialkov}
  A.,  {Dowell} J.,   {Price} D.~C.,  2021, \mn@doi [\mnras]
  {10.1093/mnras/stab1363}, \href
  {https://ui.adsabs.harvard.edu/abs/2021MNRAS.tmp.1362S} {}

\bibitem[\protect\citeauthoryear{{Square Kilometre Array Cosmology Science
  Working Group} et~al.,}{{Square Kilometre Array Cosmology Science Working
  Group} et~al.}{2020}]{2020PASA...37....7S}
{Square Kilometre Array Cosmology Science Working Group} et~al., 2020, \mn@doi
  [\pasa] {10.1017/pasa.2019.51}, \href
  {https://ui.adsabs.harvard.edu/abs/2020PASA...37....7S} {37, e007}

\bibitem[\protect\citeauthoryear{{Switzer} et~al.,}{{Switzer}
  et~al.}{2013}]{2013MNRAS.434L..46S}
{Switzer} E.~R.,  et~al., 2013, \mn@doi [\mnras] {10.1093/mnrasl/slt074}, \href
  {https://ui.adsabs.harvard.edu/abs/2013MNRAS.434L..46S} {434, L46}

\bibitem[\protect\citeauthoryear{{Tan} et~al.,}{{Tan}
  et~al.}{2021}]{2021ApJS..255...26T}
{Tan} J.,  et~al., 2021, \mn@doi [\apjs] {10.3847/1538-4365/ac0533}, \href
  {https://ui.adsabs.harvard.edu/abs/2021ApJS..255...26T} {255, 26}

\bibitem[\protect\citeauthoryear{{Tegmark}}{{Tegmark}}{1997}]{1997PhRvD..55.5895T}
{Tegmark} M.,  1997, \mn@doi [\prd] {10.1103/PhysRevD.55.5895}, \href
  {https://ui.adsabs.harvard.edu/abs/1997PhRvD..55.5895T} {55, 5895}

\bibitem[\protect\citeauthoryear{{Tegmark}}{{Tegmark}}{1998}]{1998tsra.conf..270T}
{Tegmark} M.,  1998, in {Olinto} A.~V.,  {Frieman} J.~A.,   {Schramm} D.~N.,
  eds, Eighteenth Texas Symposium on Relativistic Astrophysics. p.~270
  (\mn@eprint {arXiv} {astro-ph/9702019})

\bibitem[\protect\citeauthoryear{{The HERA Collaboration} et~al.,}{{The HERA
  Collaboration} et~al.}{2021}]{2021arXiv210802263T}
{The HERA Collaboration} et~al., 2021, arXiv e-prints, \href
  {https://ui.adsabs.harvard.edu/abs/2021arXiv210802263T} {p. arXiv:2108.02263}

\bibitem[\protect\citeauthoryear{{Thyagarajan} et~al.,}{{Thyagarajan}
  et~al.}{2015a}]{2015ApJ...804...14T}
{Thyagarajan} N.,  et~al., 2015a, \mn@doi [\apj] {10.1088/0004-637X/804/1/14},
  \href {https://ui.adsabs.harvard.edu/abs/2015ApJ...804...14T} {804, 14}

\bibitem[\protect\citeauthoryear{{Thyagarajan} et~al.,}{{Thyagarajan}
  et~al.}{2015b}]{2015ApJ...807L..28T}
{Thyagarajan} N.,  et~al., 2015b, \mn@doi [\apjl]
  {10.1088/2041-8205/807/2/L28}, \href
  {https://ui.adsabs.harvard.edu/abs/2015ApJ...807L..28T} {807, L28}

\bibitem[\protect\citeauthoryear{{Thyagarajan}, {Parsons}, {DeBoer}, {Bowman},
  {Ewall-Wice}, {Neben}  \& {Patra}}{{Thyagarajan}
  et~al.}{2016}]{2016ApJ...825....9T}
{Thyagarajan} N.,  {Parsons} A.~R.,  {DeBoer} D.~R.,  {Bowman} J.~D.,
  {Ewall-Wice} A.~M.,  {Neben} A.~R.,   {Patra} N.,  2016, \mn@doi [\apj]
  {10.3847/0004-637X/825/1/9}, \href
  {https://ui.adsabs.harvard.edu/abs/2016ApJ...825....9T} {825, 9}

\bibitem[\protect\citeauthoryear{{Tinker}, {Kravtsov}, {Klypin}, {Abazajian},
  {Warren}, {Yepes}, {Gottl{\"o}ber}  \& {Holz}}{{Tinker}
  et~al.}{2008}]{2008ApJ...688..709T}
{Tinker} J.,  {Kravtsov} A.~V.,  {Klypin} A.,  {Abazajian} K.,  {Warren} M.,
  {Yepes} G.,  {Gottl{\"o}ber} S.,   {Holz} D.~E.,  2008, \mn@doi [\apj]
  {10.1086/591439}, \href
  {https://ui.adsabs.harvard.edu/abs/2008ApJ...688..709T} {688, 709}

\bibitem[\protect\citeauthoryear{{Tinker}, {Robertson}, {Kravtsov}, {Klypin},
  {Warren}, {Yepes}  \& {Gottl{\"o}ber}}{{Tinker}
  et~al.}{2010}]{2010ApJ...724..878T}
{Tinker} J.~L.,  {Robertson} B.~E.,  {Kravtsov} A.~V.,  {Klypin} A.,  {Warren}
  M.~S.,  {Yepes} G.,   {Gottl{\"o}ber} S.,  2010, \mn@doi [\apj]
  {10.1088/0004-637X/724/2/878}, \href
  {https://ui.adsabs.harvard.edu/abs/2010ApJ...724..878T} {724, 878}

\bibitem[\protect\citeauthoryear{{Trott} et~al.,}{{Trott}
  et~al.}{2016}]{2016ApJ...818..139T}
{Trott} C.~M.,  et~al., 2016, \mn@doi [\apj] {10.3847/0004-637X/818/2/139},
  \href {https://ui.adsabs.harvard.edu/abs/2016ApJ...818..139T} {818, 139}

\bibitem[\protect\citeauthoryear{{Villaescusa-Navarro}
  et~al.,}{{Villaescusa-Navarro} et~al.}{2018}]{2018ApJ...866..135V}
{Villaescusa-Navarro} F.,  et~al., 2018, \mn@doi [\apj]
  {10.3847/1538-4357/aadba0}, \href
  {https://ui.adsabs.harvard.edu/abs/2018ApJ...866..135V} {866, 135}

\bibitem[\protect\citeauthoryear{{Virtanen} et~al.,}{{Virtanen}
  et~al.}{2020}]{2020NatMe..17..261V}
{Virtanen} P.,  et~al., 2020, \mn@doi [Nature Methods]
  {10.1038/s41592-019-0686-2}, \href
  {https://ui.adsabs.harvard.edu/abs/2020NatMe..17..261V} {17, 261}

\bibitem[\protect\citeauthoryear{{Wolz}, {Tonini}, {Blake}  \& {Wyithe}}{{Wolz}
  et~al.}{2016}]{2016MNRAS.458.3399W}
{Wolz} L.,  {Tonini} C.,  {Blake} C.,   {Wyithe} J.~S.~B.,  2016, \mn@doi
  [\mnras] {10.1093/mnras/stw535}, \href
  {https://ui.adsabs.harvard.edu/abs/2016MNRAS.458.3399W} {458, 3399}

\bibitem[\protect\citeauthoryear{{Wolz}, {Murray}, {Blake}  \& {Wyithe}}{{Wolz}
  et~al.}{2019}]{2019MNRAS.484.1007W}
{Wolz} L.,  {Murray} S.~G.,  {Blake} C.,   {Wyithe} J.~S.,  2019, \mn@doi
  [\mnras] {10.1093/mnras/sty3142}, \href
  {https://ui.adsabs.harvard.edu/abs/2019MNRAS.484.1007W} {484, 1007}

\bibitem[\protect\citeauthoryear{{Wolz} et~al.,}{{Wolz}
  et~al.}{2022}]{2022MNRAS.510.3495W}
{Wolz} L.,  et~al., 2022, \mn@doi [\mnras] {10.1093/mnras/stab3621}, \href
  {https://ui.adsabs.harvard.edu/abs/2022MNRAS.510.3495W} {510, 3495}

\bibitem[\protect\citeauthoryear{{Wyithe} \& {Loeb}}{{Wyithe} \&
  {Loeb}}{2009}]{2009MNRAS.397.1926W}
{Wyithe} J. S.~B.,  {Loeb} A.,  2009, \mn@doi [\mnras]
  {10.1111/j.1365-2966.2009.15019.x}, \href
  {https://ui.adsabs.harvard.edu/abs/2009MNRAS.397.1926W} {397, 1926}

\bibitem[\protect\citeauthoryear{{Zheng} et~al.,}{{Zheng}
  et~al.}{2005}]{2005ApJ...633..791Z}
{Zheng} Z.,  et~al., 2005, \mn@doi [\apj] {10.1086/466510}, \href
  {https://ui.adsabs.harvard.edu/abs/2005ApJ...633..791Z} {633, 791}

\bibitem[\protect\citeauthoryear{{Zheng} et~al.,}{{Zheng}
  et~al.}{2017}]{2017MNRAS.464.3486Z}
{Zheng} H.,  et~al., 2017, \mn@doi [\mnras] {10.1093/mnras/stw2525}, \href
  {https://ui.adsabs.harvard.edu/abs/2017MNRAS.464.3486Z} {464, 3486}

\bibitem[\protect\citeauthoryear{Zonca, Singer, Lenz, Reinecke, Rosset, Hivon
  \& Gorski}{Zonca et~al.}{2019}]{Zonca2019}
Zonca A.,  Singer L.,  Lenz D.,  Reinecke M.,  Rosset C.,  Hivon E.,   Gorski
  K.,  2019, \mn@doi [Journal of Open Source Software] {10.21105/joss.01298},
  4, 1298

\bibitem[\protect\citeauthoryear{{d'Amico}, {Gleyzes}, {Kokron}, {Markovic},
  {Senatore}, {Zhang}, {Beutler}  \& {Gil-Mar{\'\i}n}}{{d'Amico}
  et~al.}{2020}]{2020JCAP...05..005D}
{d'Amico} G.,  {Gleyzes} J.,  {Kokron} N.,  {Markovic} K.,  {Senatore} L.,
  {Zhang} P.,  {Beutler} F.,   {Gil-Mar{\'\i}n} H.,  2020, \mn@doi [\jcap]
  {10.1088/1475-7516/2020/05/005}, \href
  {https://ui.adsabs.harvard.edu/abs/2020JCAP...05..005D} {2020, 005}

\makeatother
\end{thebibliography}




\appendix

\section{Brightness Temperature Power Spectrum}
\label{sec:apdxps}
In the section we present a detailed derivation of Eq. (\ref{eq:mab}). We start first with the Dirac $\delta-$function in comoving space for transverse and line of sight directions. We define Dirac $\delta-$function $\tilde{\delta}_D(k_\parallel)$ along the $k_\parallel$ axis such that
\begin{equation}
    \Delta X \int \frac{{\rm d}k}{2\pi} \tilde{\delta}_{\rm D}(k-k_\parallel) \tilde{f}(k) = \tilde{f}(k_\parallel)
\end{equation}
for arbitrary function $\tilde{f}$, where $\Delta X$ is the line-of-sight length of the survey volume and can be written as $\Delta X = Y \delta f N_{\rm ch}$ with $\delta f$ being the frequency bandwidth and $N_{\rm ch}$ the number of channels.

The survey volume $\mathbb{V}$ can be written as $\mathbb{V} = \mathbb{S}\cdot \Delta X$ where $\mathbb{S}$ is the survey area. As we will see later, both $\mathbb{V}$ and $\mathbb{S}$ are just abstract quantities to account for the physical units, and will be cancelled out in the final equation. We can then write down the definition for 2D Dirac delta function:
\begin{equation}
    \mathbb{S} \int \frac{{\rm d}^2k_\perp}{(2\pi)^2} \tilde{\delta}^2_{\rm D}(\bm{k}_\perp-\bm{k}'_\perp) \tilde{f}(\bm{k}_\perp) = \tilde{f}(\bm{k}'_\perp)
\end{equation}
for arbitrary function $\tilde{f}(\bm{k_\perp})$.

For a frequency independent beam response $A(l,m,f) \equiv A(l,m)$, we can write its Fourier transform as
\begin{equation}
    \tilde{A}(\bm{k}) = \frac{X^2 \Delta X}{\mathbb{V}} \tilde{\delta}_{\rm D}(k_\parallel) \tilde{A}_\perp \Big(\bm{k}_\perp\Big).
\label{eq:fbeam2d}
\end{equation}

Using the above equation we can express Eq. (\ref{eq:psdelay}) as 2-D integrals:
\begin{equation}
\begin{split}
P_{\rm d} =&\Big(\frac{2k_B}{\lambda^2}\Big)^2\frac{\mathbb{V}^4}{X^4Y^2}\int \frac{{\rm d}^3k'}{(2\pi)^3}\frac{{\rm d}^3k''}{(2\pi)^3} \tilde{A}(\bm{k}-\bm{k}')\tilde{A}^{*}(\bm{k}-\bm{k}'')\\
&\times\langle\tilde{T}(\bm{k}')\tilde{T}^{*}(\bm{k}'')\rangle\\
=& \Big(\frac{2k_B}{\lambda^2}\Big)^2\frac{\mathbb{V}^2}{Y^2} \int \frac{{\rm d}^2k'_{\perp}}{(2\pi)^2}\frac{{\rm d}^2k''_{\perp}}{(2\pi)^2} \tilde{A}_\perp \Big(\bm{k}_\perp -\bm{k}'_\perp\Big) \tilde{A}^*_\perp\Big(\bm{k}_\perp-\bm{k}''_\perp\Big)\\
&\times \langle\tilde{T}(\bm{k}'_\perp,k_\parallel)\tilde{T}^{*}(\bm{k}''_\perp,k_\parallel)\rangle
\end{split}
\end{equation}

The brightness power spectrum can be written as
\begin{equation}
    \mathbb{V} \langle\tilde{T}(\bm{k}'_\perp,k_\parallel)\tilde{T}^{*}(\bm{k}''_\perp,k_\parallel)\rangle = \tilde{\delta}^2_{\rm D}(\bm{k}'_\perp-\bm{k}''_\perp) P_{\rm T}(\bm{k}'_\perp,k_\parallel),
\end{equation}
which leads to
\begin{equation}
\begin{split}
    P_{\rm d} =& \Big(\frac{2k_B}{\lambda^2}\Big)^2\frac{\mathbb{V}}{Y^2} \int \frac{{\rm d}^2k'_{\perp}}{(2\pi)^2}\frac{{\rm d}^2k''_{\perp}}{(2\pi)^2} \tilde{A}_\perp \Big(\bm{k}_\perp -\bm{k}'_\perp\Big) \tilde{A}^*_\perp\Big(\bm{k}_\perp-\bm{k}''_\perp\Big)\\
    &\times \tilde{\delta}^2_{\rm D}(\bm{k}'_\perp-\bm{k}''_\perp) P_{\rm T}(\bm{k}'_\perp,k_\parallel)\\
    = & \Big(\frac{2k_B}{\lambda^2}\Big)^2\frac{\mathbb{V}}{\mathbb{S}Y^2}\int \frac{{\rm d}^2k'_{\perp}}{(2\pi)^2}\bigg|\tilde{A}_\perp \Big(\bm{k}_\perp -\bm{k}'_\perp\Big)\bigg|^2 P_{\rm T}(\bm{k}'_\perp,k_\parallel)\\
    = & \Big(\frac{2k_B}{\lambda^2}\Big)^2\frac{\Delta X}{Y^2}\int \frac{{\rm d}^2k'_{\perp}}{(2\pi)^2}\bigg|\tilde{A}_\perp \Big(\bm{k}_\perp -\bm{k}'_\perp\Big)\bigg|^2 P_{\rm T}(\bm{k}'_\perp,k_\parallel).
\end{split}
\label{eq:pdperp}
\end{equation}

{Equation \ref{eq:pdperp} shows that the delay power spectrum mixes different modes of $P_{\rm T}$ through the beam response. For wide-FOV arrays, the large beam corresponds to a narrow Fourier pair $\tilde{A}$ so that $P_{\rm T}$ can be extracted out of the integral. Alternatively, one can also deconvolve the beam response together with the w-projection kernel \citep{2008ISTSP...2..647C}, as shown for example in \cite{2016ApJ...818..139T}. Here, we are dealing with dish arrays with narrow beams so we take the mode-mixing effects of beam response into account explicitly later in this section while avoiding the computationally-consuming deconvolution.}

Recalling Eq. (\ref{eq:2dvisps}), we can substitute the $|\tilde{\mathbf{V}}^i|^2$ with $P_{\rm d}$ in the above equation to get 
\begin{equation}
\begin{split}
    &\hat{p}^{\rm d}_\alpha =\frac{\sum_i \chi_\alpha(\bm{u}_i,\eta_i) |\tilde{\mathbf{V}}^i|^2}{\sum_i \chi_\alpha(\bm{u}_i,\eta_i)}\\
    &= \Big(\frac{2k_B}{\lambda^2}\Big)^2\frac{\Delta X}{Y^2\sum_i \chi^{i}_\alpha}\sum_i \chi^{i}_\alpha \int \frac{{\rm d}^2k_{\perp}}{(2\pi)^2}\bigg|\tilde{A}_\perp \Big(\bm{k}_\perp^i -\bm{k}_\perp\Big)\bigg|^2 P_{\rm T}(\bm{k}_\perp,k_\parallel^i),
\end{split}
\end{equation}
where $\bm{k}_\perp^i = 2\pi \bm{u}_i/X$ and $i$ loops over all Fourier transformed $u$-$v$ grids. Now recalling Eq. (\ref{eq:tbandps}), we can further expand the above equation
\begin{equation}
\begin{split}
    \hat{p}^{\rm d}_\alpha = &\Big(\frac{2k_B}{\lambda^2}\Big)^2\frac{\Delta X}{Y^2\sum_i \chi_\alpha^i} \sum_i \int \frac{{\rm d}^2\bm{k}_{\perp}}{(2\pi)^2} \chi_\alpha^i\,\bigg|\tilde{A}_\perp \Big(\bm{k}_\perp^i -\bm{k}_\perp\Big)\bigg|^2\\
    &\times \sum_\beta \chi_\beta^i \hat{P}_{\rm T}\Big(|\bm{k}_\perp|_\beta,k_\parallel^i\Big)\\
    = & \big(\sum_{\beta}\big) \mathcal{M}_{\alpha\beta}\,\hat{p}^{\rm T}_\beta.
\end{split}
\end{equation}
Finally, using $\Delta X = Y\delta f N_{\rm ch}$, we show that
\begin{equation}
\begin{split}
    \big(\mathcal{M}\big)_{\alpha\beta}  = & \Big(\frac{2k_{\rm B}}{\lambda^2}\Big)^2\frac{N_{\rm ch}\,\delta f}{Y\sum_i \chi_\alpha(\bm{k}_i)} \sum_i  \int \frac{{\rm d^2}\bm{k}_\perp}{(2\pi)^2} \chi_\alpha(\bm{k}_i) \\&\times \Big|\tilde{A}_\perp \Big(\bm{k}^i_\perp-\bm{k}_\perp\Big)\Big|^2\,\chi_\beta(\bm{k}_\perp,k_\parallel^i),
\end{split}
\end{equation}
which is Eq. (\ref{eq:mab}).
Note that the above equation depends on the summation of each baseline $i$. It takes into account of the sampling of $u$-$v$ plane assuming equal data weights for each baseline, similar to ``natural weighting'' in interferometric imaging. One can also define the selection function $\chi_{\alpha,\beta}$ differently to change the weighting. Therefore, the matrix formalism can effectively deconvolves the primary beam response and point spread function simultaneously.

\begin{figure}
    \centering
    \includegraphics[width=\linewidth]{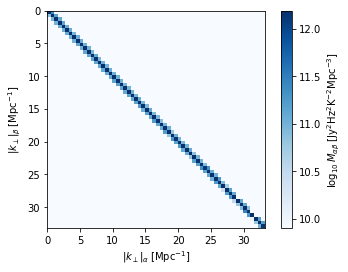}
    \caption{The mode-mixing matrix $\mathcal{M}_{\alpha\beta}$ used in this paper calculated according to Eq. (\ref{eq:mab}) with the choice of annulus $u$-$v$ bins mentioned in Section \ref{sec:estform}. Values below $10^{10} {\rm Jy^2 Hz^2 K^{-2}Mpc^{-3}}$ are masked for better presentation. }
    \label{fig:mab}
\end{figure}

As mentioned in Section \ref{sec:skysim}, we grid $u$-$v$ plane into annulus bins with the edges of the bin being $[0,100,200,...,6000]$. The resulting $\mathcal{M}_{\alpha\beta}$ is shown in Figure \ref{fig:mab}. As shown, the mixture of different angular modes due to the size of the beam mainly affects bins near the diagonal, with the width of $\Delta |\bm{k}_\perp| \sim 0.5\,{\rm Mpc^{-1}}$.

\section{Converting \althi\ mass to flux density}
\label{sec:hiflux}
In this section, we briefly derive the conversion between \hi\ mass to flux density for cosmological simulations.

Suppose we grid the sky with equal area pixels, each with an area of $\Omega_{\rm pix}$ (this area is purely pedagogical). In one frequency channel with comoving distance $X$ and line-of-sight scale of $\Delta X$, we have
\begin{equation}
    T_{\rm \hi}^{\rm pix} = \sum_i \frac{C_{\rm HI}^i M_{\rm \hi}^i}{V_{\rm pix}},\; V_{\rm pix} = X^2 \Omega_{\rm pix} \Delta X,
\end{equation}
where $i$ loops over the \hi\ sources within the pixel.

The flux density from one pixel can then be calculated as
\begin{equation}
    I_{\rm pix} = \sum_i \frac{C_{\rm HI}^i M_{\rm \hi}^i}{X^2 \Omega_{\rm pix} \Delta X} \frac{2 k_{\rm B}}{\lambda_i^2} \Omega_{\rm pix} = \sum_i \frac{2 k_{\rm B}}{\lambda_i^2} \frac{C_{\rm HI}^i M_{\rm \hi}^i}{X^2 \Delta X},
\end{equation}
and the flux density of each source is
\begin{equation}
    I_{i} =  \frac{2 k_{\rm B}}{\lambda_i^2} \frac{C_{\rm HI}^i M_{\rm \hi}^i}{X^2 \Delta X}.
\label{eq:mass2flux}
\end{equation}

In reality, the flux density of the sources depends on the peculiar velocity (see e.g. \citealt{2017PASA...34...52M}). Our result instead depends on the bandwidth of the frequency channel. It is assumed that the frequency displacement caused by peculiar velocity does not misplace \hi\ galaxies into other frequency bins and that the width of the emission profile is negligible comparing to the frequency channel bandwidth. 


\bsp	
\label{lastpage}
\end{document}